\documentclass[
  journal=largetwo,
  manuscript=research-paper,
  style=apa
  year=2023,
  volume=xx,
]{cup-journal}

\usepackage{amsmath}
\usepackage[nopatch]{microtype}
\usepackage{booktabs}
\usepackage{comment}
\usepackage{longtable}
\usepackage{supertabular}
\usepackage{setspace}
\usepackage[euler]{textgreek}
\usepackage{graphicx}
\usepackage{amsmath,amssymb}
\usepackage{fancyhdr}
\usepackage{placeins}

\title{On More than Two Decades of Celestial Reference Frame VLBI Observations in the Deep South: IVS-CRDS (1995 - 2021)}

\author{S. Weston}
\affiliation{Institute for Radio Astronomy \& Space Research (IRASR)
School of Engineering, Computer and Mathematical Sciences
Faculty of Design and Creative Technologies
Auckland University of Technology, New Zealand.}
\alsoaffiliation{Joint first authors}
\email[S. Weston]{sweston@aut.ac.nz}

\author{A. de Witt}
\affiliation{The South African Radio Astronomy Observatory, South Africa}
\alsoaffiliation{Joint first authors}
\author{Hana Kr\'asn\'a}
\affiliation{Technische Universit\"at Wien (TU Wien),
Department f\"ur Geod\"asie und Geoinformation, Wiedner Hauptstraße 8, A-1040 Wien, Austria}

\author{Karine Le Bail}
\affiliation{NVI, Inc. at NASA Goddard Space Flight Center, Code 61A, Greenbelt, MD 20771, USA}
\alsoaffiliation{Chalmers University of Technology/Onsala Space Observatory, 439 92 Onsala, Sweden}

\author{Sara Hardin}
\affiliation{U.S. Naval Observatory, Washington, DC 20392, USA}

\author{David Gordon}
\affiliation{U.S. Naval Observatory, Washington, DC 20392, USA}

\author{Shu Fengchun}
\affiliation{Shanghai Astronomical Observatory, Chinese Academy of Sciences, 80 Nandan Road, Shanghai 200030, China}

\author{Alan Fey}
\affiliation{U.S. Naval Observatory, Washington, DC 20392, USA}
\alsoaffiliation{Chalmers University of Technology/Onsala Space Observatory, 439 92 Onsala, Sweden}

\author{Matthias Schartner}
\affiliation{ETH Z\"urich, Institute of Geodesy and Photogrammetry, Zurich, Switzerland}

\author{Sayan Basu}
\affiliation{The South African Radio Astronomy Observatory, South Africa}

\author{Oleg Titov}
\affiliation{Geoscience Australia, PO Box 378, Canberra, ACT 2601, Australia}

\author{Dirk Behrend}
\affiliation{NVI, Inc. at NASA Goddard Space Flight Center, Code 61A, Greenbelt, MD 20771, USA}

\author{Christopher S. Jacobs}
\affiliation{Jet Propulsion Laboratory, California Institute of Technology, 
4800 Oak Grove Dr., Pasadena CA}

\author{Warren Hankey}
\affiliation{University of Tasmania, Hobart, Australia}

\author{Federico Salguero}
\affiliation{National Scientific and Technical Research Council, Argentina, (CONICET)}

\author{John E. Reynolds}
\affiliation{CSIRO Space and Astronomy, Australia Telescope National Facility, P.O. Box 76, Epping, NSW 1710, Australia}

\keywords{VLBI, IVS, Geodesy} %% First letter not capped

\begin{document}

\begin{abstract}
The International VLBI Service for Geodesy \& Astrometry (IVS) regularly provides high-quality data to produce Earth Orientation Parameters (EOP), and for the maintenance and realization of the International Terrestrial and Celestial Reference Frames, ITRF and ICRF. The first iteration of the celestial reference frame (CRF) at radio wavelengths, the \mbox{ICRF{\small 1}}, was adopted by the International Astronomical Union (IAU) in 1997 to replace the FK5 optical frame. Soon after, the IVS began official operations and in 2009 there was a significant increase in data sufficient to warrant a second iteration of the CRF, \mbox{ICRF{\small 2}}. The most recent \mbox{ICRF{\small 3}}, was adopted by the IAU in 2018. However, due to the geographic distribution of observing stations being concentrated in the Northern hemisphere, CRFs are generally weaker in the South due to there being fewer Southern Hemisphere observations. To increase the Southern Hemisphere observations, and the density, precision of the sources, a series of deep South observing sessions was initiated in 1995. This initiative in 2004 became the IVS Celestial Reference Frame Deep South \mbox{(IVS-CRDS)} observing program. This paper covers the evolution of the CRDS observing program for the period 1995 to 2021, details the data products and results, and concludes with a summary of upcoming improvements to this ongoing project.
%\footnote{Copyright \textcopyright 2023, All Rights Reserved.}
\end{abstract}

%===================================================================
\setcounter{secnumdepth}{3}
\section{Introduction}

%DB : Perhaps add a short history of the ICRF, that is, the three incarnations including when they became effective.

Very Long Baseline Interferometry (VLBI) is a collaborative and cooperative endeavour involving many individuals and institutions around the world. Part of this cooperation and collaboration, using VLBI observations of extragalactic radio sources to conduct geodesy and astrometry, has been formalised under the umbrella of the International VLBI Service for Geodesy and Astrometry %\citep[IVS,][] 
\citep{Nothnagel2017}, which has organized this cooperation over the past 23 years. The IVS provides products derived from the analysis of VLBI data to the scientific community, of which some of the main products are:

% https://itrf.ign.fr/en/solutions/ITRF2020
% https://meetingorganizer.copernicus.org/EGU21/EGU21-2056.html
% https://presentations.copernicus.org/EGU21/EGU21-2056_presentation.pdf
\begin{enumerate}
\item Accurate measurements of station positions and velocities which contribute significantly to the realization of the International Terrestrial Reference Frame \citep[ITRF,][]{itrf2020}, in particular to the definition of its scale
\item Accurate measurements of the angular positions of extragalactic radio sources which define and realize the International Celestial Reference Frame \citep[ICRF,][]{Charlot_2020}
\item The five daily Earth Orientation Parameters \citep[EOP,][]{Eubanks1993}, including Earth rotation parameter $UT1-UTC$ and nutation which are provided uniquely, which provides the link between the ICRF and the ITRF
\end{enumerate}

%\begin{itemize}
% \item[\ding{108}] Terrestrial Reference Frame (TRF)
%  \item[\ding{108}] International Celestial Reference Frame (ICRF)
%  \item[\ding{108}] Earth Orientation Parameters (EOP)
%\end{itemize}

The ICRF is a catalogue of positions of extragalactic radio sources with the highest precision, which are crucial to many applications. For example, the ICRF provides sources that are observed in the geodetic VLBI observations conducted by the IVS to measure the orientation of Earth in space and station motions \citep{Hellmers2021}, which in turn contributes to the realization of the ITRF and allows studies of the motion of the tectonic plates and the interior of the Earth \citep{Carter1993}, and measurements of the ionosphere and troposphere which allows for atmospheric studies \citep{Heinkelmann2011J}. High-accuracy celestial reference frames are also crucial for many other applications such as satellite tracking, orbit determination, and deep-space navigation \citep{Thornton2000}, alignment of the planetary ephemerides \citep{Park2021}, studies of special and general relativity \citep{Fomalont2009} such as Galacto-centric acceleration \citep{MacMillan2019}, and for phase-referencing observations in astronomy \citep{Beasley1995, Rioja2020}. 

The first iteration of the ICRF \citep[ICRF{\small 1},][]{Ma1998} was based on positions of 608 extragalactic radio sources, of these a subset of 212 sources uniformly distributed on the sky were identified as "defining sources", and served to define the axes of the reference frame. This iteration used data obtained through VLBI observations at the standard geodetic and astrometric S/X-band (2.3/8.4~GHz) frequencies. The ICRF{\small 1} was adopted by the International Astronomical Union (IAU) as the new fundamental CRF in 1997, and replaced IAU's Fundamental Catalogue FK5 \citep{Fricke1988}, which was based on optical observations of galactic stars. The IVS began operations in 1999, and coordinated continued S/X-band observations and improvements on the original ICRF. As a result a second iteration, the ICRF{\small 2} \citep{Fey2015} with 3414 sources of which 295 are defining sources, was adopted by the IAU in 2009 and officially replaced the ICRF{\small 1} on 01 January 2010. Since the first release of ICRF, the ongoing observing programs run by the IVS together with other specific independent projects have ensured the continued expansion of the VLBI database, and as such a third iteration, the ICRF{\small 3} \citep{Charlot_2020} was adopted by the IAU in 2018 (see Resolution B2\footnote{See Resolution B2 at \url{https://iau.org/static/resolutions/IAU2018_ResolB2_English.pdf}.}) and replaced the ICRF{\small 2} on 01 January 2019. The ICRF{\small 3} is based on nearly 40 years of data and contains S/X positions of 4536 extragalactic sources, including 303 uniformly distributed defining sources. The ICRF{\small 3} has a noise floor of 30~\textmu as, with median position uncertainties at S/X-band of about 127~\textmu as in right ascension and 218~\textmu as in declination -- more than a factor of three improvement over the \mbox{ICRF{\small 2}}. In addition, the ICRF{\small 3} added two complementary catalogues at K-band (24 GHz) and X/Ka-band (32 GHz), making it the first multi-wavelength frame ever realized. Since ICRF{\small 3} there have been many dedicated efforts to further improve the CRF at radio wavelengths, and current CRF solutions \citep[e.g. sx-usno-220422,][]{Gordon2022} already show a remarkable improvement over the \mbox{ICRF{\small 3}} 
\citep{deWitt2022, Gordon2022}.
% \citep{deWitt2022, Gordon2022}.

\begin{figure}[h]
\centering
\includegraphics[width=\textwidth]{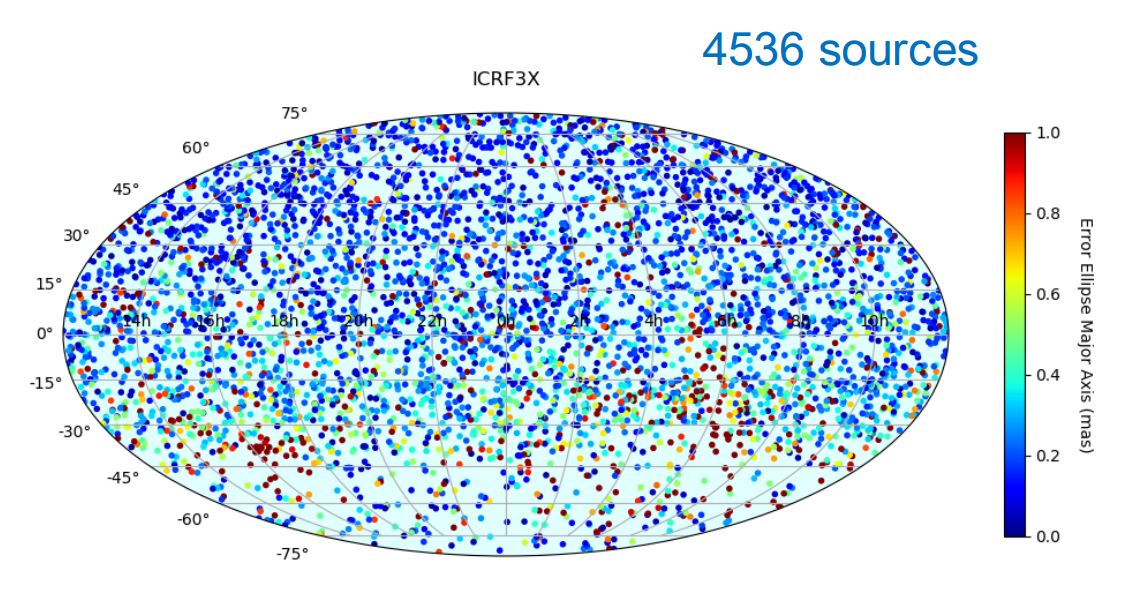}
\caption{The ICRF3-SX (2018) has 4536 sources, but quite clearly the number of sources become sparse below a declination of $-30^{\circ}$ \citep{Charlot_2020}.}
\label{fig:icrf3-sx}
\end{figure}

\begin{figure}[hbt!]
\centering
%trim=left botm right top
  \includegraphics[trim=2.4cm 2.5cm 6cm 6cm,clip,width=0.99\textwidth]{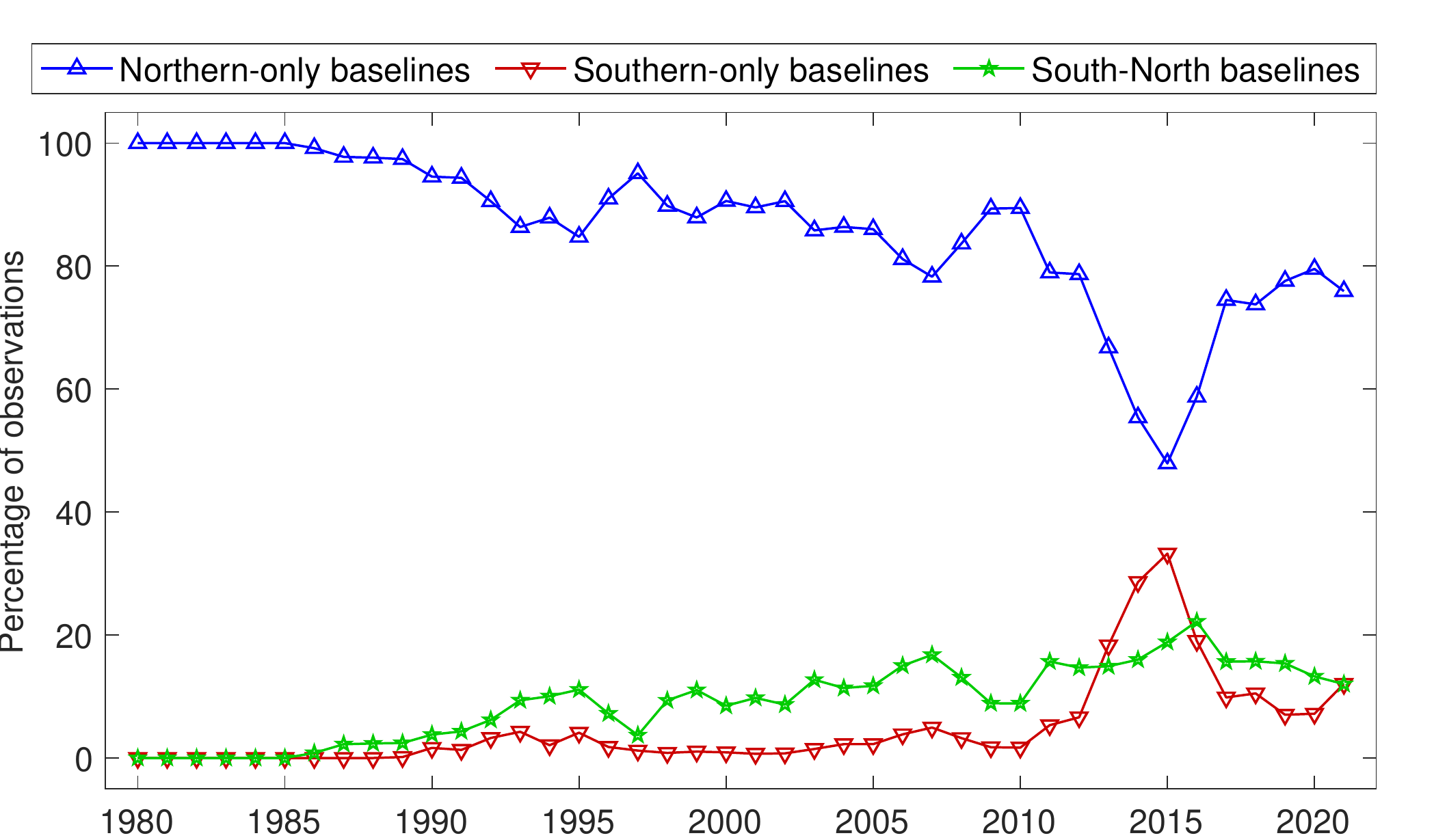}
  \hspace{1cm}
\\
  \includegraphics[trim=2.4cm 2.5cm 6cm 6cm,clip,width=0.99\textwidth]{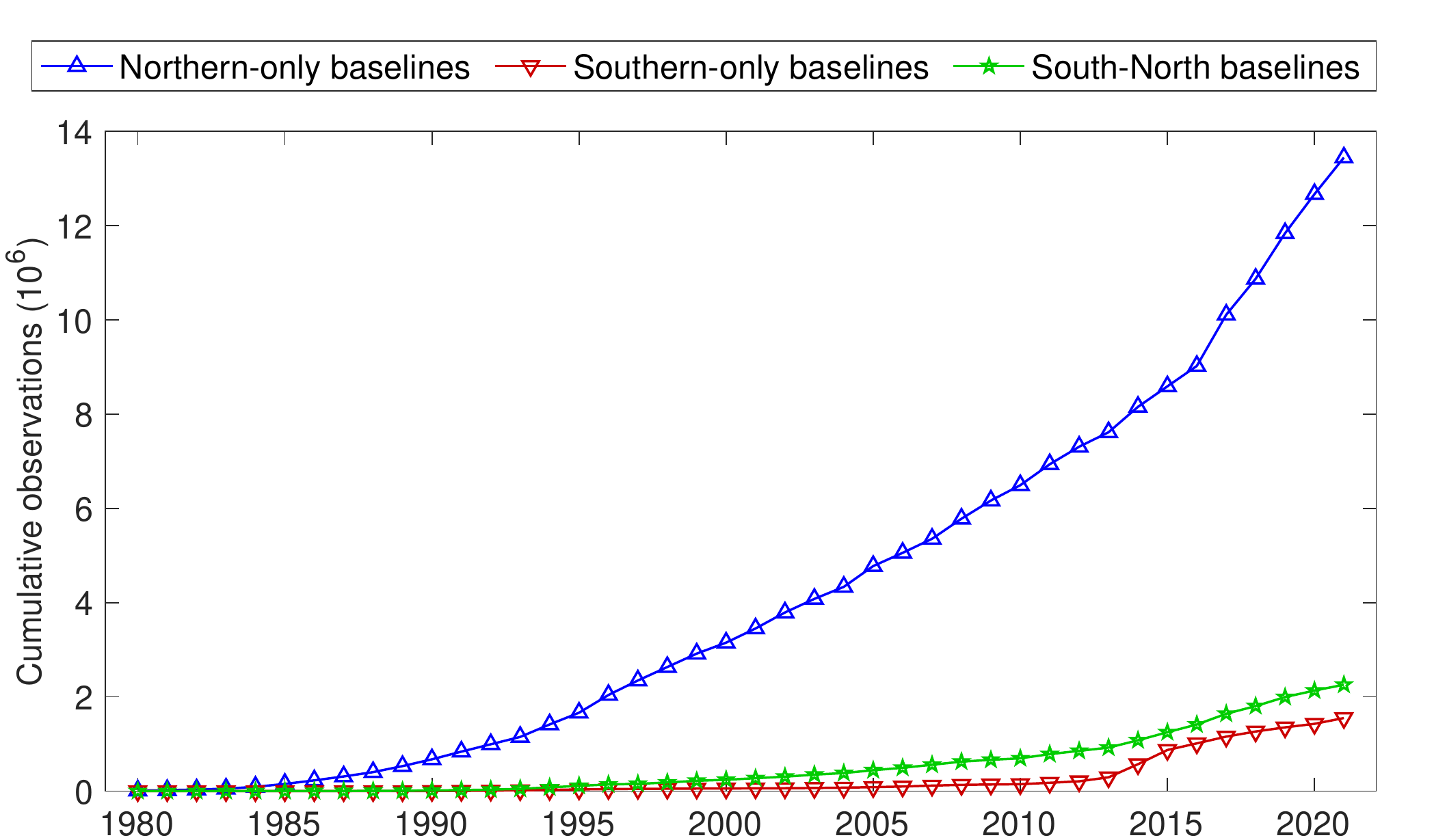}
\caption{The plot on the top shows the percentage of astrometric and geodetic VLBI observations for the period April 1980 to June 2021 on Northern only (blue), Southern only (red), and North-South (green) baselines. The  plot on the bottom shows the cumulative totals for the same baselines over the same period.}
\label{fig:crds-South-vs-North-baselines}
\end{figure}

Catalogues of compact radio sources, including the \mbox{ICRF{\small 3}}, are generally weaker in the South by factors of 2 or more in both density and precision \citep{deWitt2022}. This is because the stations contributing to geodetic and astrometric VLBI observations have an uneven distribution between the hemispheres, currently $\sim$~80$\%$ of the stations are in the Northern Hemisphere and $\sim$~20$\%$ of the stations are in the Southern Hemisphere. As a result, and despite the significant improvements over its predecessor, the ICRF{\small 3} still has deficiencies by factors of 2--3 in the South \citep{Charlot2018, Charlot_2020}. In addition, the deficit of long North-South baselines in the ICRF{\small 3} contributes to the Declination (Dec) precision being a factor of two or more worse than the Right Ascension (RA) precision. The distribution of the \mbox{ICRF{\small 3}} S/X-band sources on a projection of the celestial sphere clearly shows the decrease in the number of sources and the number of observations per source for sources below $-30^{\circ}$ Dec, and shows that Southern sources have generally less precise positions 
\citep{Charlot_2020} (see Figure~6 reproduced here in Figure \ref{fig:icrf3-sx}\footnote{\url{https://www.aanda.org/articles/aa/full_html/2020/12/aa38368-20/F6.html}}). When looking at a world map of all 167 antennas (situated on 126 different sites) that participated in the observations used for ICRF{\small 3} 
 (see Figure~1\footnote{\url{https://www.aanda.org/articles/aa/full_html/2020/12/aa38368-20/F1.html}} in \cite{Charlot_2020}), the uneven distribution of antennas between the hemispheres is very apparent. Of those 167 antennas, only 14 antennas contributed from the Southern Hemisphere. The plots in Figure \ref{fig:crds-South-vs-North-baselines} show the number of observations for North-only, South-only and North-South baselines for all astrometric and geodetic VLBI observations between April 1980 and June 2021. The plot clearly shows the disparity between the number of Northern and Southern observations, with only $\sim$~20\% of the observations coming from North-South baselines and only $\sim$~10\% from all Southern baselines. 

%%\footnote{\url{https://www.aanda.org/articles/aa/full_html/2020/12/aa38368-20/F6.html}}

%\input{icrf3-sx}
% The Third Realization of the International Celestial Reference Frame
% https://www.iau.org/static/science/scientific_bodies/divisions/a/2018/Charlot.pdf
% Left Hand Figure from Slide 11
% Also : https://www.aanda.org/articles/aa/pdf/2020/12/aa38368-20.pdf

%\begin{figure}[hbt!]
%\begin{figure}[h]
%\centering
%\includegraphics[width=\textwidth]{icrf3x_4536_sources.jpg}
%\includegraphics{example-image.pdf}
%\caption{The ICRF3-SX (2018) has 4536 sources, but quite clearly the sources used become sparse below latitude.}
%\label{fig:icrf3-sx}
%\end{figure}
%\caption{The ICRF3-SX (2018) has 4536 sources, but quite clearly the sources used become sparse below latitude $-30^{\circ}$\citep{Charlot_2020}.}

% From Karine Lebail 2/11/2021
% These plots correspond to the latest quaterly global solution of David
% Gordon at USNO. It contains 6666 sessions (David did not consider the 1979
% sessions in his solution), including 20 VGOS sessions. First data base:
% 80APR11XC, latest data base: 21JUN09VG.
%
% David took a look at the plots. He explains the increase in Southern
% baselines in 2013, 2014 and 2015 with the AUST and AUSTRAL sessions, and the
% decrease around 2008-2010 with Hartrao being down.

The Southern Hemisphere also suffers from a lack of dedicated imaging campaigns to map and monitor the structure of individual ICRF sources. Many ICRF sources exhibit spatially extended intrinsic structure that may vary with time and frequency. It is well known that source structure can introduce significant errors in the geodetic and astrometric VLBI delay measurements and instabilities in the individual source positions \citep{Charlot1990, Gattano2018, Xu2019}. It is therefore important to map the structures of these sources on a regular basis to assess the astrometric source suitability and to track the source structure variations over time. There have, however, only been a few imaging sessions of ICRF sources in the South \citep{Hungwe2011, Petrov2019} and dedicated campaigns to map and monitor source structure have proven difficult to obtain. Nevertheless, recent investigations to image source structure from existing astrometric and geodetic observations in the South have shown that dedicated imaging campaigns, as are done in the North, may indeed be possible \citep{Basu2016,Basu2021}.
%\citep[e.g.,][]{Basu2016, Basu2021}.

Since its inception, the IVS has initiated and coordinated many observing programs intended to provide astrometric observations useful in improving and densifying the CRF. Dedicated Southern Hemisphere astrometric programs proved most successful in providing and maintaining the fundamental Southern Hemisphere reference frame for the ICRF. In particular, the Celestial Reference Frame Deep South (CRF-DS) observing sessions were initiated in 1995 and became the IVS-CRDS observing program in 2004 (henceforth called CRDS). The CRDS program aims to provide astrometric VLBI observations of sources in the deep South for improving and densifying the ICRF, and in recent years it has also contributed to the imaging and monitoring of source structure. 
%Add average number of stations and number per year sessions since 2010

This paper provides the details and evolution of the CRDS observing program for the period June 1995 to December 2021, as well as details on the data products and results. Section \ref{timeline} provides a timeline overview of the evolution and major developments of the CRDS observing program from the early sessions that started in June 1995 up to December 2021. In Section \ref{Methods} we discuss the methods employed in detail. This consists of the the participating networks and stations (\S~\ref{observations}),  source selection (\S~\ref{SourceSelection}) and scheduling strategies (\S~\ref{scheduling}). Details of the data correlation, analysis and data products are provided in \S~\ref{Correlation}. A discussion about the CRDS observing program results to date and remarks are given in Section \ref{results}. This consists of the astrometric results (\S~\ref{astr_results}) and imaging results (\S~\ref{img_results}), and finally an overview of the network performance (\S~\ref{performance}). A final discussion of the CRDS program to date and future plans to improve the observing program to further densify and strengthen the Southern CRF are discussed in Section \ref{DiscussionFuture}.

%===================================================================

\section{Detailed CRDS Observing Timeline}
\label{timeline}

This section provides an overview of the evolution and major developments of the CRDS observing program from the early sessions that started in June 1995 up to December 2021. Details include the network coverage and participating stations over this period, as well as any improvements and changes  in the frequency sequence, data rates, and scheduling strategies.

\subsection{From June 1995 (CRF-DS1) to December 2010 (CRDS49)}
The CRDS sessions started with two sessions each in 1995 and 1996. The first 9 sessions in this initial period were called CRF-DS1 to CRF-DS9. There was a gap of 7 years until 2003 then some 45 sessions were run between 2003 and 2008 (CRF-DS5 to CRDS49). These early sessions were scheduled by the United States Naval Observatory (USNO). Most were single-baseline sessions between the HartRAO 26-antenna in Hartebeesthoek, South Africa and the Hobart 26-m antenna in Tasmania, Australia, but a few included other stations, such as the Parkes 64-m antenna in New South Wales, Australia and the Deep Space Station 45 (DSS45) in Tidbinbilla, Australia. These early sessions made significant contributions to ICRF{\small 2}, being almost the only sessions to add new Southern sources. There were no CRDS sessions in 2009 or 2010 because the HartRAO 26-m antenna suffered a major bearing failure and was down for more than two years for repairs.

\subsection{From March 2011 (CRDS50) to May 2017 (CRDS88)}
In the late 2000s a group of three new 12-m VLBI antennas (Yarragadee 12-m, Katherine 12-m and Hobart 12-m), called the AuScope array \citep{Lovell2013}, were commissioned by the University of Tasmania in Australia. An identical antenna, the Warkworth 12-m, was also commissioned in Warkworth, New Zealand \citep{wark12m}. Between July 2011 and November 2012 (CRDS52-61), one or two at a time of these 12-m antennas were scheduled in the CRDS sessions. The two sessions before July 2011 (CRD50-51) consisted of single-baseline observations between the HartRAO 26-m and Hobart 26-m antenna. In 2013 all four 12-m antennas were added to the CRDS observing program. Very importantly, this increased the number of regular CRDS stations from 2 to 6, and also allowed a greater cadence of source observing within a \mbox{24-hour} geodetic or astrometric VLBI session due to the much faster slew speeds of the 12-m antennas. Between January 2013 and May 2017, twenty-six CRDS sessions were observed (CRDS62-CRDS88), of which thirteen had the full six-station network, seven had five stations participating, and six sessions observed with only 4 stations. The CRDS53 and CRDS61 sessions also included the Parkes 64-m antenna, and the CRDS58 session included the Deep Space Station 43 (DSS43) \mbox{70-m} antenna in Tidbinbilla, Australia. Up to CRDS65 the data recording rate was 128~Mbps, but in July 2013 (CRDS-66) the recording rate was increased to 256~Mbps.

Imaging of sources from CRDS sessions data started in January 2013 (CRDS-63), to analyse source structure and variability, and to update flux catalogues in the South (see \S~\ref{img_results}).

%Imaging has now been completed for 9 CRDS-sessions (63, 66, 68, 94-97, 100-102).

%From there you will see that source 0302-623 has some extended structure and the structure index that we get from this image is 3.5, which is rather high. This was an ICRF{\small 2} defining source, but it was NOT chosen as an ICRF{\small 3} defining source because of the source structure.

% The image that I sent you is from CRD100 (2019-02-18) where the source was observed with 5 stations but in only 2 scans. We are trying to improve the $u,v$-coverage for imaging and in CRD102 (2019-0507) the source was observed with 6 stations and in 6 scans. We have not imaged this session yet but the $u,v$-coverage is much better.

\subsection{From July 2017 (CRDS89) to March 2019 (CRD101)}
To be fully compliant with the VLBI Global Observing System \citep[VGOS,][]{Petrachenko2013}, the Hobart \mbox{12-m} antenna was upgraded with a new wide-band cooled receiver (2-14 GHz), and in July 2017 (CRDS-89) it was removed from the CRDS program. Between July 2017 and March 2019 (CRD101), thirteen CRDS sessions were observed of which eleven included all of the remaining five stations.  The CRDS94 session also included the HartRAO 15-m antenna. In 2017, a Southern Hemisphere astrometry group, later formalised under the IVS as the Southern VLBI Operations Centre \citep[SVOC,][]{39thIVSBoardMeeting, deWitt2019} was formed, with the aim to improve and densify the S/X-band CRF in the South. This group adopted the following measures over this period to improve the final data products:
% nb station codes changed to CRD versus CRDS
\begin{itemize}
  \item[1.] Increased sensitivity of Southern sessions for detection of weaker sources down to $\approx350$ mJy or less.
  \item[2.] Increased the data rate of Southern sessions by a factor of 8, from 128~Mbps to 1~Gbps
  \item[3.] Optimized scheduling of Southern sessions to allow for simultaneous astrometric and imaging observations
  \item[4.] Included mapping and monitoring of source structure in analysis  to quantify non-point-like structure 
  \item[5.] Improved precision of Southern source positions by a factor of 2.5
  \item[6.] Expanded the Southern source list by a factor of 2
  \item[7.] Improved the sky coverage of Southern sources
  \item[8.] Improved overlap with higher frequency radio CRFs and Gaia optical CRF
\end{itemize} 

As a result, many improvements were made to the CRDS sessions \citep{deWitt2019b}. 
From January 2018 (CRDS93), the data recording rate was increased from 256~Mbps to 1~Gbps and the frequency sequence was optimized to avoid radio frequency interference (RFI, see \S~\ref{observations}) in particular at S-band. In addition the scheduling was improved and the source list was expanded (see \S~\ref{SourceSelection}) with optimisation for both astrometry and imaging (see \S~\ref{results}).

\subsection{From May 2019 (CRD102) to July 2021 (CRD113)}
In May 2019 (CRD102) the O'Higgins 9-m antenna in Antarctica was added to the CRDS program, and in June 2019 (CRD103), the Katherine 12-m antenna was upgraded for VGOS observations and removed. In February 2020 (CRD105) the Aggo 6-m antenna in Argentina was added to the CRDS network. Between February 2020 (CRD105) and May 2021 (CRDS112), eight CRDS sessions were observed, of which half had five participating stations, and the other half had six stations. In July 2021 (CRD113) the Hobart 26-m antenna was temporarily removed from the CRDS observing program due a bearing failure, and remains out of service at the time of writing this paper. Subsequently, the CRDS113 session was observed with only four stations. The loss of the Hobart 26-m antenna, being one of only two large antennas in the CRDS network (the other being the HartRAO 26-m), has severely impacted the success rate of the CRDS sessions (see \S~\ref{results}).

\subsection{From August 2021 (CRD114) to December 2021 (CRD116)}
In 2020 the IVS-CRF Committee\footnote{https://ivscc.gsfc.nasa.gov/about/com/crfc/index.html} was formalised, also incorporating the SVOC, to make recommendations to the IVS Directing Board on observing programs and strategies for the S/X-band CRF \citep{Behrend2019, Behrend2020}. The charter\footnote{https://ivscc.gsfc.nasa.gov/about/com/crfc/crfc\_charter.pdf} for the IVS-CRF Committee was accepted by the IVS Directing Board on 14 October 2020. In 2021 the IVS-CRF Committee made recommendations to add the 25-m Very Long Baseline (VLBA) antennas in Mauna Kea (VLBA-MK) and Saint Croix (VLBA-SC), as well as the 50-m Kunming antenna in China, to the CRDS observing program. In August 2021 (CRD114) the two VLBA antennas were added to the CRDS schedule forming a seven-station network, and in October and December 2021 (CRD115 and CRD116) the 50-m Kunming antenna was also added forming an eight-station network. Unfortunately, the VLBA stations observed with a wrong LO value for S-band channels and the baselines between the VLBA stations and other antennas could not be correlated. Since then, test observations with a revised S-band setup for the VLBA stations were successful, and VLBA-MK and VLBA-SC will be added to future CRDS sessions.

%===================================================================

\section{Methods}
\label{Methods}

This section details the methods employed in the CRDS program, starting with the setup of the observing network, source selection, scheduling strategies, correlation and finally the analysis undertaken to produce the final data products.
% DB : Is this an overview table of the 61 sessions including the observing dates, recording rate, number of stations, scheduler, perhaps type of correlator?
%Table - session, antennas, nu scans \& nu sources
%Plot - date, nu scans \& nu sources, to show when improvement took place
% Multi-epoch VLBI images to study the ICRF{\small 3} Defining Sources in the Southern Hemisphere
% https://ui.adsabs.harvard.edu/abs/2018evn..confE.135B/abstract

\subsection{Observing Networks and Setup}
\label{observations}

\begin{figure*}[hbt!]               
  \centering
  \includegraphics[width=\textwidth]{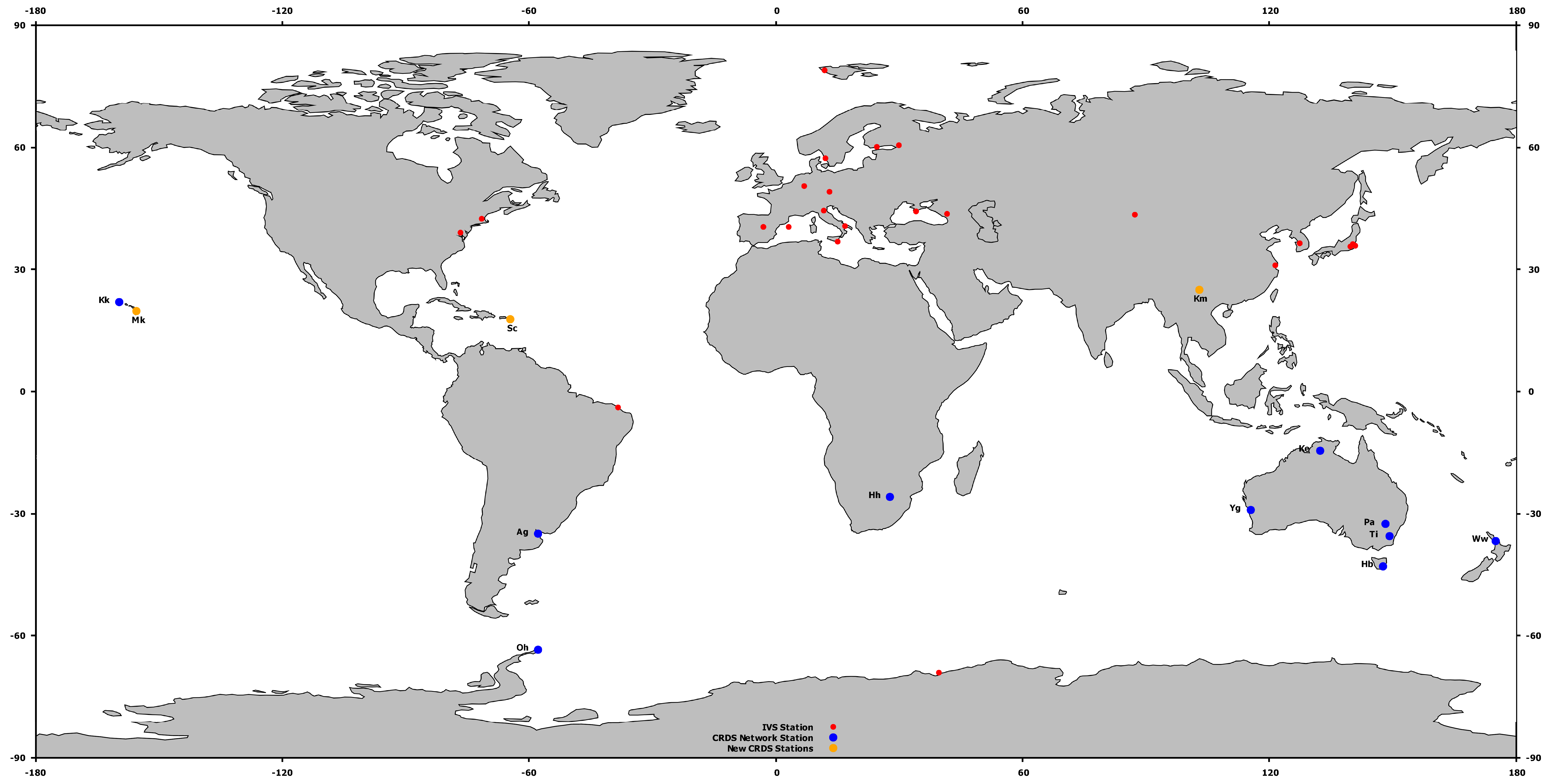}
  \caption{Network map showing IVS and cooperating stations (in red) and stations that participated in the CRDS program during the last decade (in blue and orange). Hobart hosts two stations separated by a few hundred metres, so Hb on this figure represents both Hb and Ho. This is also true for Hartebeesthoek where Hh represents both Hh and Ht.}
  \label{fig:crds-network-stations}
\end{figure*}

The early CRDS sessions consisted of mostly single-baseline observations between the HartRAO and Hobart 26-m antennas, occasionally including a few other antennas. From 2011 onward, more stations were added to the CRDS sessions and between July 2011 and March 2019 the CRDS program, for the most part, including the following six stations: HartRAO 26-m (Hh), Hobart 26-m (Ho), Hobart 12-m (Hb), Katherine 12-m (Ke), Yarragadee 12-m (Yg), and Warkworth 12-m (Ww), recalling that Hb was removed from the CRDS program in July 2017. Two of these sessions also included the Parkes 64-m antenna (Pa), while the HartRAO 15-m antenna (Ht) and DSS43 Tidbinbilla 70-m antenna (Ti) each participated in a single session. In May 2019 the O'Higgins 9-m antenna (Oh) was added to the CRDS program and in June 2019 Ke was removed. In February 2020 the AGGO 6-m antenna (Ag) was added to the CRDS network. In July 2021 (CRDS113) Ho was temporarily removed from the CRDS observing program due to a bearing failure and in October 2021 the 50-m Kunming antenna (Km) was added. In August 2021 the VLBA-MK (Mk) and VLBA-SC (Sc) antennas were added, but the data from these two antennas could not be correlated with other stations due to a mismatch in the S-band frequency setup. Details of the fourteen  IVS VLBI stations that participated in CRDS sessions between July 2011 and December 2021, are listed in Table~\ref{table:stations}. The full IVS station network is shown in Figure~\ref{fig:crds-network-stations}, highlighting the stations that participated in the CRDS sessions during this period. Note that all the stations used for CRDS sessions are IVS stations, except for Ti.

\begin{table*}[hbt!]
\footnotesize
\caption{The eleven participating CRDS VLBI stations, with their geocentric coordinates used in the session schedule files. The last three entries are the stations added very recently.}
\resizebox{0.95\textwidth}{!}{
\begin{tabular}{l|l|l|l}
\toprule
Station & Station & Station Geocentric Coordinate & Solution \\
Code &Name & XYZ (m) & (Epoch)\\
\midrule
Ag & Aggo 6-m, Observatorio Argentina-Alemán de Geodesia (AGGO), Argentina & \texttt{ 2765116.70  -4449233.81  -3626420.56} & 2020c\\
Hb & Hobart 12-m, University of Tasmania, Australia & \texttt{-3949990.58  ~2522421.17  -4311708.24} & 2011b\\
Hh & HartRAO 26-m, Hartebeesthoek, South Africa & \texttt{ 5085442.79  ~2668263.49  -2768697.04} & GLB1069\\
Ho & Hobart 26-m, University of Tasmania, Australia & \texttt{-3950236.74  ~2522347.56  -4311562.55} & GLB1069 \\
Ht & HartRAO 15-m, Hartebeesthoek, South Africa &  \texttt{ 5085490.80  ~2668161.48  -2768692.62} &  2013\\
Ke & Katherine 12-m, University of Tasmania, Australia & \texttt{-4147354.59  ~4581542.40  -1573303.30} & 2011b\\
Oh & O'Higgins 9-m, ERS/VLBI Station O'Higgins, Antarica & \texttt{ 1525833.48  -2432463.71  -5676174.48 } & 2020c\\
Pa & Parkes 64-m, CSIRO, Australia & \texttt{-4554232.05  ~2816758.99 -3454035.88} & GLB1069\\
Ti & 70-m Tidbinbilla (DSS-43), CSIRO, Australia & \texttt{-4460894.39  ~2682361.29  -3674747.98 } & GLB1069\\
Ww & Warkworth 12m, AUT University, New Zealand & \texttt{-5115324.39  ~~477843.30  -3767192.88 } & 2011b \\
Yg & Yarragadee 12m, University of Tasmania, Australia & \texttt{-2388896.06  ~5043349.97  -3078590.92} & 2011b \\
\midrule
Mk & Manua Kea VLBA Station, Hawaii, USA & \texttt{-5464075.28  -2495247.63  ~2148297.61 } & 2020c\\
Sc & St Croix VLBA Station, St Croix, USVI & \texttt{ 2607848.71  -5488069.47  ~1932739.83} & 2020c\\
Km & Kunming 40m, Yunnan Astronomical Observatory, China & \texttt{-1281153.12  ~5640864.43  ~2682653.37} & 2020c\\
\bottomrule
\end{tabular}}
%\belowtable{} % Table Footnotes
\label{table:stations}
\end{table*}

%\input{new_station_table.tex}
%
%
% New Station Mk
%INPUT =      -5464075.287|      -2495247.634|       2148297.611
%OUTPUT= 19 48 04.99806(N)|155 27 19.86003(W)|         3763.0048
% Manua Kea VLBA Station, Hawaii, USA
%--------------------------------------------------------------------------------
% New Station Sc
%INPUT =       2607848.712|      -5488069.478|       1932739.836
%OUTPUT= 17 45 23.70239(N)|064 35 01.06994(W)|          -15.0107
% St Croix VLBA Station, St Croix, USVI
%--------------------------------------------------------------------------------
% Km KUNMING   -1281153.1243    5640864.4360    2682653.3703  73670901   257.20  24.88 2020c
%
%Mk & Manua Kea VLBA Station, Hawaii, USA & 
%\texttt{-5464075.28  -2495247.63  2148297.61} \\
%Sc & St Croix VLBA Station, St Croix, USVI & 
%\texttt{ 2607848.71  -5488069.47  1932739.83} \\
%Km & Kunming 40m, Yunnan Astronomical Observatory, China   & 
%\texttt{-1281153.12  ~5640864.43  2682653.37}\\
%

\begin{figure}[hbt!]                
  \centering
  \includegraphics[width=0.96\textwidth]{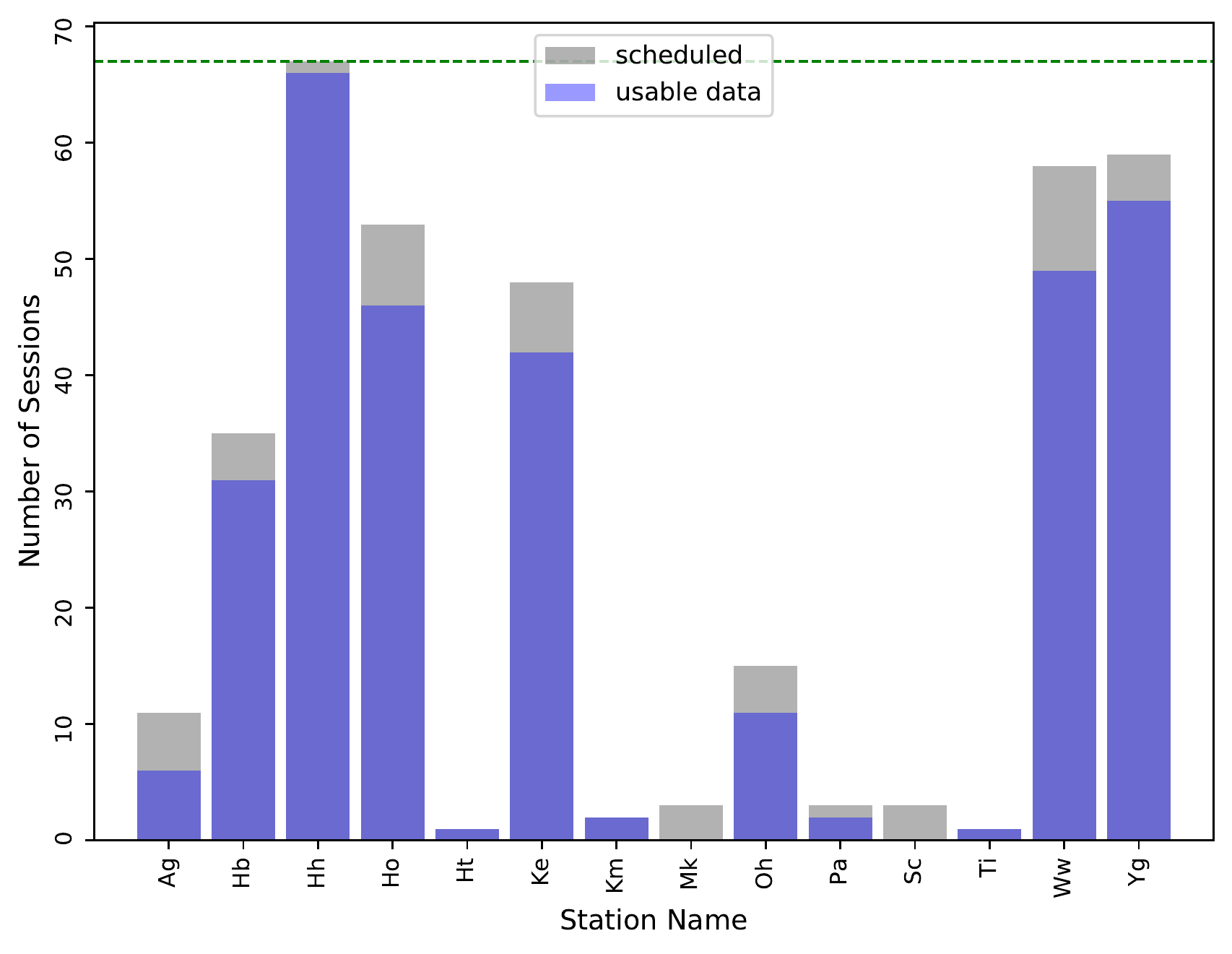}
  \caption{The total number of scheduled sessions by station for the 2011--2021 period (in grey), and the number of sessions for which an antenna had usable data (in light blue). The dashed green line shows the total number of sessions observed (67) for the 2011--2021 period.}
  \label{fig:observations_per_station}
\end{figure}

Between the first observations in 1995 and the last session included in this paper (December 2021, CRD116) the CRDS program observed a total of 116 sessions. Between March 2011 and December 2021, 67 CRDS sessions were observed and these were scheduled roughly once every two months -- giving six sessions per year, except for 2012 in which seven sessions were scheduled. Each of these sessions was 24-hours in duration and had between 2 and 8 scheduled stations, with an average of 5 stations per session.
%must fix these numbers
% Next Paragraph :  (see ref-results = \S~\ref{results})
It is important to note that not all scheduled stations necessarily participated or delivered usable data over the entire 24-hour session, due to various technical issues and outages (see \S~\ref{performance}). A summary of the number of sessions each station was scheduled in versus the number of sessions it participated in and had usable data for, during the 2011--2021 period, is shown in Figure~\ref{fig:observations_per_station}. It can be seen that the faster slewing 12-m antennas (Hb, Ke, Yg and Ww) have taken over the bulk of observing enabling a significant increase in the sources observed within a 24-hour period.

%Stuart to update this to include up to CRDS116 and ncrease label sizes

The CRDS sessions are all dual-band (S/X) observations, the S-band is a selection of frequencies from 2.2~-~2.3~GHz and X-band is a selection from 8.2~-~8.9~GHz. Ionosphere corrections are computed from the delay differences between the two bands S and X, these are then applied to the X-band solution for the final data product. Data is recorded in 14 intermediate frequency (IF) bands, consisting of eight X-band (+2 side-bands) and six S-band IFs. Up to May 2013 (CRDS65) the data was recorded with 16$\times$4~MHz bands at \mbox{1-bit}, resulting in a data recording rate of 128~Mbps. From July 2013 (CRDS66) the bands were increased to 8~MHz doubling the recording rate to 256~Mbps. In January 2018 (CRDS93) the data recording rate was increased again, from the 256~Mbps (16 $\times$ 8~MHz bands, \mbox{1-bit} recording) to 1~Gbps (16 $\times$ 16~MHz bands, \mbox{2-bit} recording) and a total bandwidth of 256~MHz. However, the VLBA antennas can only record a total bandwidth of 128~MHz (4 $\times$ 16~MHz S-band and $4 \times$ 16~MHz X-band IF's). The CRDS frequency sequence was also optimized in January 2018, mainly to avoid RFI. Test observations using different frequency setups were conducted to choose the best. In August 2021 (CRD114) the frequency sequence was updated again, to both accommodate the VLBA antennas and to avoid high side lobes and RFI. 
% David, MAtthias to check this
% should we add a table with the current frequency setup?

\subsection{Source Selection}
\label{SourceSelection}

Between 2011--2017 the sources observed in the CRDS sessions were primarily the 76 ICRF{\small 2} defining sources South of $-30^{\circ}$ Dec.
%strictly speaking  it is actually below -31 deg declination as 
%0646-306 at -30 44 19.6597807 was not included.
The ICRF{\small 2} defining sources were known to be strong ($\gtrsim500$~mJy) and relatively compact. Given the CRDS data rates during this time (128~Mbps and later 256~Mbps) they provided a pool of sufficiently bright sources detectable on even the longest baselines. In addition, there were also 20 ICRF{\small 2} non-defining sources that were observed in 2011 (CRDS51-53 and CRDS55).   
% The other sources, 11 observed 4 times, one sources 5 times and the rest only once
% Including the other sources brings the average sessions per source to 15.3
% Karine to check with Alan what was observed before CRDS50?
% Below the  21 non-def sources observed in CRDS51-53
%0008-300	0034-220	0122-514	0355-669	0633-263	0633-26B	0738-674	1012-448	1105-680	1129-580	1352-632	1505-304	1549-790	2102-659	2107-105	2117-614	2146-783	2318-087	2321-065	2333-528
%Below is the 10 non-def sources observed in CRDS92
%0010-401 0142-278 0514-459 0523-236 1039-474 1256-220 1352-632 1540-828 1646-506 1953-325
%Below is the CRF-2 Def sources observed between 2011 and 2017
%0002-478	0048-427	0104-408	0107-610	0131-522	0230-790	0235-618	0302-623	0308-611	0332-403	0334-546	0402-362	0405-385	0437-454	0454-810	0506-612	0516-621	0522-611	0524-460	0524-485	0534-340	0534-611	0537-441	0549-575	0920-397	1004-500	1022-665	1034-374	1101-536	1143-696	1144-379	1156-663	1251-713	1313-333	1325-558	1349-439	1420-679	1424-418	1448-648	1451-400	1554-643	1604-333	1611-710	1619-680	1624-617	1633-810	1657-562	1659-621	1725-795	1758-651	1806-458	1815-553	1824-582	1831-711	1925-610	1933-400	1935-692	1954-388	2002-375	2052-474	2106-413	2123-463	2142-758	2204-540	2220-351	2232-488	2236-572	2244-372	2245-328	2300-683	2326-477	2333-415	2344-514	2353-686	2355-534	2357-318
Towards the end of 2017, it was decided to revisit the strategy and review the CRDS program, in particular to expand the CRDS source list. At this point, other IVS programs had also started contributing to Southern observations, which added many more observations of sources in the South. These included for example the AUSTRAL \citep[AUA,][]{Plank2017} sessions that started in 2011 and the Asia-Oceania VLBI group for Geodesy and Astrometry \citep[AOV,][]{McCallum2019} sessions which started in 2015. It was therefore decided to use the CRDS program to re-observe all CRF sources South of $-15^{\circ}$ Dec and to give priority to sources observed in fewer than 10 sessions in order to improve the accuracy of the source positions in both coordinates \citep{deWitt2019b, deWitt2021}. Subsequently, in  November 2017 (CRDS92) ten additional bright ICRF{\small 2} non-defining sources, one of which was also observed in 2011 (CRDS53), were added to the CRDS schedule.

In January 2018 (CRDS93) the CRDS data rate was increased to 1~Gbps giving a factor of two increase in sensitivity. This allowed for weaker sources, down to $\sim$~350~mJy given an integration time of 6~min, to be included in the schedules. A list of 216 CRF sources South of $-15^{\circ}$ Dec with $\leq$~10
%observing sessions and with flux densities $\gtrsim$~350~mJy were added to the
 observing sessions and with flux densities $\sim$~350~mJy were added to the CRDS program. This list was revised as updated CRF solutions became available. A few bright K-band and X/Ka-band CRF sources that were not in the S/X-band frame ($\sim$~20 sources) were also added to the schedule. The March 2018 (CRDS94) session, by request of the ICRF{\small 3} Working Group, was dedicated to observing a list of 31 potential ICRF{\small 3} defining sources South of $-40^{\circ}$ Dec. These 31 sources had only a few observations and no available images, and thus additional observations were required ahead of finalising the ICRF{\small 3} catalogue. These observations were repeated for the August and September 2018 (CRDS97 and 98) sessions. From May 2021 (CRD112) onwards, following recommendations from the IVS-CRF committee, only ICRF{\small 3} defining sources South of $-15^{\circ}$ Dec were observed in the CRDS sessions in order to re-assess their suitability as defining sources in preparation for an ICRF-4. The sessions between 2018--2021 (CRDS93--116) added an additional 237 sources to the CRDS program. 

\begin{figure}[hbt!]
\centering
    \includegraphics[trim=4cm 10cm 4cm 10cm,clip,width=0.96\textwidth]{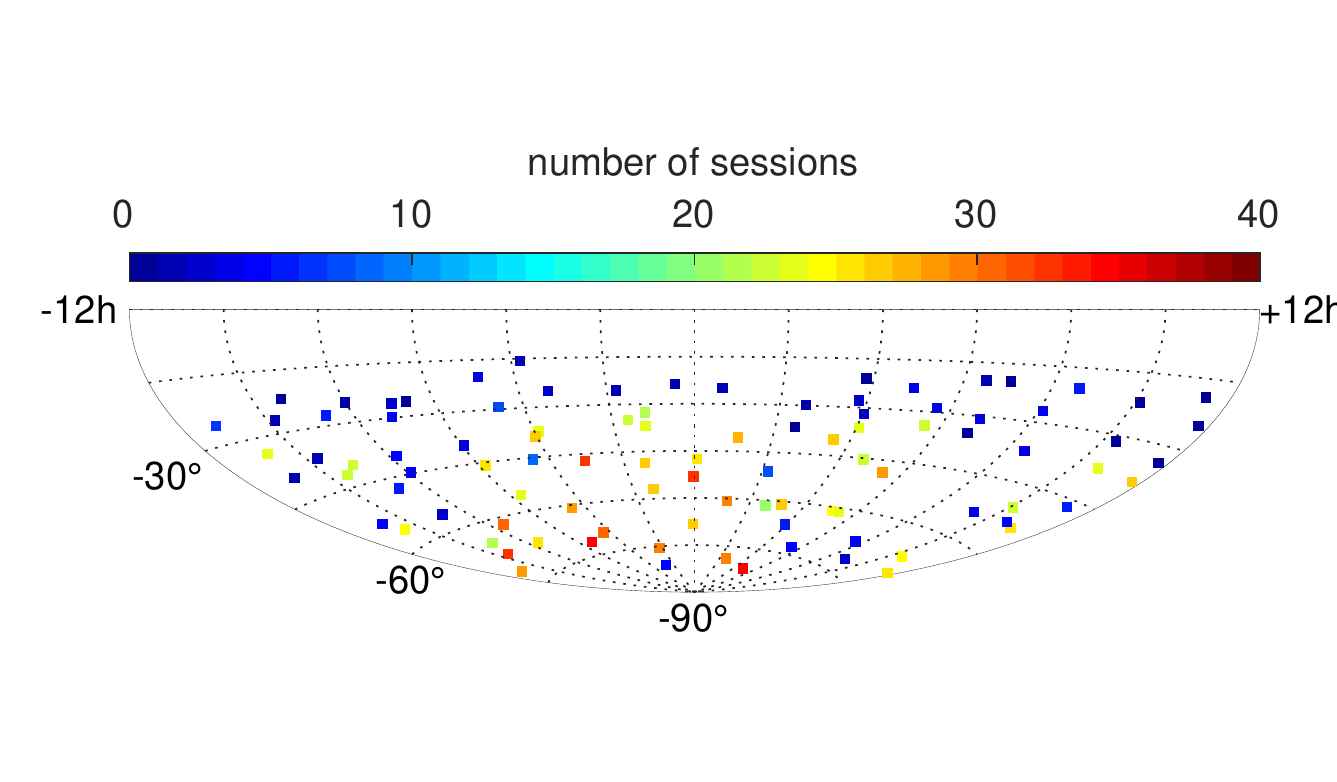}
  \\
  \includegraphics[trim=4cm 10cm 4cm 10cm,clip,width=0.96\textwidth]{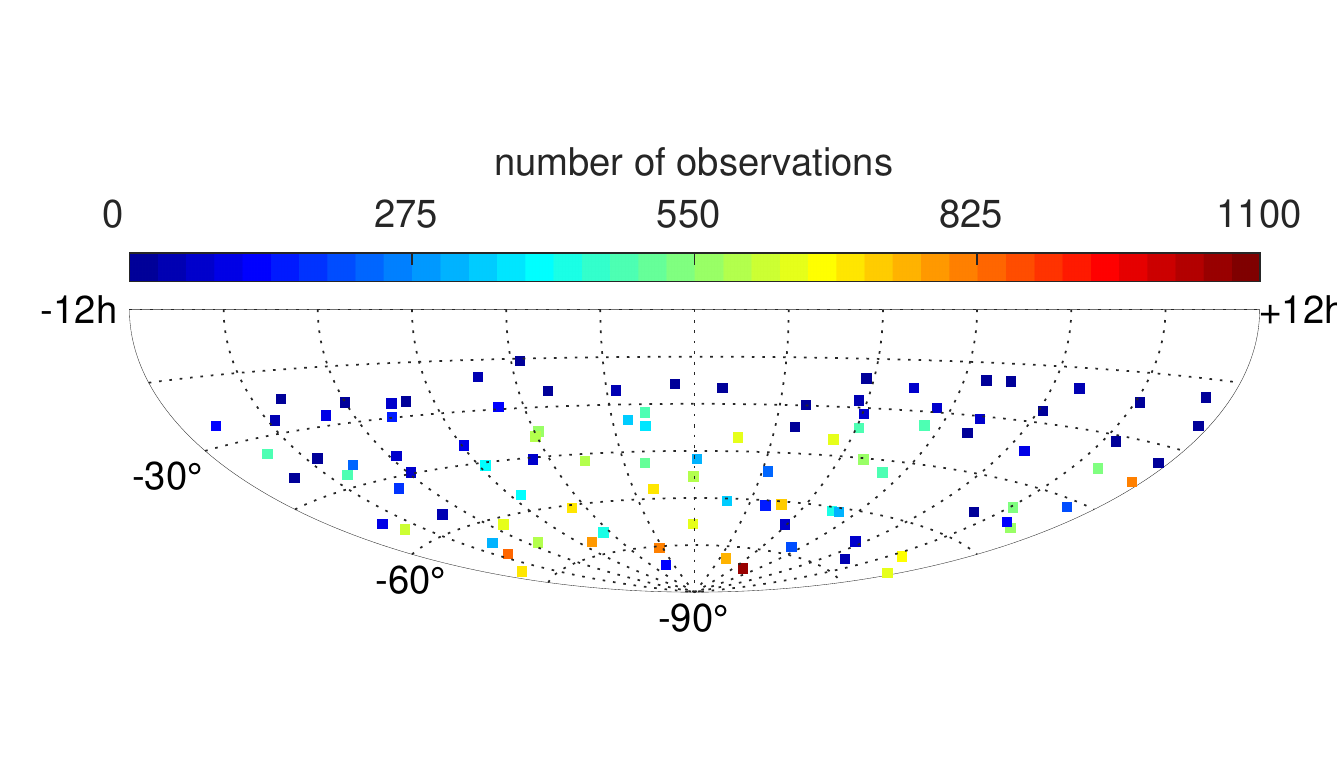}
\caption{The top plot shows the ICRF{\small 3} defining sources with a mollweide projection using a heat colour scale (on the right hand side) for the points to show the number of sessions each source was observed in, with the blue end being the least and red being the greater. The bottom plot shows the ICRF{\small 3} defining sources, also using a heat colour scale for the points to show the number of observations for each source. Both plots are for sessions CRDS50-116. }
\label{fig:crds-source-sky-distribution-defining}
\end{figure}

\begin{figure}[hbt!]
\centering
  \includegraphics[trim=4cm 10cm 4cm 10cm,clip,width=0.96\textwidth]{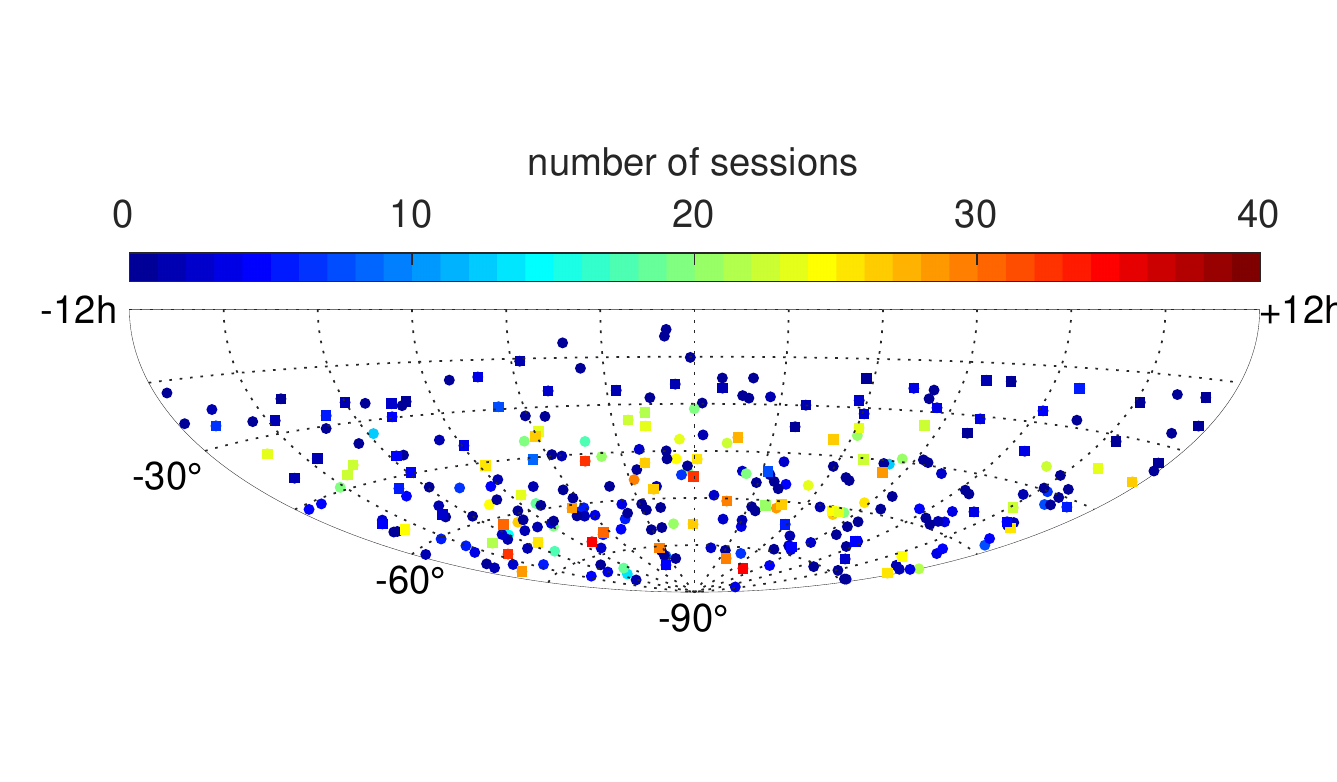}
 \\
  \includegraphics[trim=4cm 10cm 4cm 10cm,clip,width=0.96\textwidth]{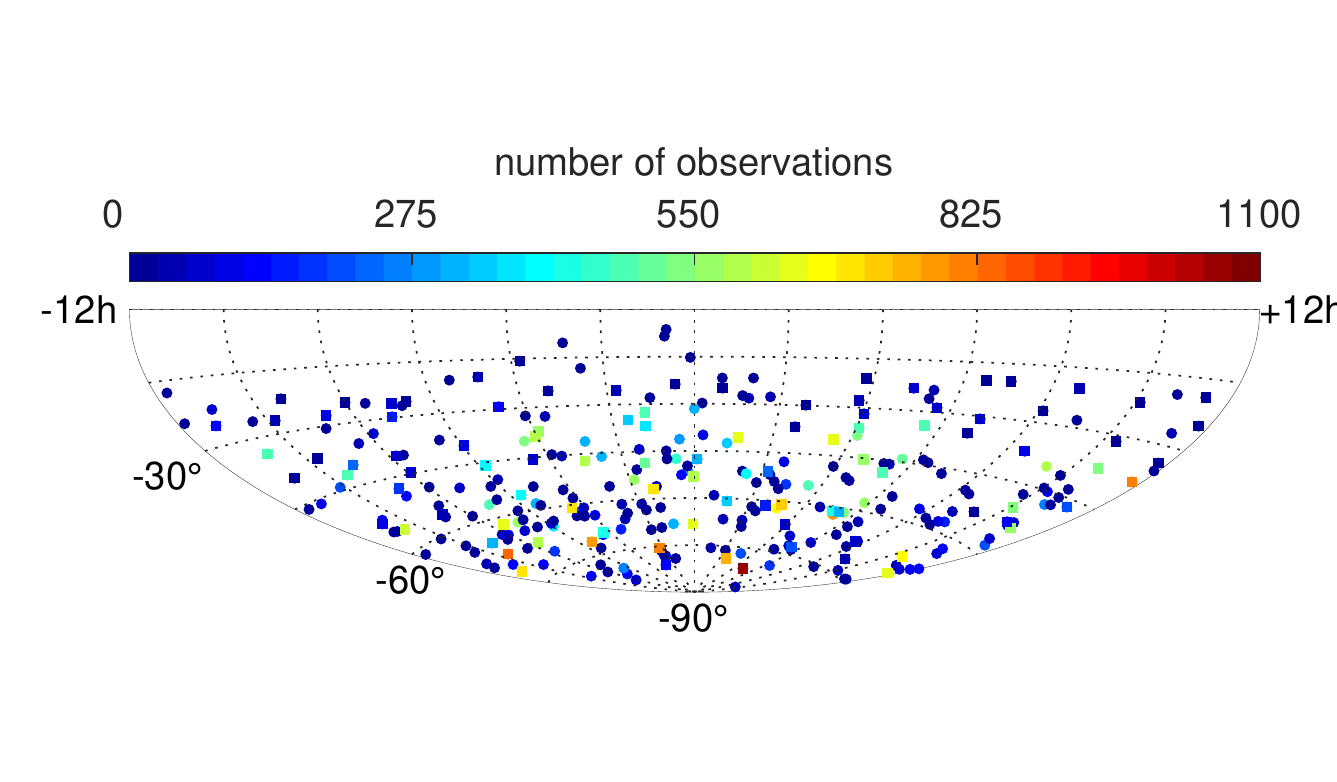}
\caption{The top plot shows all the sources with a mollweide projection using a heat colour scale for the number of sessions per source (ICRF{\small 3} sources are squares and other sources are circles). The bottom plot shows the same sources, using a heat colour scale for the number of observations per source (ICRF{\small 3} sources are squares and other sources are circles). Both plots are for sessions CRDS50-116.}
\label{fig:crds-source-sky-distribution-all}
\end{figure}

Between March 2011 and December 2021 (CRDS50--116) a total of 342 sources were observed in the CRDS sessions, including 104 of the 114 ICRF{\small 3} defining sources South of $-15^{\circ}$ Dec. From these observations 298 sources (101 ICRF{\small 3} defining sources) were successfully detected to give a detection rate of 87 \% .  Figure~\ref{fig:crds-source-sky-distribution-defining} shows the distribution of the CRDS sources used as defining sources in ICRF{\small 3} on a Mollweide projection of the celestial sphere, using a heat colour scale to show the number of sessions and the number of observations per source with the blue being the lesser and the red being the greater. Figure \ref{fig:crds-source-sky-distribution-all} depicts all the CRDS detected sources on a Mollweide projection of the celestial sphere, again using a heat colour scale to show the number of sessions and the number of observations per source. The number of sessions per source ranges between 1--35, with an average of 8 sessions and 156 observations per source. The complete source list for the 2011--2021 period is provided in \ref{Appendix_A}, where ICRF{\small 3} defining sources have been identified.

% Not sure why this is here ?
%Figure \ref{fig:crds-source-sky-distribution-all}

% Hana to confirm whether these are scheduled sources or actual "detected" sources. 
% are shown the sky distribution of the sources observed in this program, for sessions CDRS63 to CRD116. Between Figure's \ref{fig:crds-source-sky-distribution-defining} and \ref{fig:crds-source-sky-distribution-all} can be seen the improvement in the distribution of sources observed. 

% From Alet, images produced by Hana
%12/10/2021
%Hi Stuart,
%Attached are the plots from Hana, showing the sky distribution of the
%sources observed in the CRDS sessions, including sessions from CRDS63 to
%CRD107 (107 being the last correlated session at the time).
%
%crds63_107_aitoff.pdf --> shows all of the sources, but with the defining
%sources as red squares and other sources as blue circles
%
%crds63_107_numobs_all_aitoff.pdf --> shows all the sources, but colour coded
%by the number of observations per source (ICRF{\small 3} sources are squares and
%other sources are circles)
%
%crds63_107_numses_all_aitoff.pdf --> shows all the sources, but colour coded
%by the number of sessions per source (ICRF{\small 3} sources are squares and other
%sources are circles)
%
%crds63_107_numobs_def_aitoff.pdf --> shows ONLY the ICRF{\small 3} defining sources,
%colour-coded by the number of observations
%
%crds63_107_numses_def_aitoff.pdf --> shows ONLY the ICRF{\small 3} defining sources,
%colour-coded by the number of sessions
%
%A text file with the list of sources and number of sessions observations per
%source, is also attached.
%
%Best Regards,
%Alet

\subsection{Scheduling Strategies}
\label{scheduling}

%From 2011 through 2020 some 61 CRDS sessions have been run (CRDS50 - CRD110) and most have now been processed and analyzed. The Auscope antennas (Hb,Ke and Yg) joined starting in mid-2011 followed by the New Zealand Warkworth 12m antenna in 2012. The CRDS sessions over the past ten years have used between two and six Southern hemisphere antennas.  

Between 2011--2017 the schedules for the CRDS sessions were usually prepared by the USNO with some by NASA Goddard Space Flight Center (GSFC), using the geodetic VLBI scheduling software \textsc{sked} \citep{Vandenberg1997}. These sessions were all scheduled using a fixed scan length that ranged between 360--600~sec except for CRDS58 which used a scan length of 120~sec, and between 2--7 antennas were scheduled per session. This allowed for an average of 38 sources, 147 scans and 1337 observations to be scheduled per session. 
%Recalling that mostly ICRF{\small 2} defining sources South of $-30^{\circ}$ declination were observed during this period. 
The number of scans per source per session ranged between 1--14, with an average of 4 scans per source. Although the minimum number of stations per scan was set to two, which is standard for geodetic and astrometric VLBI type schedules, those sessions with four or more antennas had on average 89\% of the scans observed by four or more antennas. This made them more suitable for imaging which requires at least four stations per scan and preferably three or more scans per source.

In January 2018 the CRDS scheduling was optimised to allow for both astrometry and imaging, and from January 2018 (CRDS93) to March 2019 (CRD101) the CRDS schedules were prepared at HartRAO using the astronomical VLBI scheduling program \textsc{sched} \citep{Walker2018}. The changes that were made to the scheduling approach included the following: (1) using the full network of stations when possible for every scan, with no sub-netting as is used routinely for geodesy schedules; (2) observing at least 3--8 scans per source spread evenly over Hour Angle (HA) range to allow for optimal $u,v$-coverage for imaging without compromising the astrometric goals of the experiment; (3) including blocks with tropospheric calibrators that are also used as astrometric ties and as calibrators for imaging; and (4) scheduling sessions as part of a campaign rather than individually, to ensure that each source in the list will receive the required amount of observing time. These sessions had between 4--6 antennas and used a fixed scan length that ranged between 180--300~sec for calibrator scans ($\sim$~20 ICRF{\small 2} defining sources that were added to each session), and 300~sec for target scans ($\sim 30$~new sources that were added to each schedule). This allowed for an average of 49 sources, 227 scans and 1962 observations to be scheduled per session. The number of scans per source per session ranged between 1--8, with an average of 4.7 scans per source. The minimum number of stations per scan was set to four or higher, depending on the number of participating antennas.

\begin{figure*}[ht]
\centering
  \includegraphics[width=0.3\textwidth]{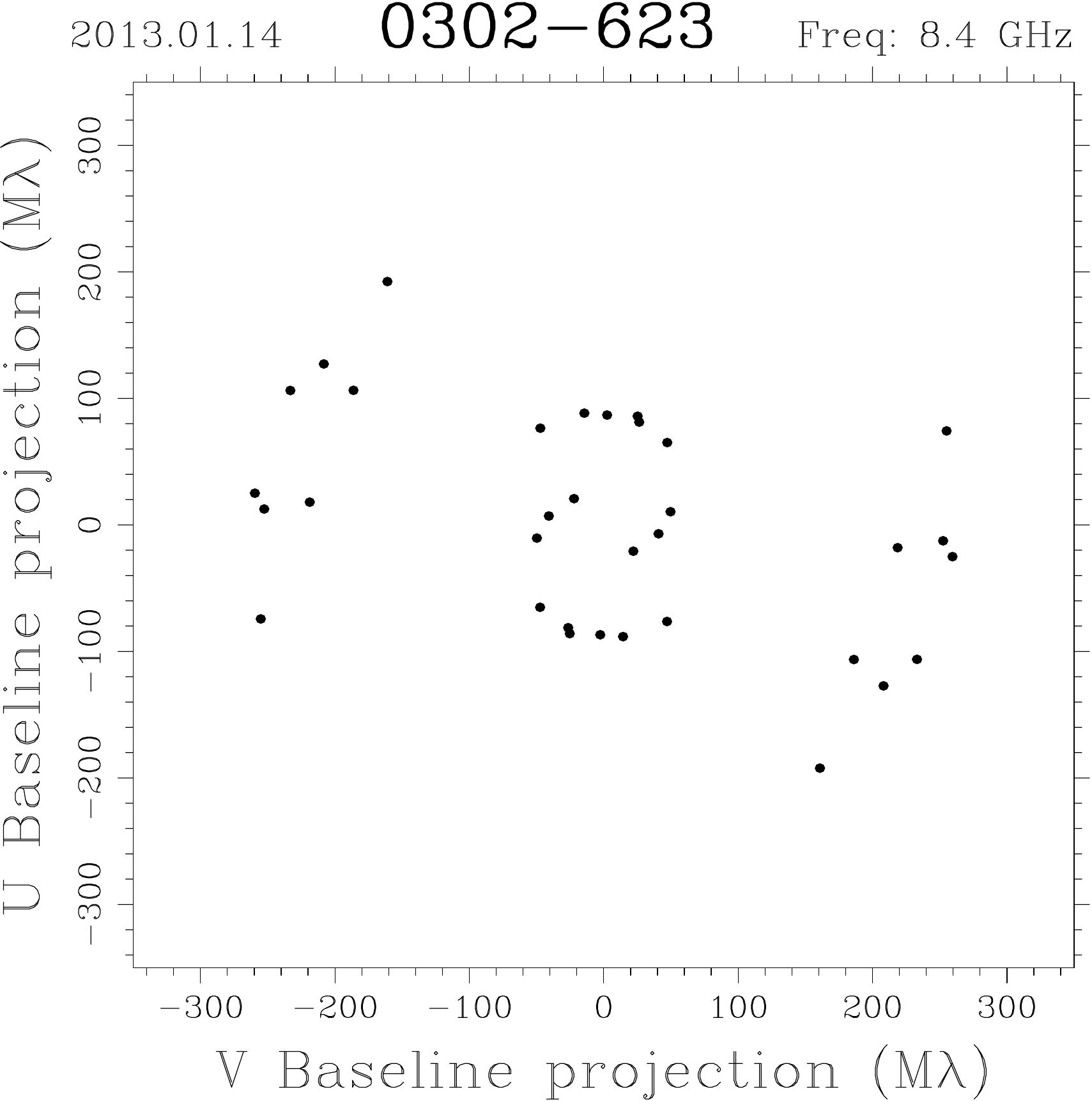}
    \hspace{0.1cm}
  \includegraphics[width=0.3\textwidth]{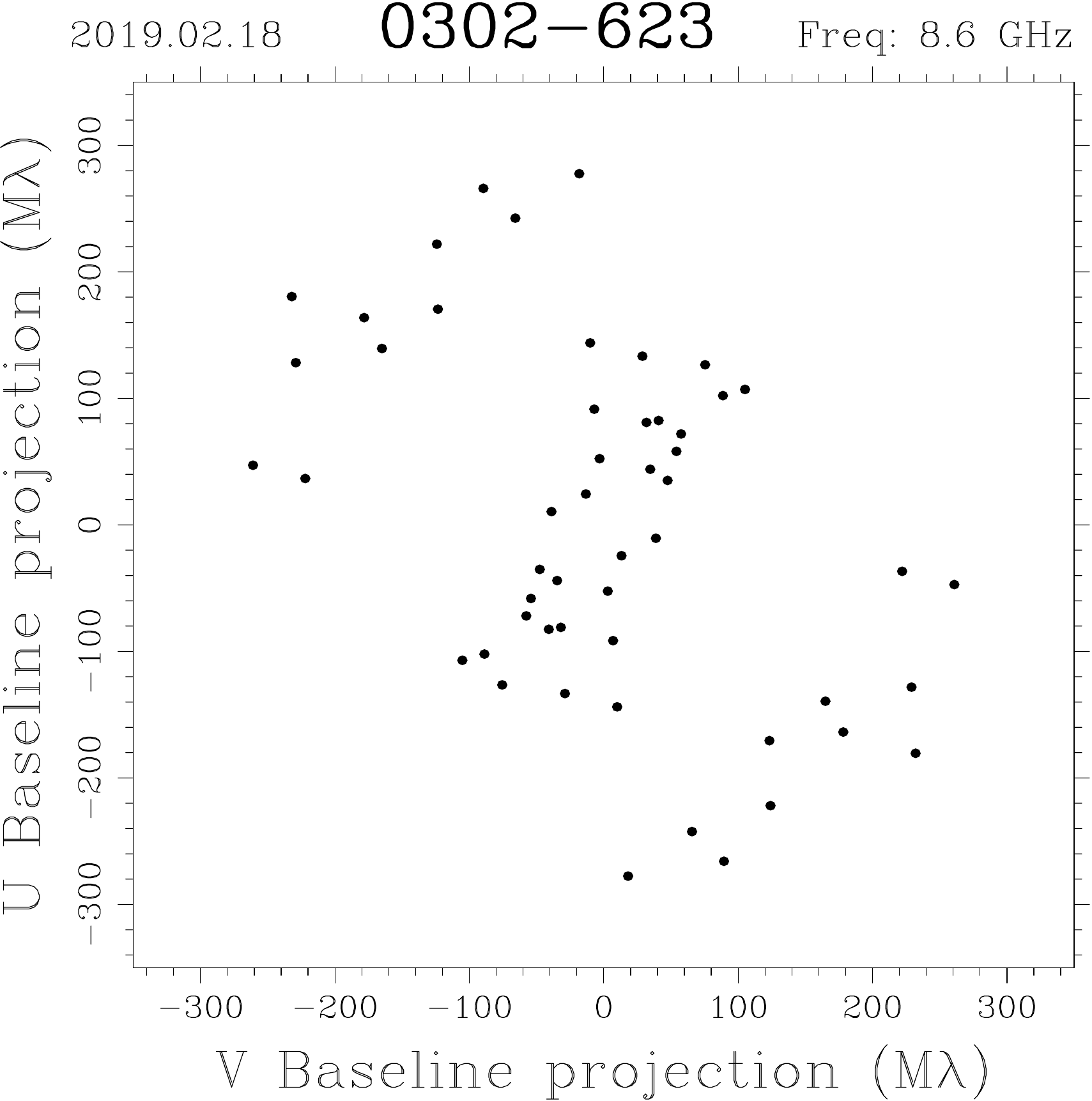}
    \hspace{0.1cm}
  \includegraphics[width=0.3\textwidth]{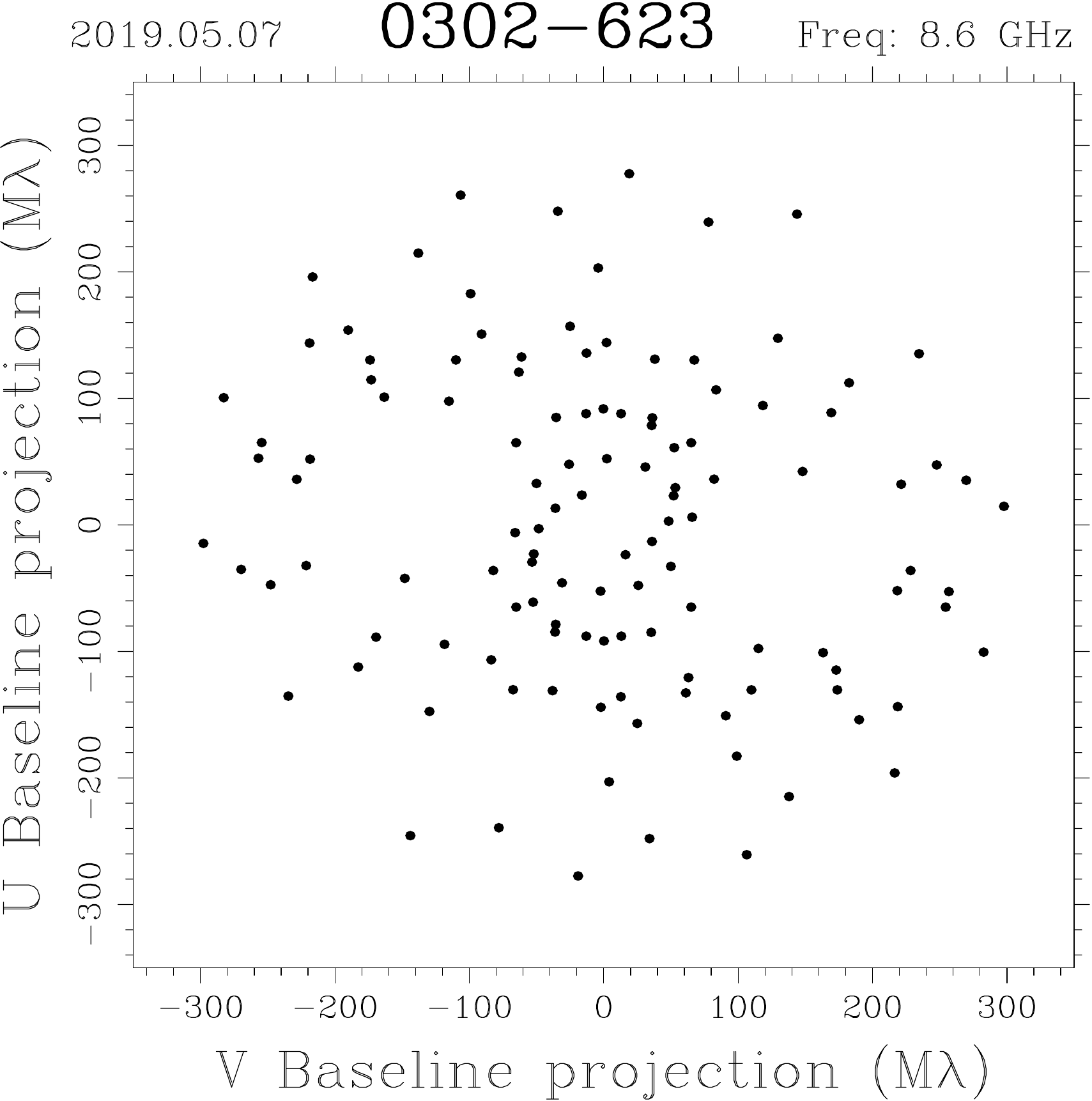}
\caption{The $u,v$-coverage plots for source 0302-623 (left to right) for session CRDS63 (January 2013), CRD100 (February 2019) and for CRD102 (May 2019).}
\label{fig:crds-improve-uv}
\end{figure*}

Figure \ref{fig:crds-improve-uv} shows the $u,v$-coverage plots for source 0302-623 and the improvement obtained starting with CRDS63 \mbox{(Jan 2014)} where the source was observed with 4 stations and in 4 scans, CRD100 \mbox{(Feb 2019)} with 5 stations but in only 2 scans and CRD102 (May 2019) where the source was observed with 6 stations and in 6 scans. This improvement in $u,v$-coverage has resulted in better imaging; important for all sources but far more important for monitoring and selecting ICRF defining sources. An example is 0302-623 which was an ICRF{\small 2} defining source, but due to the improved imaging was deselected as an ICRF{\small 3} defining source because of the source structure as seen in Figure \ref{fig:crds-images}.

From June 2019 (CRD102) the CRDS schedules were prepared at TU~Wien using the \textsc{VieSched++} \citep{Schartner2019} geodetic VLBI scheduling software. The \textsc{VieSched++} software has the capability to produce VLBI astronomical-type schedules by using a source-centric scheduling approach with special constraints for imaging (as listed in the previous section). This was done by defining individual minimum repeat times per source, representing the minimum time between two scans to the same source. The minimum repeat times were calculated based on the time the source was visible by at least four stations and a target number of scans, which was typically between five and eight. The source selection was done iteratively as discussed in \citet{Schartner2019}. Therefore, the initial source list was divided into a calibrator source group and a target source group. After each session, the initial source list was updated and the observed target sources were removed to have a proper rotation among all sources. Since the participating antennas had quite different characteristics, such as slew speeds and sensitivities, long idle times existed for some stations. To reduce these idle times and increase the Signal to Noise Ratio (SNR), the station observing times per scan were extended in case of available idle time. 

\begin{figure}[hbt!]
\centering
  \includegraphics[width=0.96\textwidth]{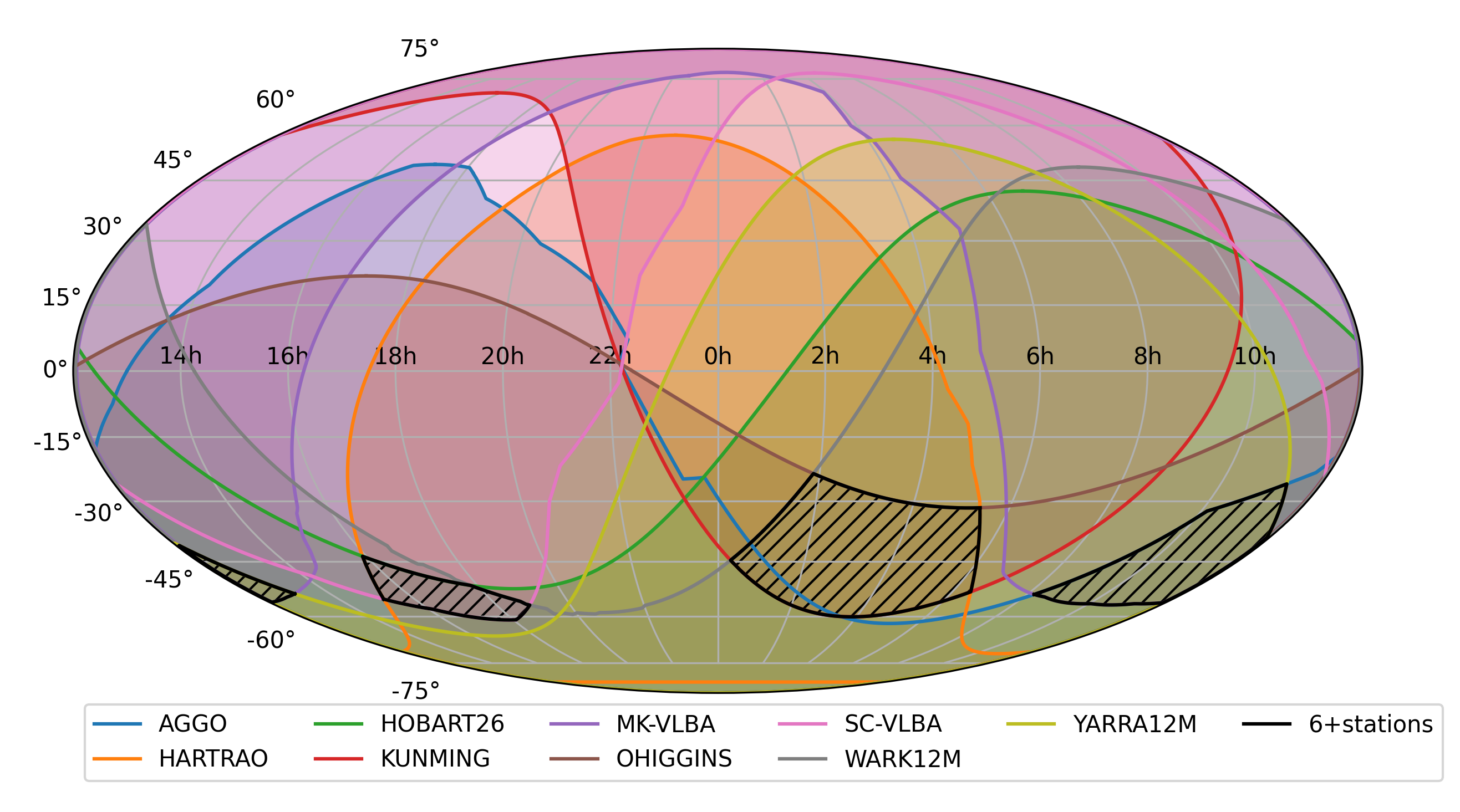}
\caption{Sky visibility per station as a function of right ascension and Dec for the CRD115 network after applying the local horizon mask and a cut-off elevation of 5 degrees. Areas visible by six or more stations are hatched in black. }
\label{fig:crds115_sky_visibility}
\end{figure}

Another significant challenge was the very inhomogeneous and sparse station network. As an example, the commonly visible sky for the October 2021 (CRD115) network is displayed in Figure \ref{fig:crds115_sky_visibility}. Areas of the highest interest, e.g. where a source can be observed by six or more stations, are highlighted with a hashed black area. It can be seen that no commonly visible sky exists between Mk and Hh and Sc and Yg, while the commonly visible sky between Km and Ag is also very small. Together, all these requirements pose significant constraints on the scheduling. To find a proper balance between these constraints, the \textsc{VieSched++} multi-scheduling feature was used to test several different optimization combinations. Out of this pool of possible schedules, the best one was selected and distributed.

% Include Matthias's network plot
% Caption Suggestion : 
%Sure, what about:
%Sky visibility per station as a function of right ascension and declination for the CRD115 network after applying the local horizon mask and a cut-off elevation of 5 degrees. Areas visible by six or more stations are hatched in black. 
%
%Not sure if "hatched" or "shaded" is more appropriate - I leave this decision to the native speakers among us ;-) 

Another challenge with the CRDS scheduling is that source flux information is often not available or not reliable. Therefore, the sessions between May 2019 (CRD102) and September 2020 (CRD110) used a fixed integration time of 280~sec per scan. However, due to the high number of non-detections, SNR-based scheduling using the \textsc{sked}-catalogue parameters \citep{Vandenberg1997} was used for the May and August 2021 (CRD112 and 113) sessions. Finally, this was changed to a strategy where the sources were divided into three groups based on their brightness and were observed in either 240, 360, or 480~sec scans. On average, the sessions scheduled with \textsc{VieSched++} (between May 2019, CRDS102 and December 2021, CRDS116) included between 6--8 antennas, which allowed for an average of 36 sources, 201 scans and 2165 observations to be scheduled per session. The number of scans per source per session ranged between 1--10, with an average of 5.7 scans per source. The schedules from the earlier sessions in this period had 95\% or more of the scans observed by four or more stations, while CRD114--116, which included Km, Mk and Sc, had only $\sim$~70\% of the scans observed by 4 or more stations. 

\begin{figure*}[hbt!]
\centering
  \includegraphics[width=0.95\textwidth]{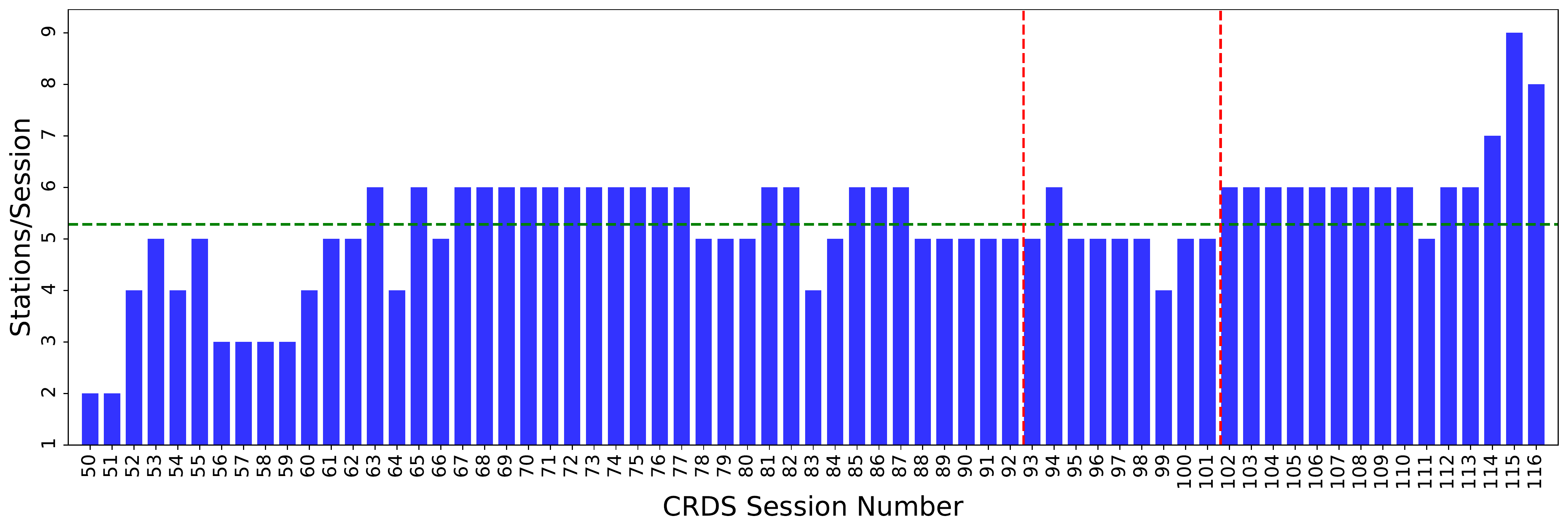}\\
  \vspace{0.1 cm}
  \includegraphics[width=0.95\textwidth]{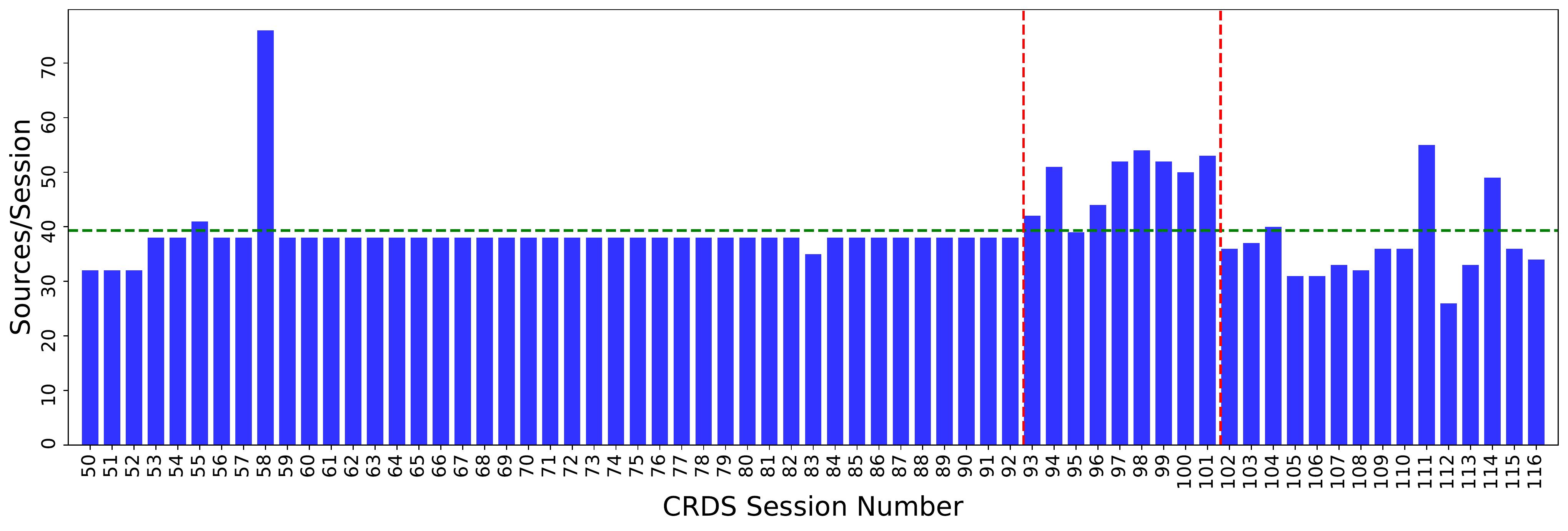}\\
  \vspace{0.1 cm}
  \includegraphics[width=0.95\textwidth]{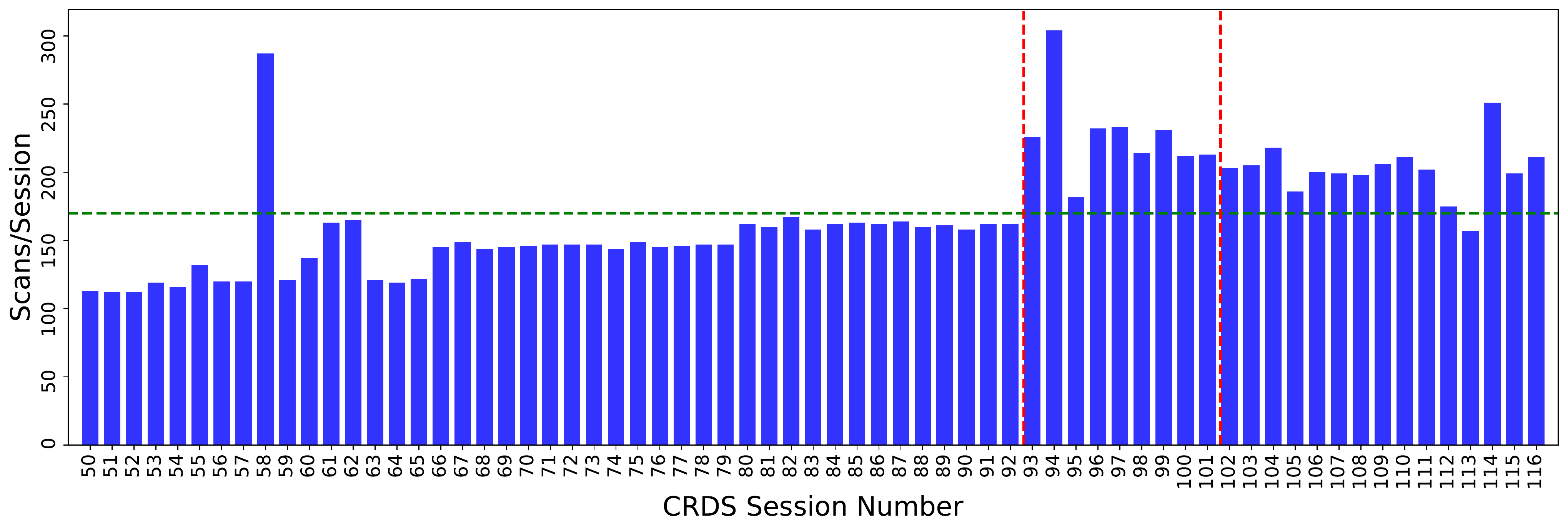}
\caption{Scheduling statistics for the CRDS sessions between 2011--2021 (CRDS50--116) showing the number of stations (top plot), the number of sessions (middle plot), and the number of scans (bottom plot). The red dashed vertical lines indicate the time frames when different scheduling software were used, and the green dashed horizontal lines show the average of the distribution.}
\label{fig:crds-scheduling-stats}
\end{figure*}

\begin{figure*}[hbt!]
\centering
  \includegraphics[width=0.96\textwidth]{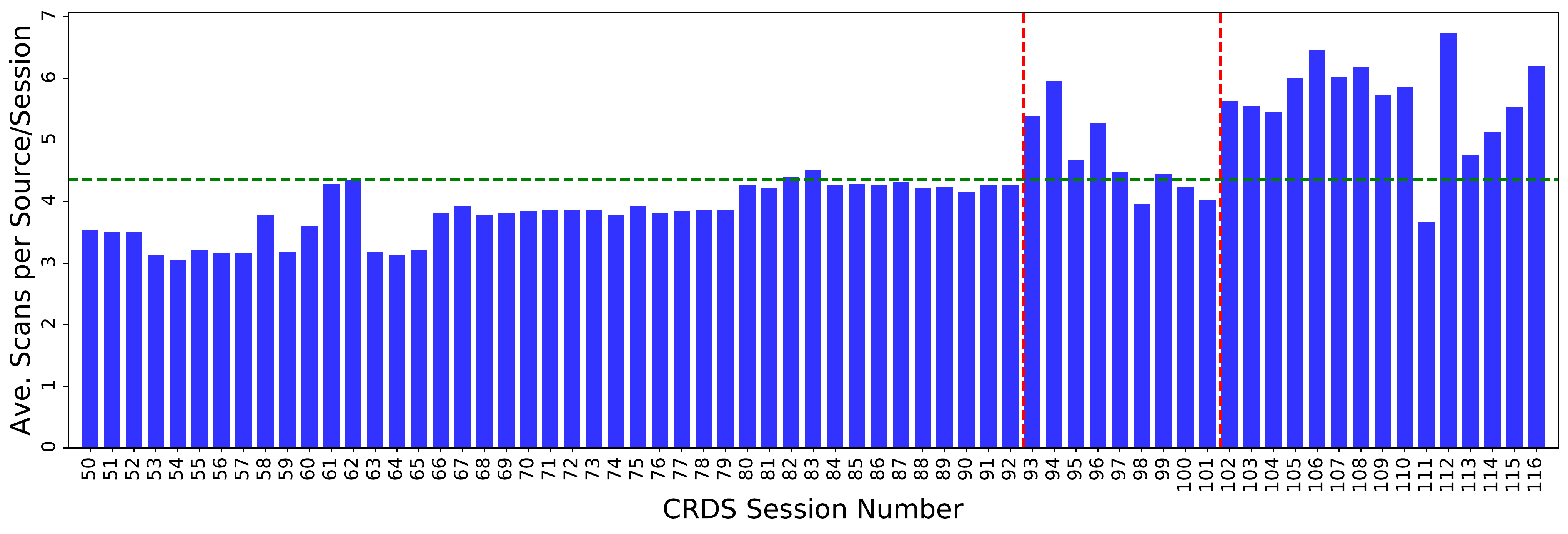}\\
  \vspace{0.1 cm}
  \includegraphics[width=0.96\textwidth]{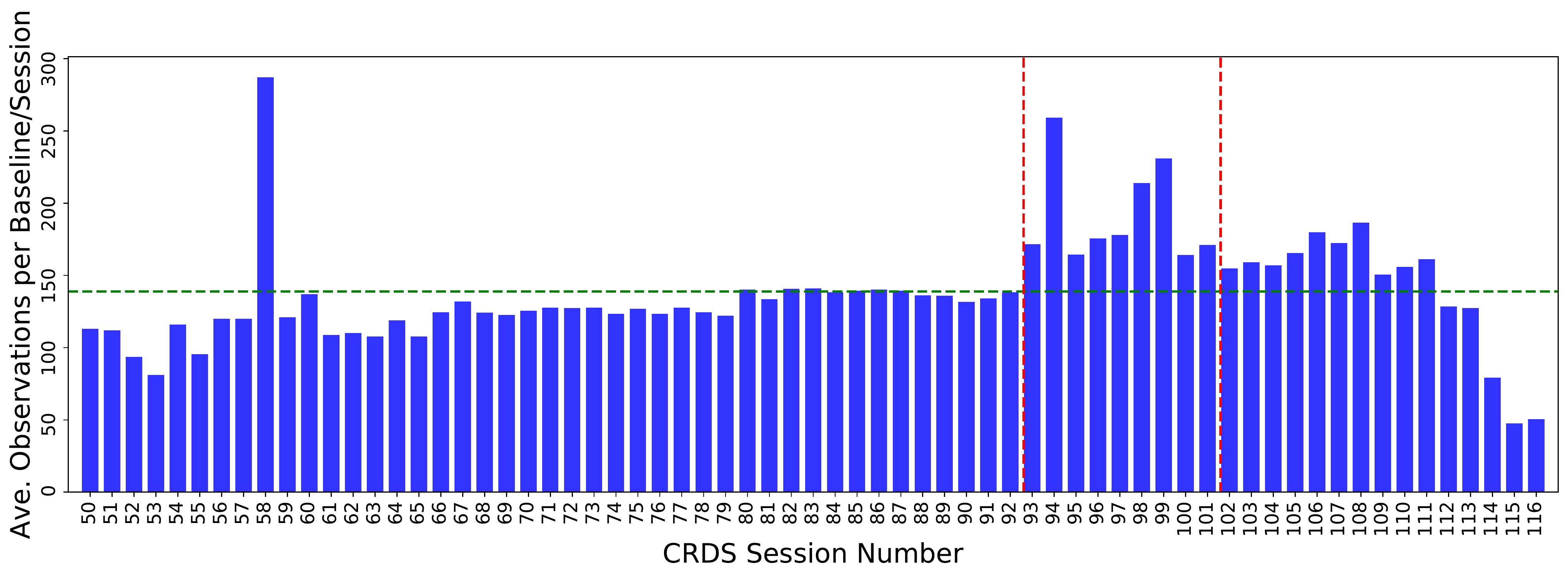}
\caption{Scheduling statistics for the CRDS sessions between 2011--2021 (CRDS50--116), showing the average number of scans per sources (top plot), and the average number of observations per baselines (bottom plot), for each session. The red dashed vertical lines indicate the time frames when different scheduling software were used, and the green dashed horizontal lines show the average of the distribution.}
\label{fig:crds-scheduling-stats-averages}
\end{figure*}

Finally, Figures~\ref{fig:crds-scheduling-stats} and~\ref{fig:crds-scheduling-stats-averages} show the overall CRDS scheduling statistics for the period 2011--2021. It is clear from these plots that, in general, the number of scheduled sources and scans per session, as well as the average number of scans per source and average number of observations per baseline, all increased from 2018 (CRDS93). However, the number of sources scheduled per session shows a decrease from May 2019 (CRD102), as both the integration time per scan and the number of scans per source were increased, allowing for fewer sources to be scheduled. The increase in the integration times was mainly due to the decrease in the sensitivity of the network at the time, owing to the loss of two of the larger and more sensitive antennas, Ke and Ho, and the addition of the smaller less sensitive antennas, Ag and Oh. The rapid decrease in the number of observations per baseline seen for the last three sessions is a result of the additional scheduling constraints imposed by the three Northern antennas (Km, Mk, and Sc) that were added to the network. It should be noted that the statistics and plots provided in this section purely reports on the numbers obtained from the scheduling and does not reflect the actual number of participating antennas nor the success rate of the observations themselves, which are discussed in \S~\ref{results}.

\subsection{Correlation, Analysis and Data Products}
\label{Correlation}
% Sara Hardin, USNO

The raw antenna data for all stations was usually correlated at the USNO using their VLBI Correlator facility. Prior to October 2014, this was done on the Mark IV hardware correlator. In 2014 USNO transitioned to the DiFX \citep{2011PASP..123..275D} software correlator (\cite{2010ivs..conf..153F}; \cite{baver2014international}). First, the raw data was correlated with the DiFX software package. Visibilities are produced and converted to the \textsc{Mk4} format for post-processing and the \textsc{fits} format for imaging. When necessary, station clock models are adjusted to reduce single-band delay residuals and the data are re-correlated. The CRDS82, 84, 86, 88, 90, 92, and 93 sessions were correlated with the Shanghai \textsc{DiFX} software correlator \citep{Shu2018} that is hosted and operated by the Shanghai Astronomical Observatory (SHAO) of the Chinese Academy of Sciences (CAS). 

After correlation with DiFX, fringe fitting was performed with the Haystack Observatory Post-processing Software, \textsc{hops} \citep{HOPS2021}. In cases where RFI degrades the fringe amplitude for a significant number of scans, the affected channels were removed. Fringes were also inspected to ensure that there was a strong phase calibration signal. Stations such as Ho and Ww had no initial phase calibration and required an artificial one applied with \textsc{hops}. Once significant RFI was removed and all stations have phase calibrations, the data was sent to the IVS Data Centres in the \textsc{vgosDB} format. However, the turnaround of the correlation of the CRDS sessions was severely affected by the arrival time of data from the Oh antenna. Since there is no high-bandwidth data connectivity to Antarctica and the Oh station, data was recorded onto hard disks and these are transported on an irregular basis when supply ships depart.
Analysis of all of the CRDS sessions was done at both USNO and GSFC. The analysis begins by computing theoretical delays using the \textsc{Calc} program \citep{Charlot_2020} and inserting them into the database. Then meteorological data (surface pressures, temperatures and humidities) and phase cal cable calibration data (if available) is added. Individual sessions were initially analysed with program \textsc{solve} \citep{Ma1978} in the interactive mode until the 2018/2019 time frame, and then with $\nu$SOLVE \citep{2014ivs..conf..253B} as it was gradually phased in. The interactive analysis first involves finding and resolving the $2\pi$ group delay ambiguities in both X and S bands and creating the ionosphere-free X/S combination. Various parameters are then solved for in a least-squares solution, such as piece-wise continuous clock and residual atmosphere terms at typically 60-minute intervals, tropospheric gradients over the entire session, antenna positions, and occasionally the coordinates of new sources. Three-sigma editing is performed and lastly the database is updated. Further analysis is then done on multiple databases with a global VLBI analysis package, such as \textsc{calc/solve} \citep{Ma1978} or various other analysis packages within IVS \citep{IVS_Analysis}. The global solutions can involve dozens or thousands of databases and solve for both global parameters (one value for the entire data span) and arc parameters (separate values for each database). Global parameters usually are the station positions, station velocities and source positions. Arc parameters are usually the piece-wise continuous clock and residual atmosphere terms (typically at 60 or 30 minute intervals), tropospheric gradients (typically at 6 hour intervals), and the five Earth orientation parameters (X and Y polar motion, UT1, and dX and dY nutation). A \textsc{solve} batch solution of the 57 available CRDS sessions from 2011–2021 shows that there were $\sim41,000$ individual baseline observations at 11 sites and that 261 sources were observed in multiple sessions.
% DG to update the above to also include 2021 sessions
% Note from DB : USNO transitioned to the DiFX software correlator in October 2014. Before they were using the Mark IV hardware correlator. Reference for the transition can be the USNO correlator report in IVS AR2014.

The databases created from the correlation and analysis as well as auxiliary data, such as schedule files and station log files, are stored in three primary IVS Data Centres (see the Data Availability Statement).

% Note from DB : USNO transitioned to the DiFX software correlator in October 2014. Before they were using the Mark IV hardware correlator. Reference for the transition can be the USNO correlator report in IVS AR2014.

%Citations:

%Deller, A. T., Brisken, W. F., Phillips, C. J., Morgan, J., Alef, W., Cappallo, R., Middelberg, E., Romney, J., Rottmann, H., Tingay, S. J., and Wayth, R.: DiFX-2: A More Flexible, Efficient, Robust, and Powerful Software Correlator, Publications of the Astronomical Society of the Pacific, 123, 275, https://doi.org/10.1086/658907, 2011.

%MIT/Haystack (2021) Haystack Observatory Postprocessing System (HOPS). https://www.haystack.mit.edu/haystack-observatory-postprocessing-system-hops/. Accessed 24 Nov 2021.

%\FloatBarrier

%===================================================================

\section{Results and Performance}
\label{results}
%Hana to expand on this section. The CRDS sessions, like all other IVS sessions contribute to EOP results etc ?? CRDS sessions were used in the ITRF2020 solution etc. Hana to provide the above details for the CRDS contribution to geodetic results.

The primary purpose of the CRDS program is astrometry, in particular to maintain and improve the S/X-band CRF in the Deep South. In recent years there have also been efforts to obtain imaging results from the CRDS sessions in order to assess the astrometric suitability of the CRF sources in the South, in particular those sources for which no images were available. Additionally, the CRDS sessions like all other IVS sessions contribute to the realization of the terrestrial reference frame. They are included in the Quarterly combination of 24-hour IVS sessions\footnote{\url{https://ivscc.gsfc.nasa.gov/IVS_AC/ITRF2020/itrf2020_sx_sessionTable_v2021Feb10.txt}} computed by the BKG/DGFI-TUM Combination Centre which releases the estimated TRF as an official IVS product \citep{bachmann2021}. They are used for the station coordinate estimation but are not used for EOP determination. The USNO Quarterly solution todate uses CRDS59,69,71 \& 91 in their EOP Solution\footnote{\url{https://crf.usno.navy.mil/quarterly-vlbi-solution}}.  Furthermore, the current realization of the International Terrestrial Reference Frame ITRF2020 \citep{altamimi2022itrf2020}, includes all available CRDS sessions at that time, i.e., until CRD105.
% Link given by Warren, I see CRDS59,69,71 & 91 in the EOP
%https://crf.usno.navy.mil/quarterly-vlbi-solution
%https://ivscc.gsfc.nasa.gov/IVS_AC/ITRF2020/itrf2020_sx_sessionTable_v2021Feb10.txt
%(see Resolution B2\footnote{See Resolution B2 at \url{https://iau.org/static/resolutions/IAU2018_ResolB2_English.pdf}.})
Details of the astrometric and imaging results obtained from the CRDS sessions are provided in the following sections,  \S~\ref{astr_results} and \S~\ref{img_results}. The overall performance of the CRDS sessions is discussed in \S~\ref{performance}

\subsection{Astrometric Results}
\label{astr_results}
The early CRDS sessions from June 1995 (CRF-DS1) to December 2010 (CRDS49) made significant contributions to the ICRF{\small 2}, being almost the only sessions to add new Southern sources. The CRDS sessions between 2011 (CRDS50) and 2017 (CRDS92) primarily observed the 76 ICRF{\small 2} defining sources South of $-30^{\circ}$ Dec, and this long-term monitoring provided precise positions for the ICRF{\small 2} defining sources in the deep-South. From 2018 (CRDS93) additional sources were added to the CRDS program, as detailed in \S~\ref{SourceSelection}, in particular CRF sources South of $-15^{\circ}$ Dec observed in fewer than 10 sessions and the ICRF{\small 3} defining sources.  

The April 2022 S/X astrometric solution from the USNO \citep[e.g. sx-usno-220422,][]{Gordon2022} which includes all of the CRDS sessions up to December 2021 (CRD116), shows significant improvement over the ICRF{\small 3}. However, the number of sources is still more than a factor of 3 less in the far-South ($\leq -45^{\circ}$) compared to the far-North ($\geq 45^{\circ}$), with 269 sources in the far-South and 881 sources in the far-North. While the average number of sessions per source in the far-South is only a factor of 1.8 less than in the far-North, the average number of observations per source is more than a factor of 3 less in the far-South, with an average of 131 observations per source in the far-South and 449 in the far-North. The median formal uncertainties are a factor of 1.2 weaker in $RA \cos (Dec)$ in the far-South and a factor of 1.6 weaker in $Dec$, showing that much work is still needed in the far-South.

\begin{figure*}[ht!]
\centering
  \includegraphics[width=0.49\textwidth]{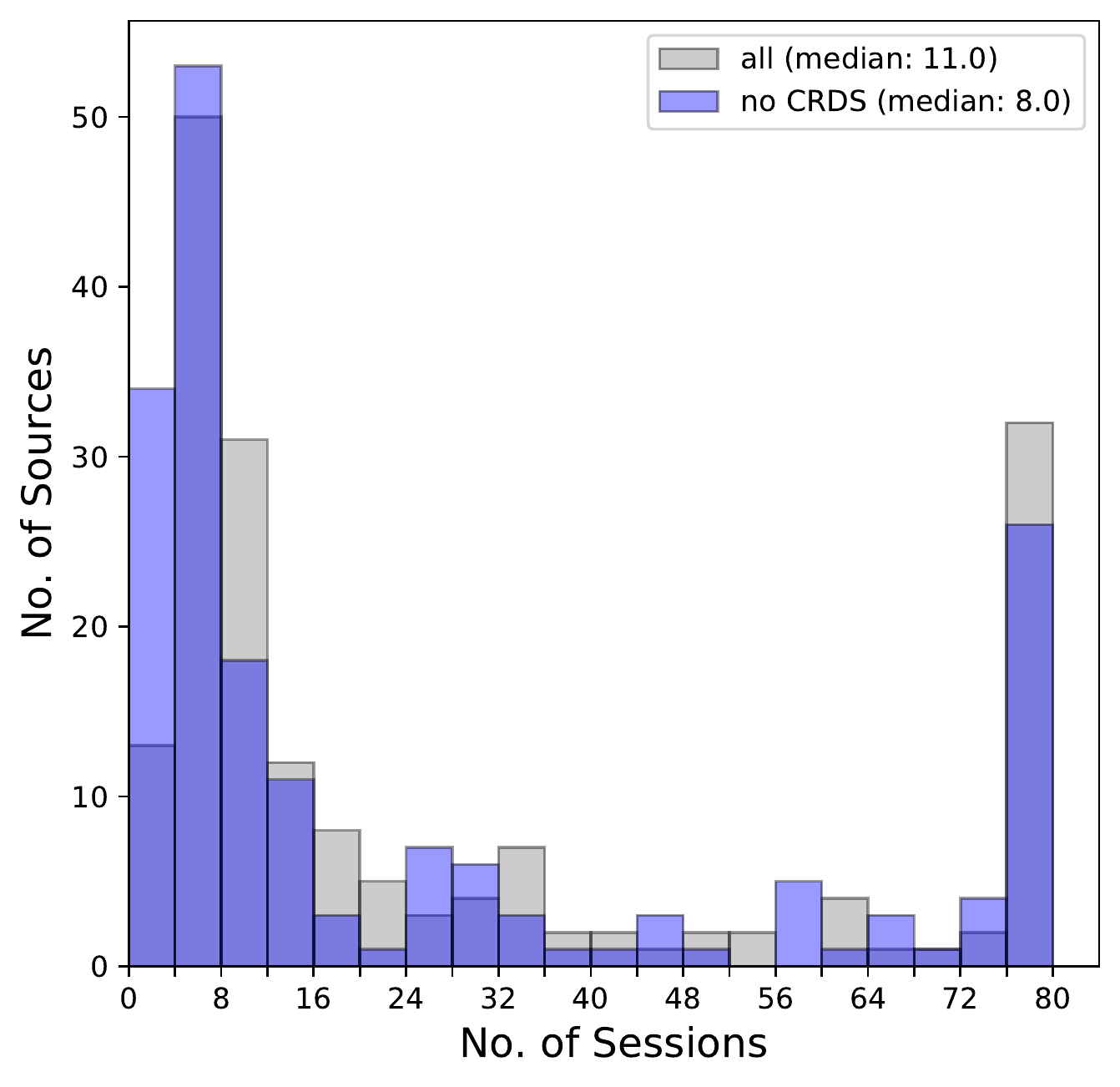} 
  \hspace{0.1cm}
  \includegraphics[width=0.49\textwidth]{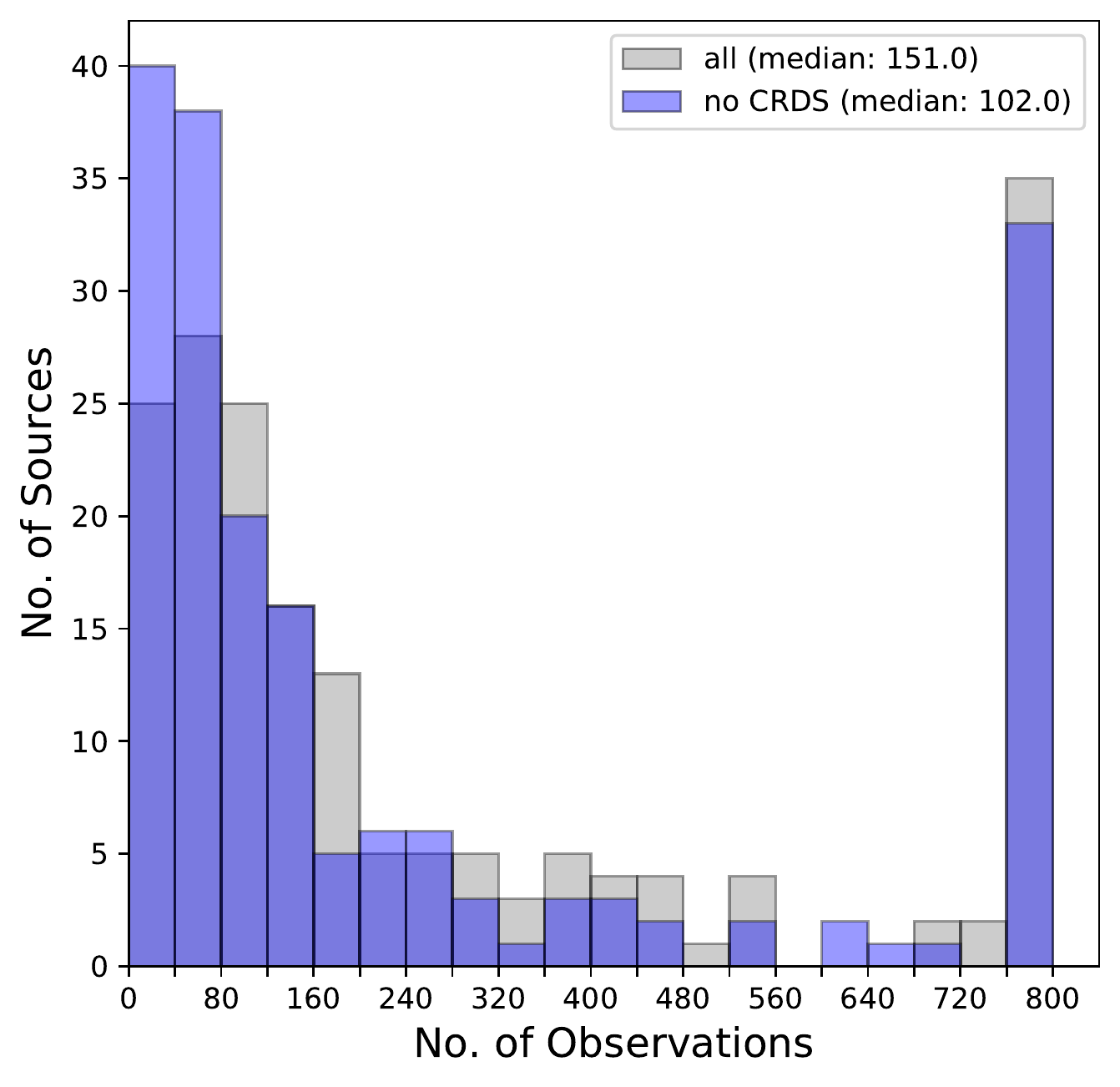} \\
  \includegraphics[width=0.49\textwidth]{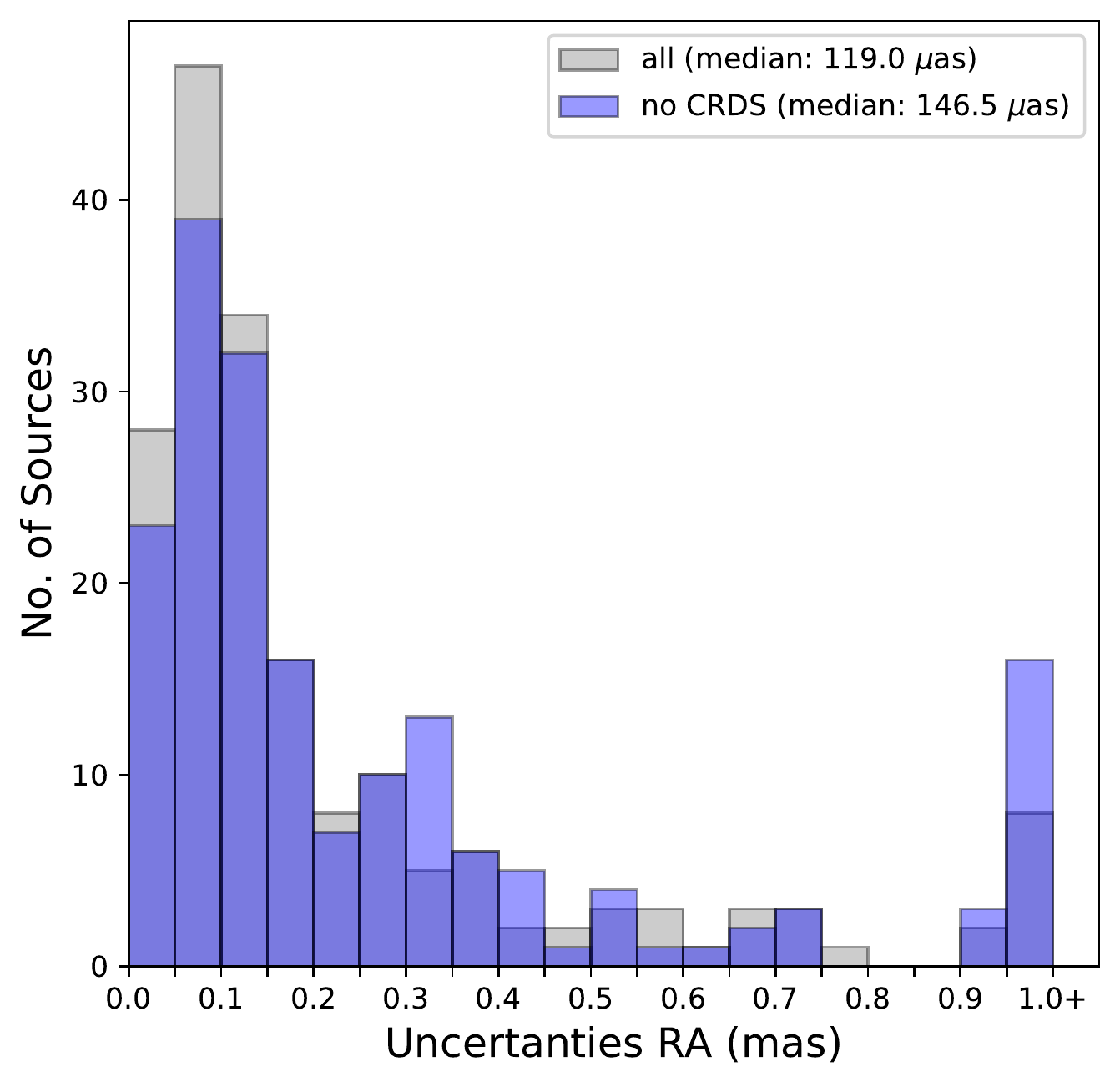} 
  \hspace{0.1cm}
  \includegraphics[width=0.49\textwidth]{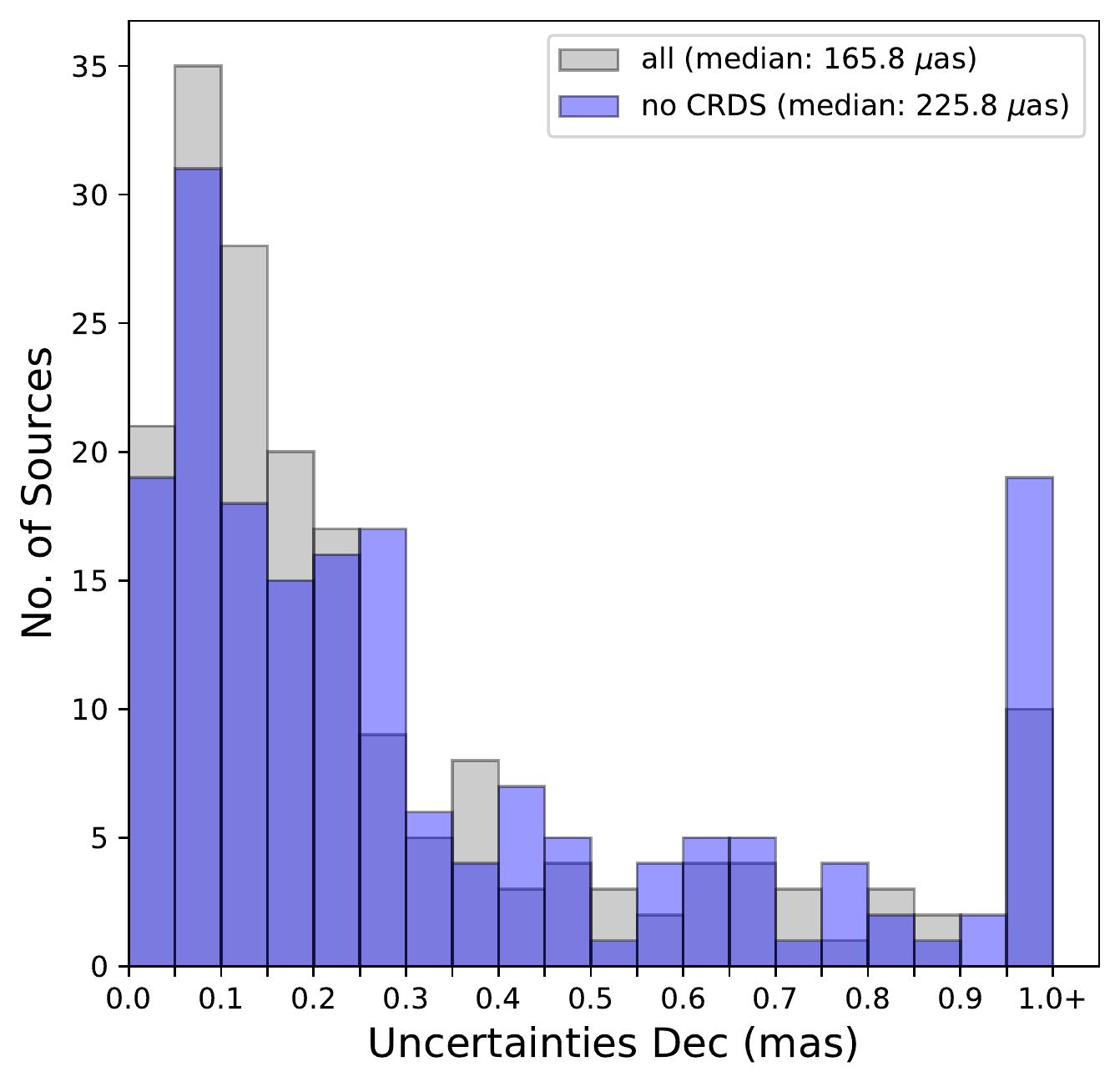}\\
\caption{Comparison of statistics between a CRF solution of the 182 deep-South sources observed in CRDS sessions with all sessions (in grey) and one without the CRDS sessions between 2018--2021 (in blue). Median values for each distribution are listed in parenthesis.}
\label{fig:crds-astrometric-stats}
\end{figure*}

In order to measure the impact of the more recent CRDS observations (January 2018, CRDS93--December 2021,CRD116), we compare the April 2022 S/X astrometric solution (sx-usno-220422) to a solution where we exclude all CRDS sessions from January 2018 (CRDS93) onward. The April 2022 astrometric solution includes 266 sources that were observed and detected in CRDS sessions between 2018 and 2021, but for this comparison, we consider only the 182 of those sources that are in the far-South. Figure~\ref{fig:crds-astrometric-stats} shows the number of sessions and number of observations, as well as the $RA \cos (Dec)$ and $Dec$ uncertainties from the April 2022 S/X astrometric solution (sx-usno-220422) for each of the 182 sources, with and without the CRDS sessions. The median number of sessions per source increases from 8 to 11, when CRDS sessions are included, and the number of observations increases by a factor of 1.5. The median uncertainties are a factor of 1.2 lower in $RA \cos (Dec)$ and 1.4 lower in $Dec$, when the CRDS sessions are included in the solution. Thus, the CRDS sessions have significantly improved the precision of the 182 sources observed in the far-South since 2018.  

%\begin{figure*}[ht]
%\centering
%  \includegraphics[width=0.3\textwidth]{0302-623_2013_01_14_uv_crds63.pdf}
%    \hspace{0.1cm}
%  \includegraphics[width=0.3\textwidth]{0302-623_2019_02_18_uv_crd100.pdf}
%    \hspace{0.1cm}
%  \includegraphics[width=0.3\textwidth]{0302-623_2019_05_06_uv_crd102.pdf}
%\caption{The $u,v$-coverage plots for source 0302-623 (left to right) for session CRDS63 (January 2013), CRD100 (February 2019) and for CRD102 (May 2019).}
%\label{fig:crds-improve-uv}
%\end{figure*}

\subsection{Imaging Results}
\label{img_results}

A campaign to image CRF sources in the deep-South using the CRDS sessions started in January 2013 \citep{Basu2016}, and the imaging of twelve CRDS sessions (CRDS63, 66, 68, 94, 95, 96, 97, 98, 100, 101, 102, and 103) has been completed to date \citep{Basu2018, Basu2021}. Imaging of the older sessions, prior to CRDS93, was challenging because of poor sensitivity and poor $u,v$-coverage, and for this reason only three of these sessions were imaged. CRDS93 was not imaged as it had no usable data for Hh which provides the long baselines, and CRDS99 was not imaged as it had only three participating stations.

The correlated visibility data was calibrated with the NRAO’s Astronomical Imaging Processing System \citep[\textsc{aips,}][]{greisen2003}. The data were first split into single-frequency (S- and X-band) files and each band was then calibrated separately using the standard calibration and editing steps and utilities available in \textsc{aips}. Data for sessions prior to CRDS93 were not available in Flexible Image Transport System (\textsc{fits}) file format for use in \textsc{aips}, and an interface program called \textsc{mk4in} \citep{alef2002} was used to import the raw \textsc{mkiv} correlator data into \textsc{aips} and then convert and export it into \textsc{fits} format. This process was carried out at the Bonn correlator. Amplitude calibration, in general, has been a challenge, since system temperature (T$_{\rm sys}$) and gain curve information is not readily available for IVS sessions. The T$_{\rm sys}$ and gain curve information had to be extracted from the respective session log files, and for some stations nominal values had to be used as no T$_{\rm sys}$ was available. An automated pipeline was used for self-calibration and imaging using the Caltech Difference Mapping software \citep[\textsc{difmap},][]{shepherd1997}.

\begin{figure*}[hbt!]
\centering
  \includegraphics[width=0.45\textwidth]{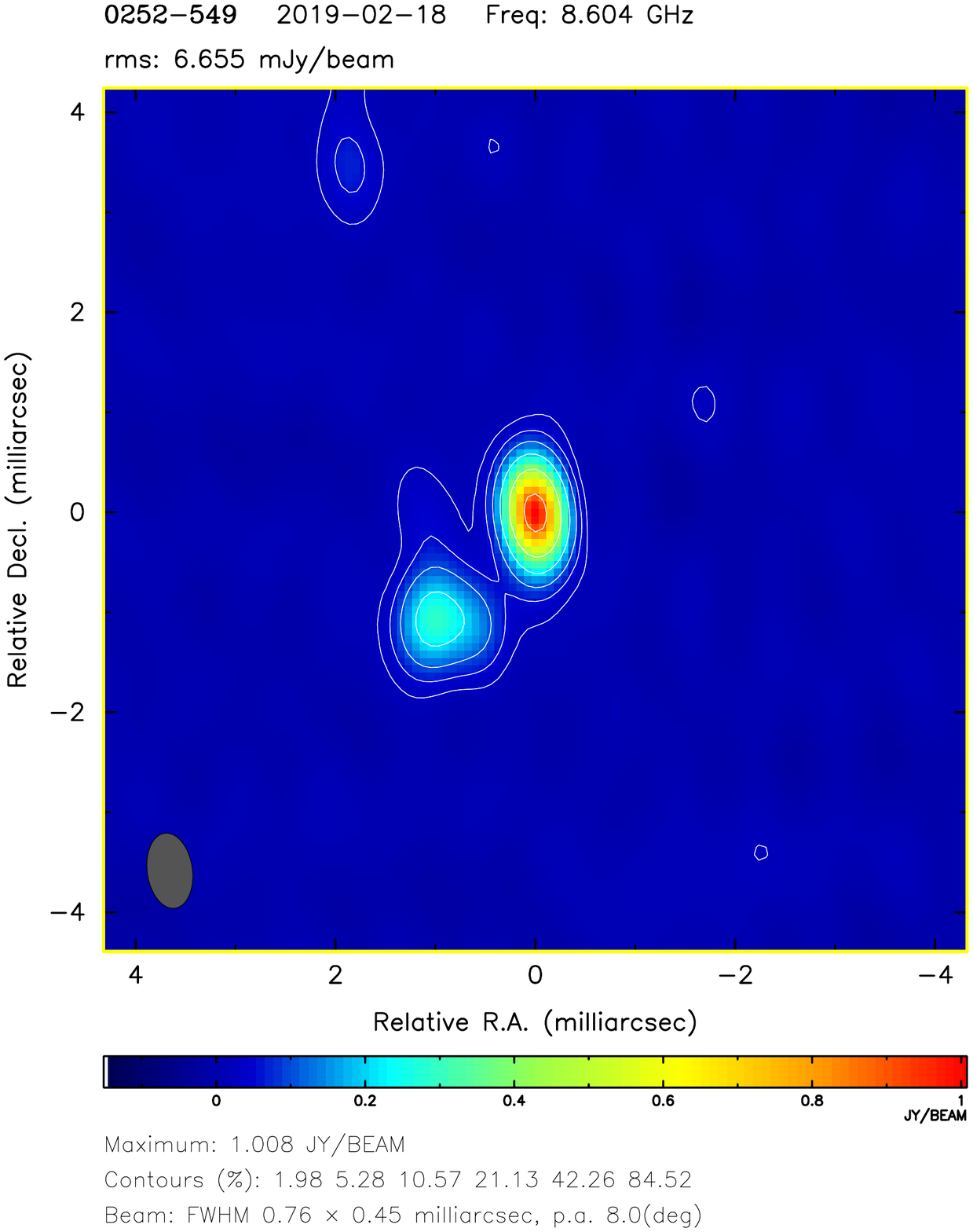}
  \hspace{0.1cm}
  \includegraphics[width=0.45\textwidth]{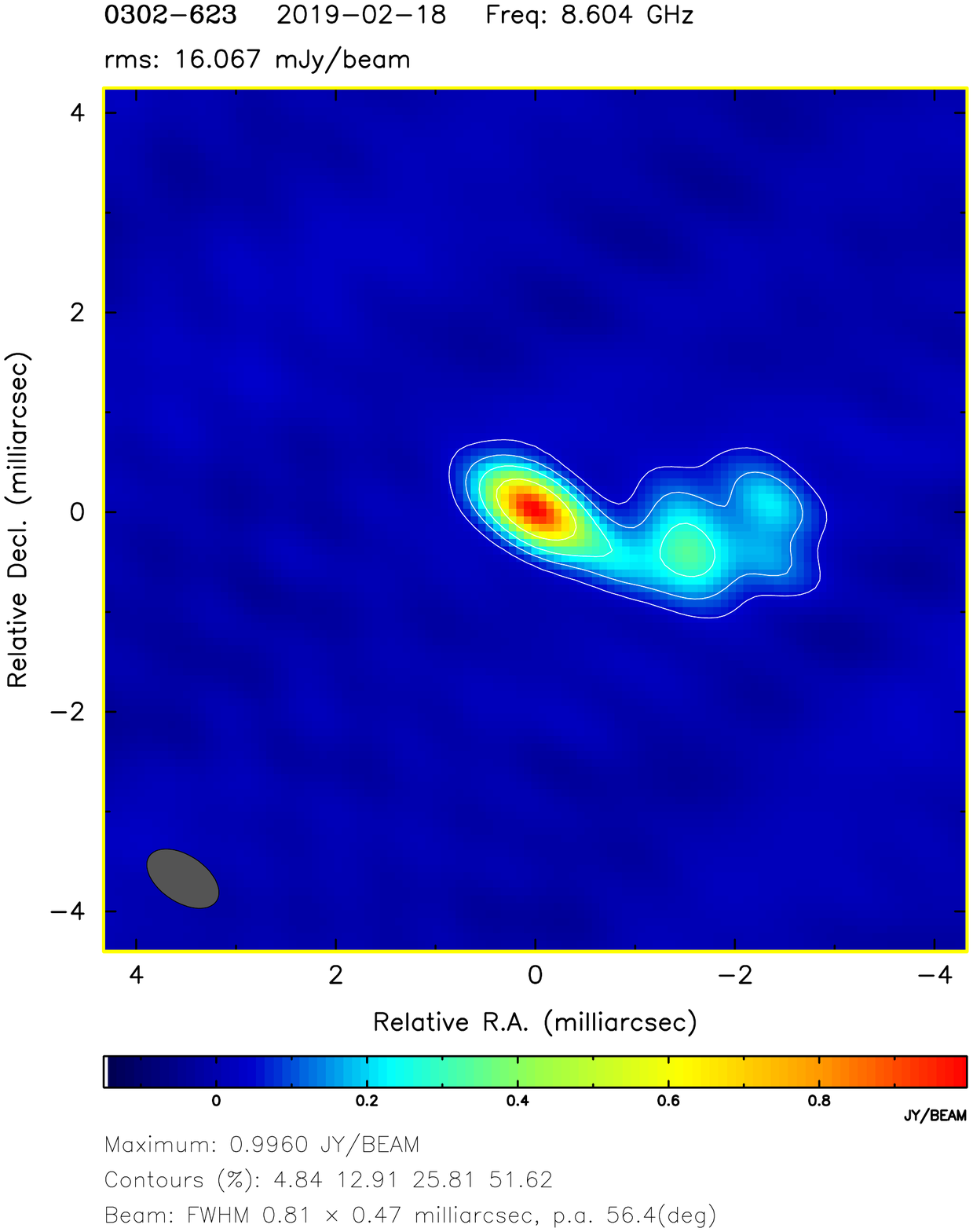}\\
  \vspace{0.2cm}
  \includegraphics[width=0.45\textwidth]{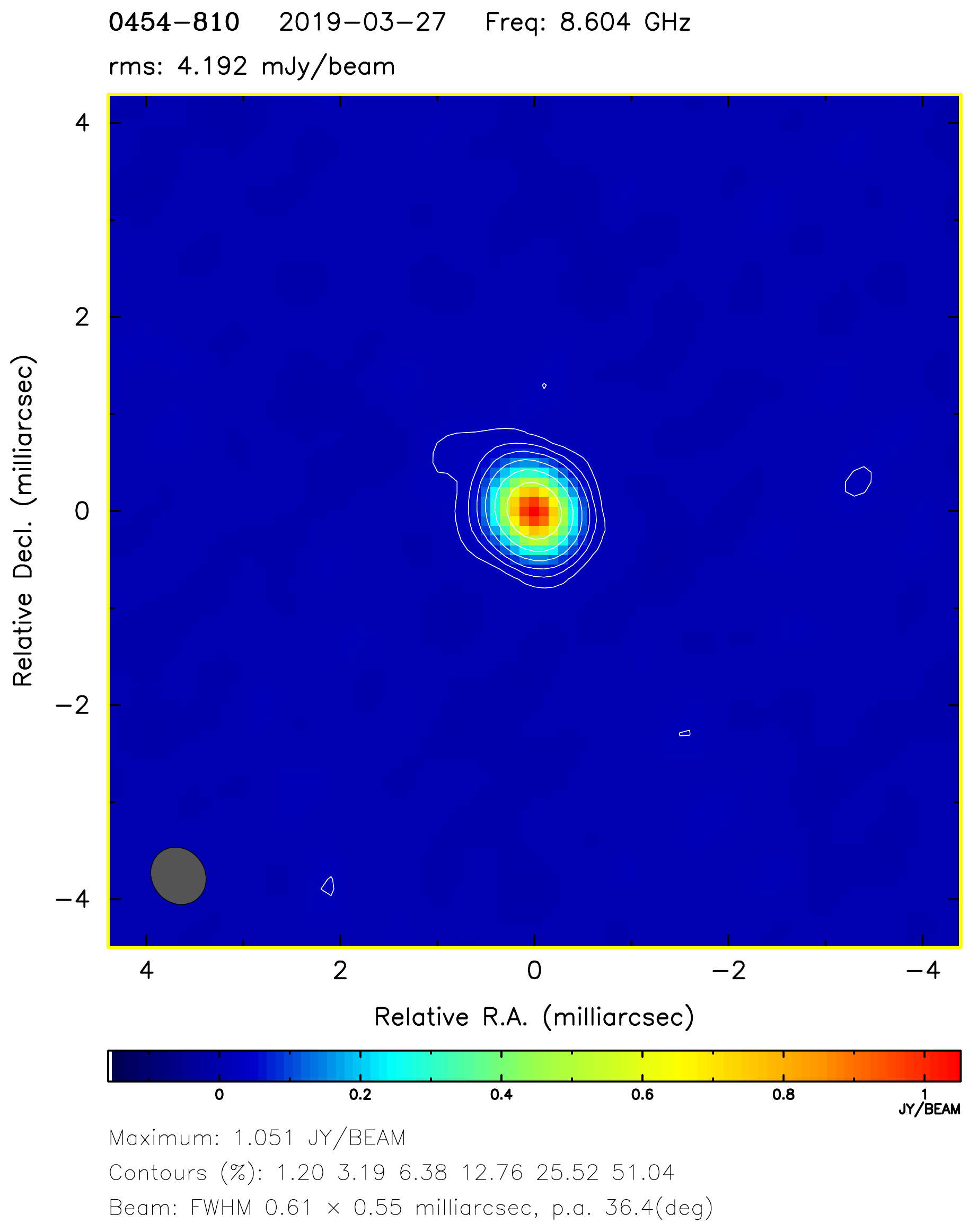}
  \hspace{0.1cm}
  \includegraphics[width=0.45\textwidth]{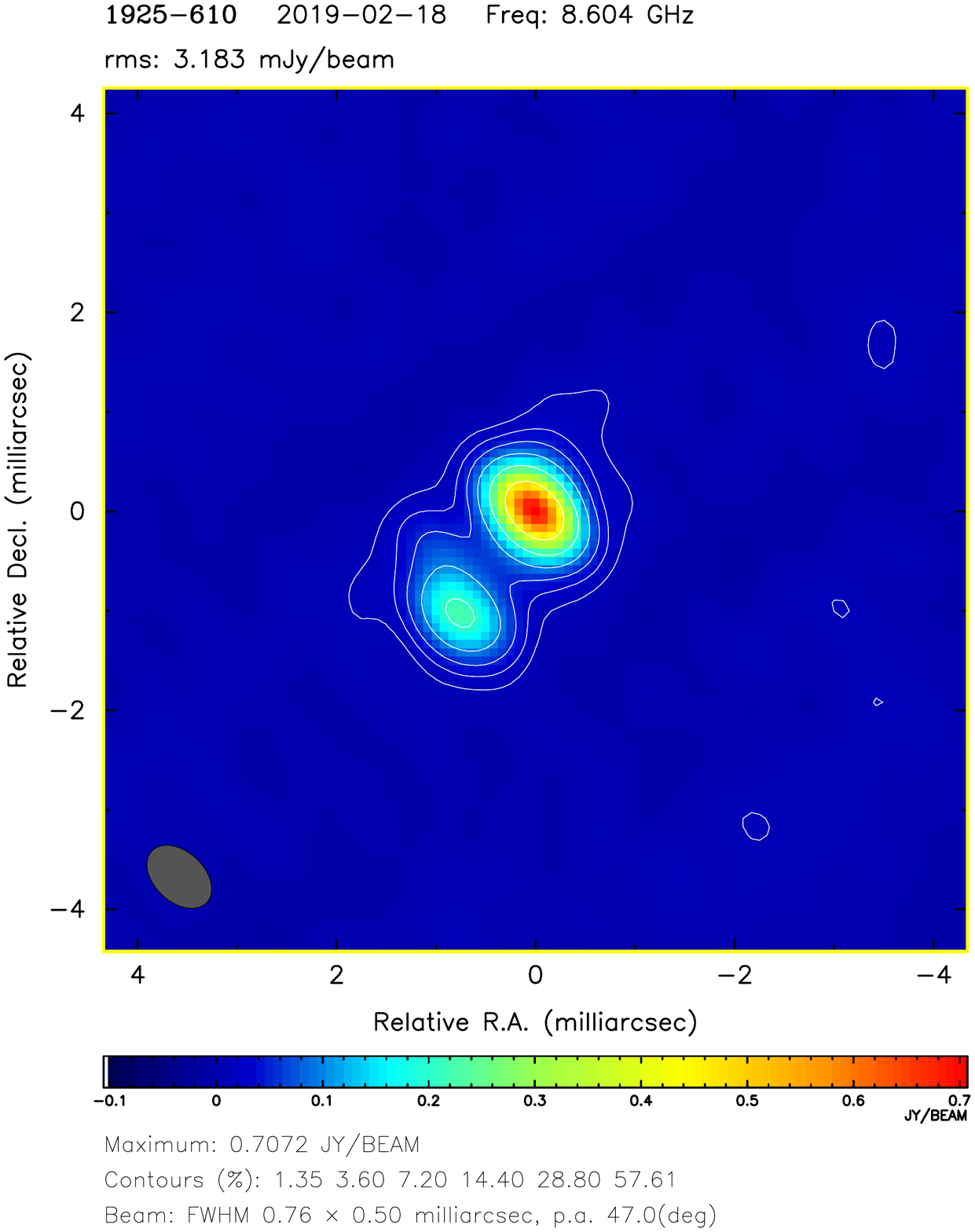}\\
\caption{Colour maps with contours for sources 0252-549 (top left), 0302-623 (top right), 0454-810 (bottom left) and 1925-610 (bottom right) at 8.6~GHz from CRDS sessions between 2018 and 2019. North is Up and East is to the Left. The Full Width Half Maximum (FWHM) beamsize is graphically indicated in the bottom left corner.}
\label{fig:crds-images}
\end{figure*}

A total of 185 sources were imaged from twelve CRDS sessions between 2013--2019 at both S- and X-band, with some sources imaged at multiple epochs. Table~\ref{table:images} lists the participating stations and the number of sources observed and imaged in each of the twelve sessions. Figure~\ref{fig:crds-images} shows images for a representative sample of four sources observed in CRDS sessions between 2018--2019. The source 0252-549 shows a bright second component in the CRDS image in Figure~\ref{fig:crds-images}. No images of this source are available from the Astrogeo Radio Fundamental Catalogue \citep[RFC,][]{Petrov2022}. Three of the sources, 0302-623, 0454-810 and 1925-610 are ICRF{\small 2} defining sources. The source 0302-623 shows bright extended emission in the CRDS image in Figure~\ref{fig:crds-images} and was removed as a defining source for the ICRF{\small 3}. There are no images of this source available in the Astrogeo RFC. The source 0454-810 is an ICRF{\small 3} defining source. The CRDS image in Figure~\ref{fig:crds-images} as well as images available from the Astrogeo RFC, all show a very compact structure for 0454-810. The source 1925-610 is an ICRF{\small 3} defining source and only one image, from 2010, is available from the Astrogeo RFC. The CRDS image in Figure~\ref{fig:crds-images} shows a bright second component and the status of 1925-610 as an ICRF{\small 3} defining source should be revisited in future.  

%\resizebox{\textwidth}{!}{\begin{tabular}{rlrrrrrrr}
\begin{table}[hbt!]
\footnotesize
\caption{The twelve CRDS sessions for which images have been produced.}
\resizebox{0.95\textwidth}{!}{
\begin{tabular}{l|c|c|c|c}
\toprule
Session  & Date & Participating & Sources & Sources \\
Name & yyyy-mm-dd & Stations & Observed & Imaged \\
\midrule
CRDS63 & 2013-01-14 & Hh–Ho–Ke–Yg & 38 & 34\\
CRDS66 & 2013-07-30 & Hb–Hh–Ho–Ke–Ww & 38 & 13\\
CRDS68 & 2013-11-27 & Hb–Hh–Ho–Ke–Ww–Yg & 38 & 37\\
CRDS94 & 2018-03-21 & Hh–Ht–Ke–Ww–Yg & 51 & 26\\
CRDS95 & 2018-05-07 & Hh–Ho–Ke–Ww–Yg & 39 & 38\\
CRDS96 & 2018-06-20 & Hh–Ho–Ke–Ww–Yg & 44 & 43\\
CRDS97 & 2018-08-14 & Hh–Ho–Ke–Ww–Yg & 52 & 52\\
CRDS98 & 2018-09-26 & Hh–Ho–Ke–Ww–Yg & 54 & 51\\
CRD100 & 2019-02-18 & Hh–Ho–Ke–Ww–Yg & 50 & 50\\
CRD101 & 2019-03-27 & Hh–Ke–Ww–Yg & 53 & 45\\
CRD102 & 2019-05-06 & Hh–Ho–Ke–Oh–Ww–Yg & 36 & 36\\
CRD103 & 2019-06-25 & Hh–Ho–Oh–Ww–Yg & 37 & 37\\
\bottomrule
\end{tabular}}
%\belowtable{} % Table Footnotes
\label{table:images}
\end{table}

%\FloatBarrier

\subsection{Performance}
\label{performance}

%Unfortunatly there is not have a complete set of analysis reports for the CRDS sessions, so the statistics presented here are not complete and are missing CRDS94,95 \& 100. 
%Table \ref{session_performance} summarises the data set used for this overview of the CRDS network performance: the first coumn is the session code, column 2 is the number of IVS stations scheduled,  column 3 is the number of scheduled observations in the session, column 4 is the number of observations that were able to be correlated, column 5 are the number of recoverable observations, column 6 are the number of observations actualy used for the analysis and column 7 is the post-fit residuals in picoseconds (ps) for the session.

%\input{table3.tex}

The performance of VLBI relies on a global network of antennas, owned and operated by many diverse organisations. The loss of any one antenna can have a huge knock-on effect by removing multiple baselines, which has a compounded impact on the observations available for analysis. It is important to note that for an array of $N$ antennas there will be $N(N-1)/2$ baselines. Thus for example taking an array of 5 antennas gives 10 baselines, losing one antenna reduces the number of baselines to 6; a 40 \% reduction. An antenna or multiples can be lost due to mechanical, electrical or other issues for a whole or part of a 24-hr session. 

\begin{figure*}[hbt!]%                 use [hb] only if necceccary!
  \centering
  \includegraphics[width=0.9\textwidth]{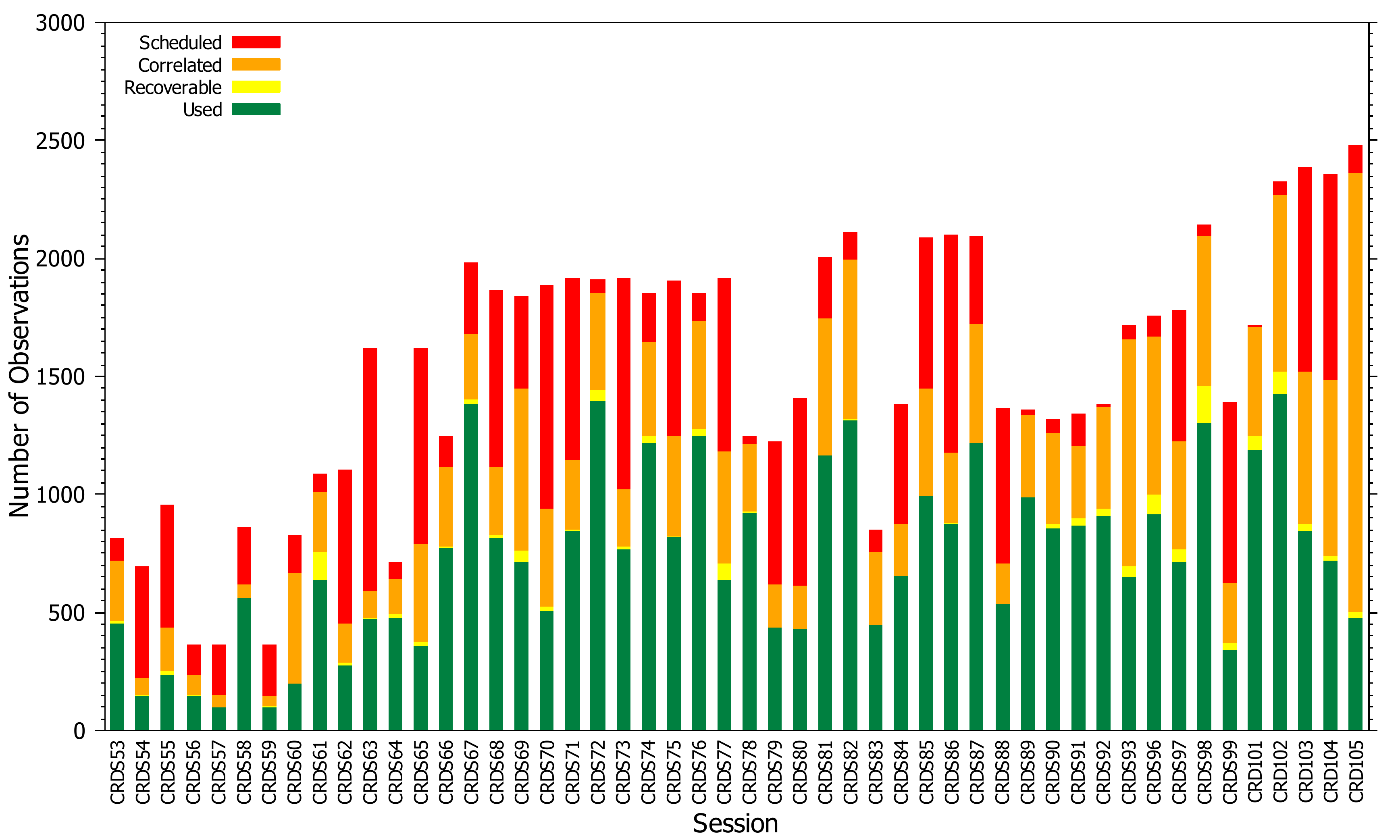}
  \caption{The per session statistics from the analysis reports ordered by time. In red is the total number of observations per session scheduled; in orange are those that were able to be correlated; in yellow is the number able to be recovered for analysis and in green are the number of observations used in the analysis solution.}
  \label{fig:sessions-statistics}
\end{figure*}

Even though scans within observations can be correlated, poor performance by one antenna for many reasons will also impact those actually used in the analysis. For example, scheduling an antenna with the wrong system equivalent flux density (SEFD) or a physical issue such as a cooled receiver having a fault and running hot/warm. A warm receiver will greatly reduce the sensitivity of an antenna. The resulting performance will be in some sense equivalent to the station having a cold receiver but observing for (typically) only one-third of the nominal time and therefore recording the equivalent of only one-third of the expected bits. Also, poor pointing can be converted into an equivalent lost sensitivity and then equivalent fraction of lost bits. RFI is becoming much more prevalent and an issue for many stations at S-Band (2-3 GHz), which also causes scans to be excluded by the correlator and be unavailable for the analysis. The per session statistics from the analysis reports ordered by session number are shown in Figure \ref{fig:sessions-statistics} and show the result of some of the above issues (A compilation of the CRDS53-116 session statistics are presented in \ref{Appendix_B}). In the figure and the appendix scheduled/correlated/recoverable/used observations are defined thus. Within a 24 hour IVS session multiple observations of sources with different baselines and stations are schedule, these observations are defined as \textbf{scheduled}. Of these observations not all are able to be correlated for various reasons, so the number able to be correlated is less than the number scheduled and is presented here as \textbf{Correlated}. Even though a scheduled observation is able to be correlated it might not be usable for analysis, so we have the number \textbf{Recoverable}. Then number \textbf{Used} are those able to be actually applied for the final analysis and data solution.

%In red is the total number of observations per session scheduled; in orange are those that were able to be correlated; in yellow is the number able to be recovered for analysis and in green are the number of observations used in the analysis solution.
%\input{sessions-statistics.tex}
% See plot-stations.gnu for how to generate this plot
% cd 'D:\OneDrive - AUT University\IVS\CRDS Data Paper\Earth System Science Data\db'
 
\begin{figure*}[hbt!]%                 use [hb] only if necceccary!
  \centering
  \includegraphics[width=0.9\textwidth]{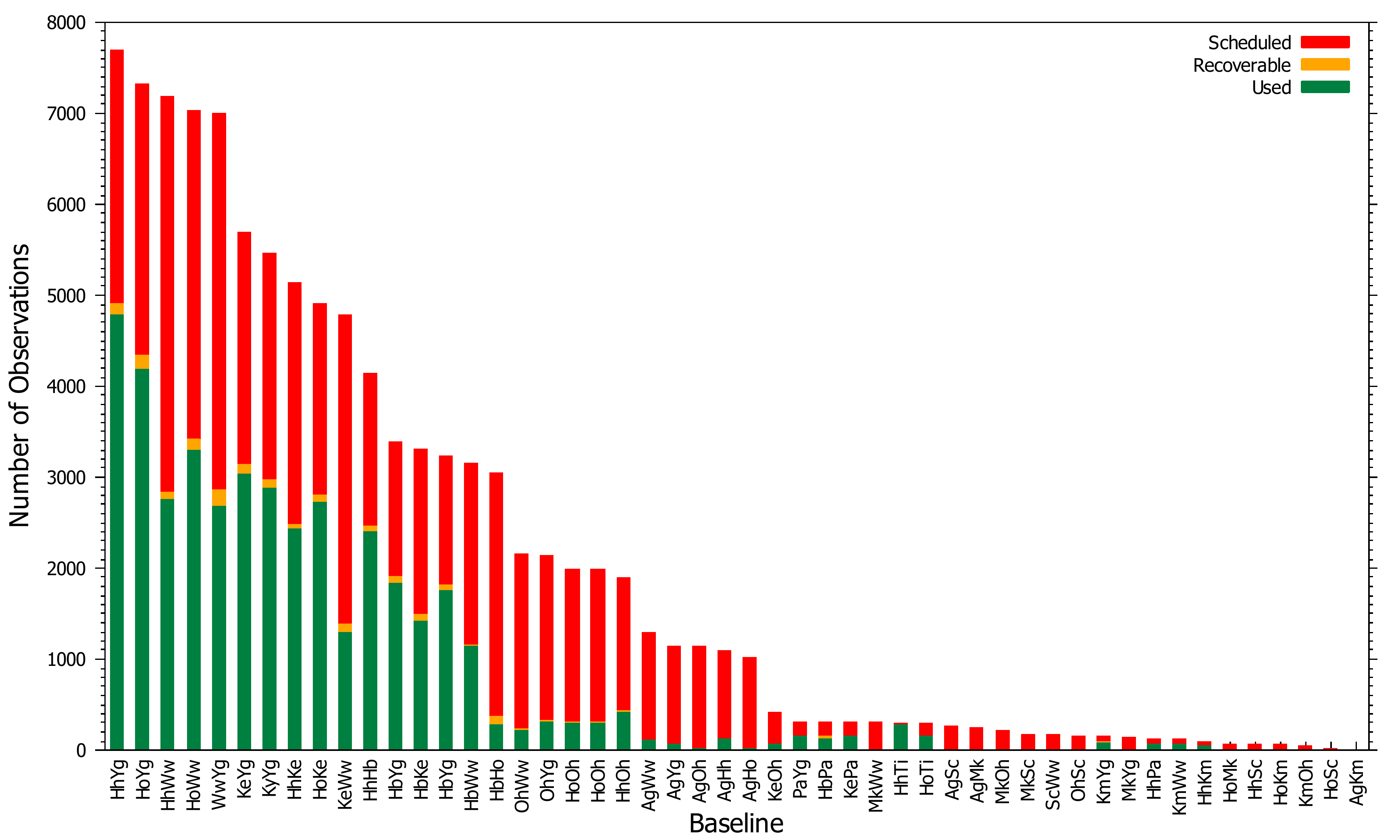}
  \caption{The number of observations made by each baseline from the analysis reports, ordered by the number of scheduled observations greatest to lowest from right to left. In red is the total number of observations scheduled; orange are those that were able to be recovered; and green are the number of observations used in the analysis solution.}
  \label{fig:baseline-performance-linear}
\end{figure*}

\begin{figure*}[hbt!]%                 use [hb] only if necceccary!
  \centering
  \includegraphics[width=0.9\textwidth]{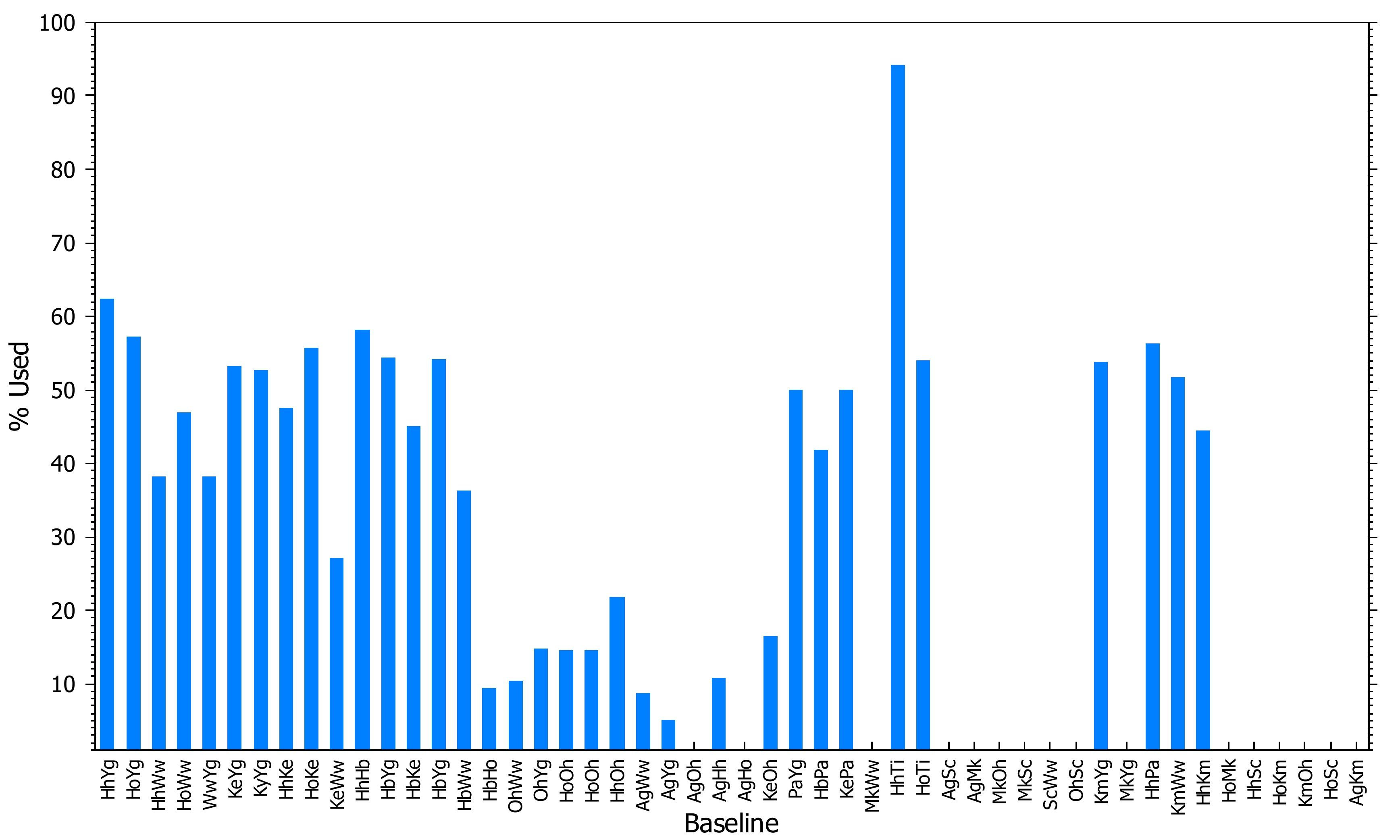}
  \caption{This figure shows the percentage of observations out of those scheduled which were used in an analysis solution by baseline. The order of the baselines is the same as in Figure \ref{fig:baseline-performance-linear} to aid comparing the two figures.}
  \label{fig:baseline-performance-percent-used}
\end{figure*}

One of the open problems, evident from the session statistics, is that only a small percentage of the observations from the CRDS sessions are actually used for analysis (on average $\approx 44$\%). However, closer inspection of the analysis reports shows that the main contributor to this high failure rate is the long baseline observations by the smaller antennas. The Ag and Oh antennas are both small dishes with large SEFDs ($> 10000$ Jy) and are mostly tagged-along in the CRDS schedules. Stations that are scheduled in tagged-along mode are not essential as scheduled for the analysis, but can provide additional data that might be useful. This is common practice for new stations or after a major station upgrade or change until the data quality has been assessed and is acceptable to become a fully participating network station. In Figure \ref{fig:baseline-performance-linear} the actual number of scans scheduled by baseline is shown. In Figure \ref{fig:baseline-performance-percent-used} there are the \% of scheduled versus used by baseline.

%\input{baseline-performance.tex}
% See db/baseline_performance.gnu for how to generate this plot
% D:\OneDrive - AUT University\IVS\CRDS Data Paper\Earth System Science Data\db
 
\begin{figure*}[hbt!]%                 use [hb] only if necceccary!
  \centering
   \includegraphics[width=0.9\textwidth]{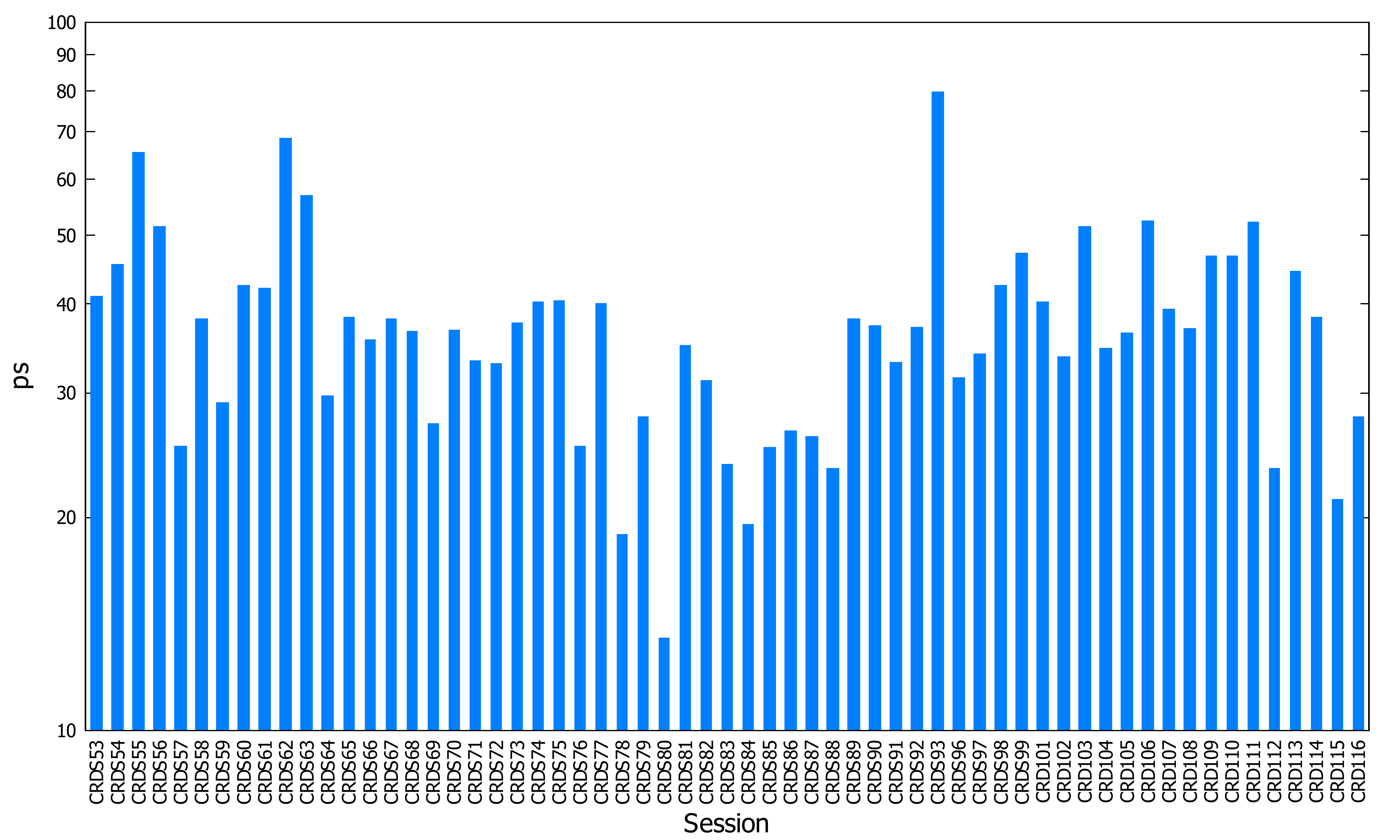}
  \caption{The post-fit residual delay rms values over all observations of a CRDS schedule, ordered by session reference code.}
  \label{fig:sessions-statistics-ps}
\end{figure*}

The post-fit residual delay rms values over all observations of the CRDS schedules are presented in Figure \ref{fig:sessions-statistics-ps}. We find a median session fit of 37 ps over all the CRDS sessions we have data for; this is comparable with the IVS R1/R4 observations with a median value of 34 ps \citep{Plank2017}.

%\input{sessions-statistics-ps.tex}
% See plot-stations.gnu for how to generate this plot
% cd 'D:\OneDrive - AUT University\IVS\CRDS Data Paper\Earth System Science Data\db'
 
%\FloatBarrier

%===================================================================
%\section{Conclusion}
\section{Discussion and Future Plans}  %% \conclusions[modified heading if necessary]
\label{DiscussionFuture}

The primary purpose of the CRDS program is to improve the celestial reference frame in the Southern Hemisphere.  Since January 2018, various improvements were made to CRDS sessions, e.g., the data rate was increased from 256 Mbps to 1 Gbps, the frequency sequence was optimised to avoid RFI, the scheduling was improved and optimised for both astrometry and imaging, and the source list was expanded. In addition, in 2021, three large antennas Km, Sc, and Mk, were added to the CRDS sessions to increase the number of stations and sensitivity of the existing network. However, the addition of these northern stations will only improve the results for sources down to about $-50^{\circ}$ Dec and not for sources in the Deep-South. Initially there were scheduling issues for Sc and Mk as the hardware setup was incorrect in the distributed VEX file, causing these stations to be unable to participate in CRD114, 115 and 116. Furthermore, scheduling of these CRDS sessions which include Northern antennas is problematic. In addition, the transitioning of the Australian 12-m telescopes to VGOS broad-band and the temporary removal of the Hobart 26-m telescope in Australia and then the Warkworth 12-m in New Zealand due to repairs, has had a major impact on the CRDS sessions and the CRF in the Deep-South.

For future CRDS observations into the next decade, we hope to increase the data of the CRDS sessions, further improving the sensitivity. For existing stations an increase in the data recording rate from 1 to 2 Gbps is possible.  An improvement in sensitivity can be used in two possible ways, one to observe sources for a shorter period and allowing for an increase in the number of sources and/or observations per session; second to add further weaker sources that will improve the coverage and density of sources in the Southern Hemisphere.

We anticipate the repair and return of the Ho and Ww antennas in the near future, and potentially adding additional southern stations to the CRDS network, such as the antennas forming part of the Long Baseline Array (LBA). However, most of the LBA antennas are not currently capable of dual-band S/X observations. Another possibility is adding one of the Tidbinbilla antennas in Australia to the CRDS network. However, the Tidbinbilla antennas are only available for a few hours per 24-hour session. The 34-m Tidbinbilla antennas may be available for longer, while time on the 70-m antenna is difficult to get and the slew rates are slow (0.25 deg/s). On the other hand, the Hb and Ke 12-m VGOS antennas are available for use in mixed-mode observations, and correlation of mixed-mode CRDS sessions needs to be investigated further.

We have also investigated the possibility of adding large sensitive antennas operating in single band (X) mode only, e.g., the 30 m Warkworth antenna in New Zealand. However, the possible shutdown of the Warkworth observatory may result in the loss of both the 12-m and 30-m Warkworth antennas. We are also investigating the possibility of including larger more sensitive stations in South America, The Zapala station is equipped with a 35-m antenna but there are no plans to install a wider S/X band receiver in the near future. There is also a plan to built a 40-m general purpose radio telescope in San Juan (China-Argentina Radio Telescope, CART). As of 2020, the antenna foundation has been completed, but the antenna delivery and installation was postponed due to the COVID-19 pandemic. This antenna, once completed, could be a promising candidate for CRF work.

It is expected that with the addition of new stations, better knowledge of sources such as flux density and source structure from imaging should lead to improved scheduling. As a result of these and other steps, further improvements to the ICRF will be obtained for the Southern Hemisphere in the years ahead.

\FloatBarrier
%===================================================================

\begin{acknowledgement}
We wish to acknowledge the International VLBI Service for Geodesy \& Astrometry (IVS) for providing the organizational infrastructure that was essential to this work. In addition we would like to recognise the many organisations that under the IVS umbrella provide the Network Stations, Operation Centres, Correlators, Data Centres , Analysts Centres and the Technology Development Centres. 

The research was carried out in part at the Jet Propulsion Laboratory, California Institute of Technology, under a contract with the National Aeronautics and Space Administration (80NM0018D0004).

From the 1st July 2023 operation of Warkworth will be transferred from AUT University to Space Operations New Zealand Ltd with new funding from Land Information New Zealand (LINZ).
\end{acknowledgement}

\paragraph{Data Availability Statement}

The final data products from this work are stored in three primary IVS Data Centers hosted at the Paris Observatory Data Center, OPAR  \citep{barache2021}; the Federal Agency for Cartography and Geodesy, Germany Data Center, BKG \citep{girdiuk2021}; and the GSFC's Crustal Dynamics Data Information System, CDDIS \citep{Michael2021}. 

The three primary IVS Data Centres mirror each other several times during the day to ensure common consistent holdings. Access is free for users and data can be accessed from the url's provided in the following list: 

\begin{itemize}
  \item[] \textbf{OPAR} using FTP at \url{ftp://ivsopar.obspm.fr}
  \item[] \textbf{BKG}   using FTP at \url{ftp://ivs.bkg.bund.de/pub/vlbi} and HTTP at \url{https://ivs.bkg.bund.de/data_dir/vlbi/}
  \item[] \textbf{GSFC} via the Crustal Dynamics Data Information System (CDDIS) at \url{https://cddis.nasa.gov/archive/vlbi/}
\end{itemize}

How to access the data and products is different for each data centre, and users will have to familiarise themselves with the methods and processes to extract the required data.

%\endnote in some journals will behave like \footnote; and \printendnotes will not output anything. 
\printendnotes

%===================================================================

\printbibliography

%===================================================================

\clearpage
\appendix

%\input{appendix_a.tex}
%\section{Appendix A : Source list for sessions CRDS50-116} 
%\input{supertabular}
%\input{crds50_116_sources_table}
\clearpage
%\onecolumn
\section{Source list for sessions CRDS50-116}    %% Appendix A
\label{Appendix_A}
This appendix contains a list of the 298 sources detected in the CRDS sessions between January 2011 and December 2021 (CRDS50--116). The format: column~1 is the IVS source name, columns~2 and 3 are the right ascension and declination coordinates (J2000), column~4 is the number of CRDS sessions that a source was scheduled in and column~5 is the total number of observations of that source in all the scheduled sessions. ICRF-3 defining sources are highlighted in boldface. 

%\begin{center}
\begin{table}[hbt!]
\begin{tabular}{|l|l|l|c|c|}
\toprule
\textbf{IVS Name} & \textbf{RA} &\textbf{Dec} &\textbf{$N_{\rm sess}$} &\textbf{$N_{\rm obs}$}\\
 & (h) (min) (s) & ($^{\circ}$) (') ('') & & \\ 
\midrule
0008-300 & 0 10 45.18 & -29 45 13.18 & 1 & 2 \\
0010-401 & 0 12 59.91 & -39 54 26.06 & 2 & 86 \\
0034-220 & 0 37 14.83 & -21 45 24.71 & 1 & 5 \\
\textbf{0035-252}&\textbf{0 38 14.74}&\textbf{-24 59 2.24}&\textbf{2}&\textbf{16} \\
0037-593 & 0 40 7.85 & -59 3 52.76 & 3 & 34 \\
0048-427 & 0 51 9.50 & -42 26 33.29 & 23 & 353 \\
0100-760 & 1 2 18.66 & -75 46 51.73 & 3 & 55 \\
0104-275 & 1 6 26.08 & -27 18 11.83 & 1 & 26 \\
\textbf{0104-408}&\textbf{1 6 45.11}&\textbf{-40 34 19.96}&\textbf{28}&\textbf{649} \\
\textbf{0107-610}&\textbf{1 9 15.48}&\textbf{-60 49 48.46}&\textbf{30}&\textbf{355} \\
0110-668 & 1 12 18.91 & -66 34 45.19 & 3 & 51 \\
0113-283 & 1 15 23.88 & -28 4 55.22 & 1 & 46 \\
0116-219 & 1 18 57.26 & -21 41 30.14 & 1 & 11 \\
0122-514 & 1 24 57.39 & -51 13 16.17 & 3 & 27 \\
0131-522 & 1 33 5.76 & -52 0 3.95 & 20 & 405 \\
0142-278 & 1 45 3.39 & -27 33 34.33 & 2 & 48 \\
0155-549 & 1 56 49.71 & -54 39 48.50 & 1 & 10 \\
0159-668 & 2 1 7.74 & -66 38 12.67 & 2 & 35 \\
0202-765 & 2 2 13.69 & -76 20 3.06 & 1 & 3 \\
0206-625 & 2 8 1.17 & -62 16 35.53 & 1 & 13 \\
0206-689 & 2 7 50.93 & -68 37 55.16 & 4 & 25 \\
\textbf{0208-512}&\textbf{2 10 46.20}&\textbf{-51 1 1.89}&\textbf{8}&\textbf{237} \\
0214-522 & 2 16 3.20 & -52 0 12.48 & 1 & 1 \\
0219-637 & 2 20 54.17 & -63 30 19.39 & 1 & 14 \\
\textbf{0227-369}&\textbf{2 29 28.45}&\textbf{-36 43 56.82}&\textbf{1}&\textbf{2} \\
0227-542 & 2 29 12.79 & -54 3 24.03 & 1 & 7 \\
0229-479 & 2 31 11.80 & -47 46 11.58 & 3 & 95 \\
\textbf{0230-790}&\textbf{2 29 34.95}&\textbf{-78 47 45.60}&\textbf{30}&\textbf{751} \\
\textbf{0234-301}&\textbf{2 36 31.17}&\textbf{-29 53 55.54}&\textbf{2}&\textbf{24} \\
\textbf{0235-618}&\textbf{2 36 53.25}&\textbf{-61 36 15.18}&\textbf{21}&\textbf{159} \\
0241-564 & 2 43 26.53 & -56 12 42.44 & 1 & 1 \\
0252-549 & 2 53 29.18 & -54 41 51.44 & 5 & 179 \\
0302-623 & 3 3 50.63 & -62 11 25.55 & 29 & 667 \\
\textbf{0308-611}&\textbf{3 9 56.10}&\textbf{-60 58 39.06}&\textbf{27}&\textbf{727} \\
0312-770 & 3 11 55.25 & -76 51 50.85 & 7 & 194 \\
\textbf{0332-403}&\textbf{3 34 13.65}&\textbf{-40 8 25.40}&\textbf{27}&\textbf{643} \\
0334-546 & 3 35 53.92 & -54 30 25.11 & 24 & 486 \\
\textbf{0346-279}&\textbf{3 48 38.14}&\textbf{-27 49 13.57}&\textbf{3}&\textbf{33} \\
\textbf{0347-211}&\textbf{3 49 57.83}&\textbf{-21 2 47.74}&\textbf{1}&\textbf{16} \\
0355-483 & 3 57 21.92 & -48 12 15.16 & 1 & 17 \\
\end{tabular}
\end{table}

\begin{table}[hbt!]
\begin{tabular}{|l|l|l|c|c|}
\textbf{0355-669}&\textbf{3 55 47.88}&\textbf{-66 45 33.82}&\textbf{6}&\textbf{71} \\
\textbf{0400-319}&\textbf{4 2 21.27}&\textbf{-31 47 25.95}&\textbf{3}&\textbf{64} \\
\textbf{0402-362}&\textbf{4 3 53.75}&\textbf{-36 5 1.91}&\textbf{24}&\textbf{480} \\
0405-385 & 4 6 59.04 & -38 26 28.04 & 21 & 542 \\
0431-512 & 4 32 21.18 & -51 9 25.19 & 1 & 5 \\
0432-606 & 4 33 34.11 & -60 30 13.77 & 2 & 29 \\
\textbf{0437-454}&\textbf{4 39 0.85}&\textbf{-45 22 22.56}&\textbf{23}&\textbf{574} \\
0440-520 & 4 41 58.28 & -51 54 54.17 & 1 & 13 \\
0441-699 & 4 40 47.76 & -69 52 18.09 & 2 & 72 \\
0450-743 & 4 48 48.56 & -74 17 31.25 & 1 & 25 \\
\textbf{0454-234}&\textbf{4 57 3.18}&\textbf{-23 24 52.02}&\textbf{4}&\textbf{63} \\
\textbf{0454-810}&\textbf{4 50 5.44}&\textbf{-81 1 2.23}&\textbf{35}&\textbf{1064} \\
\textbf{0506-612}&\textbf{5 6 43.99}&\textbf{-61 9 40.99}&\textbf{25}&\textbf{419} \\
0507-611 & 5 7 54.67 & -61 4 43.12 & 1 & 15 \\
0514-459 & 5 15 45.25 & -45 56 43.20 & 1 & 12 \\
0516-621 & 5 16 44.93 & -62 7 5.39 & 27 & 776 \\
0517-726 & 5 16 37.72 & -72 37 7.47 & 1 & 40 \\
0521-262 & 5 23 18.47 & -26 14 9.55 & 1 & 26 \\
\textbf{0522-611}&\textbf{5 22 34.43}&\textbf{-61 7 57.13}&\textbf{24}&\textbf{303} \\
0523-236 & 5 25 6.51 & -23 38 10.81 & 1 & 44 \\
0524-460 & 5 25 31.40 & -45 57 54.69 & 13 & 48 \\
\textbf{0524-485}&\textbf{5 26 16.67}&\textbf{-48 30 36.79}&\textbf{29}&\textbf{472} \\
\textbf{0530-727}&\textbf{5 29 30.04}&\textbf{-72 45 28.51}&\textbf{5}&\textbf{193} \\
\textbf{0534-340}&\textbf{5 36 28.43}&\textbf{-34 1 11.47}&\textbf{23}&\textbf{478} \\
0534-611 & 5 34 35.77 & -61 6 7.07 & 20 & 535 \\
\textbf{0537-286}&\textbf{5 39 54.28}&\textbf{-28 39 55.95}&\textbf{4}&\textbf{75} \\
0537-441 & 5 38 50.36 & -44 5 8.94 & 21 & 507 \\
0542-735 & 5 41 50.78 & -73 32 15.35 & 4 & 71 \\
0549-575 & 5 50 9.58 & -57 32 24.40 & 26 & 567 \\
0601-706 & 6 1 11.25 & -70 36 8.79 & 3 & 77 \\
0610-436 & 6 12 28.61 & -43 37 48.38 & 1 & 8 \\
0621-787 & 6 18 30.16 & -78 43 2.14 & 5 & 170 \\
0622-441 & 6 23 31.79 & -44 13 2.55 & 1 & 6 \\
0622-645 & 6 23 7.70 & -64 36 20.72 & 1 & 6 \\
0624-546 & 6 25 52.23 & -54 38 50.71 & 1 & 22 \\
\textbf{0627-199}&\textbf{6 29 23.76}&\textbf{-19 59 19.72}&\textbf{2}&\textbf{12} \\
0628-627 & 6 28 57.49 & -62 48 44.74 & 1 & 33 \\
0628-671 & 6 28 39.61 & -67 12 47.41 & 1 & 8 \\
0634-584 & 6 35 40.83 & -58 27 10.28 & 1 & 4 \\
\textbf{0642-349}&\textbf{6 44 25.28}&\textbf{-34 59 41.95}&\textbf{1}&\textbf{6} \\
\textbf{0646-306}&\textbf{6 48 14.10}&\textbf{-30 44 19.66}&\textbf{4}&\textbf{69} \\
\textbf{0700-197}&\textbf{7 2 42.90}&\textbf{-19 51 22.04}&\textbf{1}&\textbf{8} \\
\textbf{0700-465}&\textbf{7 1 34.55}&\textbf{-46 34 36.63}&\textbf{3}&\textbf{6} \\
\textbf{0738-674}&\textbf{7 38 56.50}&\textbf{-67 35 50.83}&\textbf{5}&\textbf{61} \\
\textbf{0742-562}&\textbf{7 43 20.49}&\textbf{-56 19 32.96}&\textbf{5}&\textbf{84} \\
0743-673 & 7 43 31.61 & -67 26 25.55 & 1 & 24 \\
0744-691 & 7 44 20.39 & -69 19 7.16 & 1 & 18 \\
\textbf{0804-267}&\textbf{8 6 12.72}&\textbf{-26 52 33.31}&\textbf{5}&\textbf{25} \\
\textbf{0809-493}&\textbf{8 11 8.80}&\textbf{-49 29 43.51}&\textbf{1}&\textbf{2} \\
0820-578 & 8 21 20.53 & -58 0 18.75 & 1 & 6 \\
0823-500 & 8 25 26.87 & -50 10 38.49 & 1 & 3 \\
\end{tabular}
\end{table}

\begin{table}[hbt!]
\begin{tabular}{|l|l|l|c|c|}
\textbf{0826-373}&\textbf{8 28 4.78}&\textbf{-37 31 6.28}&\textbf{4}&\textbf{105} \\
\textbf{0834-201}&\textbf{8 36 39.22}&\textbf{-20 16 59.50}&\textbf{6}&\textbf{52} \\
0842-754 & 8 41 27.04 & -75 40 27.87 & 1 & 8 \\
0843-547 & 8 45 2.48 & -54 58 8.54 & 4 & 47 \\
0848-588 & 8 49 14.88 & -59 5 0.77 & 1 & 4 \\
0851-577 & 8 52 38.73 & -57 55 29.81 & 2 & 98 \\
\textbf{0855-716}&\textbf{8 55 11.77}&\textbf{-71 49 6.46}&\textbf{3}&\textbf{35} \\
0858-279 & 9 0 40.04 & -28 8 20.34 & 1 & 11 \\
0903-573 & 9 4 53.18 & -57 35 5.78 & 4 & 138 \\
\textbf{0918-534}&\textbf{9 19 44.04}&\textbf{-53 40 6.45}&\textbf{4}&\textbf{22} \\
0920-397 & 9 22 46.42 & -39 59 35.07 & 23 & 584 \\
0936-853 & 9 30 32.57 & -85 33 59.70 & 4 & 54 \\
0944-469 & 9 46 51.34 & -47 7 59.23 & 2 & 26 \\
0956-409 & 9 58 38.30 & -41 10 33.17 & 1 & 19 \\
0959-443 & 10 1 59.91 & -44 38 0.60 & 1 & 8 \\
\textbf{1004-217}&\textbf{10 6 46.41}&\textbf{-21 59 20.41}&\textbf{1}&\textbf{4} \\
\textbf{1004-500}&\textbf{10 6 14.01}&\textbf{-50 18 13.47}&\textbf{23}&\textbf{534} \\
1005-739 & 10 6 4.15 & -74 9 44.09 & 1 & 29 \\
1012-448 & 10 14 50.35 & -45 8 41.15 & 7 & 117 \\
\textbf{1016-311}&\textbf{10 18 28.75}&\textbf{-31 23 53.85}&\textbf{1}&\textbf{5} \\
\textbf{1022-665}&\textbf{10 23 43.53}&\textbf{-66 46 48.72}&\textbf{25}&\textbf{664} \\
\textbf{1034-374}&\textbf{10 36 53.44}&\textbf{-37 44 15.07}&\textbf{24}&\textbf{541} \\
1036-431 & 10 38 14.70 & -43 25 45.89 & 1 & 14 \\
\textbf{1036-529}&\textbf{10 38 40.66}&\textbf{-53 11 43.27}&\textbf{5}&\textbf{126} \\
1039-474 & 10 41 44.65 & -47 40 0.07 & 8 & 253 \\
1041-452 & 10 43 51.42 & -45 30 9.56 & 1 & 8 \\
1045-188 & 10 48 6.62 & -19 9 35.73 & 1 & 12 \\
1045-620 & 10 47 42.95 & -62 17 14.63 & 5 & 126 \\
1048-470 & 10 50 53.40 & -47 19 4.55 & 1 & 9 \\
1048-526 & 10 50 38.03 & -52 49 48.33 & 2 & 56 \\
1049-534 & 10 51 9.10 & -53 44 46.54 & 1 & 51 \\
1059-631 & 11 1 54.38 & -63 25 22.60 & 4 & 69 \\
\textbf{1101-536}&\textbf{11 3 52.22}&\textbf{-53 57 0.70}&\textbf{26}&\textbf{573} \\
1105-680 & 11 7 12.70 & -68 20 50.73 & 2 & 40 \\
1109-567 & 11 12 7.27 & -57 3 39.75 & 5 & 56 \\
\textbf{1116-462}&\textbf{11 18 26.96}&\textbf{-46 34 15.00}&\textbf{6}&\textbf{195} \\
1117-270 & 11 20 16.19 & -27 19 6.37 & 1 & 40 \\
\textbf{1124-186}&\textbf{11 27 4.39}&\textbf{-18 57 17.44}&\textbf{1}&\textbf{9} \\
1129-580 & 11 31 43.29 & -58 18 53.44 & 8 & 216 \\
1130-741 & 11 32 19.11 & -74 25 9.02 & 1 & 1 \\
1133-681 & 11 36 2.10 & -68 27 5.82 & 2 & 80 \\
1133-739 & 11 36 9.66 & -74 15 45.27 & 2 & 15 \\
\textbf{1143-245}&\textbf{11 46 8.10}&\textbf{-24 47 32.90}&\textbf{1}&\textbf{7} \\
\textbf{1143-332}&\textbf{11 46 28.45}&\textbf{-33 28 42.63}&\textbf{1}&\textbf{4} \\
\textbf{1143-696}&\textbf{11 45 53.62}&\textbf{-69 54 1.80}&\textbf{26}&\textbf{641} \\
\textbf{1144-379}&\textbf{11 47 1.37}&\textbf{-38 12 11.02}&\textbf{27}&\textbf{824} \\
1148-671 & 11 51 13.43 & -67 28 11.09 & 4 & 136 \\
1151-348 & 11 54 21.79 & -35 5 29.06 & 1 & 11 \\
1156-663 & 11 59 18.31 & -66 35 39.43 & 22 & 124 \\
\end{tabular}
\end{table}

\begin{table}[hbt!]
\begin{tabular}{|l|l|l|c|c|}

1206-238 & 12 9 2.45 & -24 6 20.76 & 1 & 17 \\
1213-172 & 12 15 46.75 & -17 31 45.40 & 1 & 12 \\
1214-609 & 12 17 5.63 & -61 13 25.43 & 2 & 4 \\
1215-457 & 12 18 6.25 & -46 0 28.99 & 5 & 20 \\
1236-684 & 12 39 46.65 & -68 45 30.89 & 2 & 8 \\
\textbf{1244-255}&\textbf{12 46 46.80}&\textbf{-25 47 49.29}&\textbf{7}&\textbf{137} \\
\textbf{1245-454}&\textbf{12 48 28.50}&\textbf{-45 59 47.18}&\textbf{5}&\textbf{110} \\
1249-673 & 12 52 43.21 & -67 37 38.75 & 1 & 44 \\
\textbf{1251-713}&\textbf{12 54 59.92}&\textbf{-71 38 18.44}&\textbf{29}&\textbf{705} \\
1256-220 & 12 58 54.48 & -22 19 31.13 & 2 & 56 \\
1300-554 & 13 3 49.22 & -55 40 31.61 & 1 & 23 \\
\textbf{1306-395}&\textbf{13 9 48.49}&\textbf{-39 48 33.09}&\textbf{2}&\textbf{21} \\
1307-556 & 13 10 43.36 & -55 52 11.53 & 1 & 9 \\
\textbf{1312-533}&\textbf{13 15 4.18}&\textbf{-53 34 35.87}&\textbf{5}&\textbf{94} \\
\textbf{1313-333}&\textbf{13 16 7.99}&\textbf{-33 38 59.17}&\textbf{24}&\textbf{488} \\
1323-527 & 13 26 49.23 & -52 56 23.63 & 3 & 115 \\
\textbf{1325-558}&\textbf{13 29 1.14}&\textbf{-56 8 2.67}&\textbf{25}&\textbf{629} \\
1326-698 & 13 30 11.08 & -70 3 13.08 & 3 & 128 \\
1334-649 & 13 37 52.45 & -65 9 24.90 & 4 & 25 \\
1336-260 & 13 39 19.89 & -26 20 30.50 & 1 & 18 \\
1343-601 & 13 46 49.04 & -60 24 29.36 & 6 & 12 \\
1349-439 & 13 52 56.53 & -44 12 40.39 & 20 & 210 \\
1352-632 & 13 55 46.61 & -63 26 42.57 & 6 & 14 \\
\textbf{1406-267}&\textbf{14 9 50.17}&\textbf{-26 57 36.98}&\textbf{2}&\textbf{52} \\
\textbf{1412-368}&\textbf{14 15 26.02}&\textbf{-37 5 26.97}&\textbf{2}&\textbf{8} \\
1417-782 & 14 23 43.55 & -78 29 34.90 & 5 & 125 \\
\textbf{1420-679}&\textbf{14 24 55.56}&\textbf{-68 7 58.09}&\textbf{33}&\textbf{833} \\
1420-725 & 14 24 52.24 & -72 41 17.09 & 6 & 119 \\
\textbf{1424-418}&\textbf{14 27 56.30}&\textbf{-42 6 19.44}&\textbf{23}&\textbf{469} \\
\textbf{1435-218}&\textbf{14 38 9.47}&\textbf{-22 4 54.75}&\textbf{1}&\textbf{8} \\
\textbf{1448-648}&\textbf{14 52 39.68}&\textbf{-65 2 3.43}&\textbf{22}&\textbf{309} \\
\textbf{1451-400}&\textbf{14 54 32.91}&\textbf{-40 12 32.51}&\textbf{23}&\textbf{243} \\
1505-304 & 15 8 52.99 & -30 36 29.43 & 1 & 11 \\
1505-496 & 15 8 38.94 & -49 53 2.31 & 5 & 117 \\
1509-564 & 15 12 55.82 & -56 40 30.64 & 1 & 1 \\
\textbf{1511-476}&\textbf{15 14 40.02}&\textbf{-47 48 29.86}&\textbf{6}&\textbf{178} \\
\textbf{1511-558}&\textbf{15 15 12.67}&\textbf{-55 59 32.84}&\textbf{3}&\textbf{43} \\
\textbf{1519-273}&\textbf{15 22 37.68}&\textbf{-27 30 10.79}&\textbf{6}&\textbf{84} \\
1528-684 & 15 33 34.49 & -68 37 19.65 & 2 & 27 \\
1530-536 & 15 34 20.66 & -53 51 13.42 & 2 & 19 \\
1531-352 & 15 34 54.69 & -35 26 23.50 & 1 & 40 \\
1533-653 & 15 38 11.92 & -65 25 51.20 & 2 & 7 \\
1540-828 & 15 50 59.14 & -82 58 6.85 & 2 & 62 \\
1544-638 & 15 48 30.40 & -64 1 34.80 & 3 & 57 \\
1549-790 & 15 56 58.87 & -79 14 4.28 & 4 & 17 \\
1554-643 & 15 58 50.28 & -64 32 29.64 & 15 & 92 \\
\textbf{1556-245}&\textbf{15 59 41.41}&\textbf{-24 42 38.83}&\textbf{2}&\textbf{4} \\
1556-580 & 16 0 12.38 & -58 11 2.97 & 1 & 6 \\
\textbf{1600-445}&\textbf{16 4 31.02}&\textbf{-44 41 31.97}&\textbf{5}&\textbf{77} \\
\end{tabular}
\end{table}

\begin{table}[hbt!]
\begin{tabular}{|l|l|l|c|c|}
1600-489 & 16 3 50.68 & -49 4 5.51 & 1 & 13 \\
1604-333 & 16 7 34.76 & -33 31 8.91 & 13 & 79 \\
\textbf{1606-398}&\textbf{16 10 21.88}&\textbf{-39 58 58.33}&\textbf{5}&\textbf{61} \\
1611-710 & 16 16 30.64 & -71 8 31.45 & 18 & 178 \\
1618-399 & 16 21 59.69 & -40 3 34.49 & 1 & 4 \\
\textbf{1619-680}&\textbf{16 24 18.44}&\textbf{-68 9 12.50}&\textbf{26}&\textbf{596} \\
1622-253 & 16 25 46.89 & -25 27 38.33 & 1 & 6 \\
\textbf{1624-617}&\textbf{16 28 54.69}&\textbf{-61 52 36.40}&\textbf{31}&\textbf{644} \\
1633-810 & 16 42 57.35 & -81 8 35.07 & 14 & 125 \\
1637-771 & 16 44 16.12 & -77 15 48.81 & 1 & 10 \\
1642-645 & 16 47 37.74 & -64 38 0.27 & 2 & 59 \\
1646-506 & 16 50 16.63 & -50 44 48.21 & 7 & 68 \\
\textbf{1647-296}&\textbf{16 50 39.54}&\textbf{-29 43 46.95}&\textbf{5}&\textbf{159} \\
\textbf{1657-261}&\textbf{17 0 53.15}&\textbf{-26 10 51.73}&\textbf{3}&\textbf{68} \\
1657-562 & 17 1 44.86 & -56 21 55.90 & 25 & 495 \\
1659-621 & 17 3 36.54 & -62 12 40.01 & 27 & 572 \\
1710-269 & 17 13 31.28 & -26 58 52.53 & 1 & 21 \\
1717-618 & 17 21 39.02 & -61 54 43.02 & 1 & 3 \\
\textbf{1718-259}&\textbf{17 21 55.98}&\textbf{-25 58 40.69}&\textbf{1}&\textbf{1} \\
1722-554 & 17 26 49.63 & -55 29 40.46 & 1 & 3 \\
1722-644 & 17 26 57.83 & -64 27 52.71 & 2 & 8 \\
1725-795 & 17 33 40.70 & -79 35 55.72 & 19 & 257 \\
1729-373 & 17 33 15.19 & -37 22 32.40 & 2 & 10 \\
1732-593 & 17 37 19.67 & -59 21 41.89 & 1 & 6 \\
1740-517 & 17 44 25.45 & -51 44 43.74 & 5 & 25 \\
1758-651 & 18 3 23.50 & -65 7 36.76 & 20 & 378 \\
\textbf{1759-396}&\textbf{18 2 42.68}&\textbf{-39 40 7.91}&\textbf{4}&\textbf{106} \\
1803-642 & 18 7 54.03 & -64 13 50.11 & 4 & 104 \\
1804-502 & 18 8 13.84 & -50 11 53.61 & 2 & 52 \\
\textbf{1806-458}&\textbf{18 9 57.87}&\textbf{-45 52 41.01}&\textbf{26}&\textbf{390} \\
1814-637 & 18 19 35.00 & -63 45 48.21 & 1 & 9 \\
\textbf{1815-553}&\textbf{18 19 45.40}&\textbf{-55 21 20.75}&\textbf{24}&\textbf{412} \\
1824-582 & 18 29 12.40 & -58 13 55.16 & 19 & 319 \\
1828-733 & 18 34 53.20 & -73 15 14.33 & 3 & 9 \\
1829-207 & 18 32 11.05 & -20 39 48.20 & 1 & 4 \\
1830-589 & 18 34 27.47 & -58 56 36.27 & 1 & 8 \\
\textbf{1831-711}&\textbf{18 37 28.71}&\textbf{-71 8 43.55}&\textbf{35}&\textbf{779} \\
1852-534 & 18 57 0.45 & -53 25 0.38 & 2 & 10 \\
\textbf{1908-201}&\textbf{19 11 9.65}&\textbf{-20 6 55.11}&\textbf{4}&\textbf{42} \\
1918-634 & 19 23 24.61 & -63 20 45.77 & 1 & 2 \\
\textbf{1921-293}&\textbf{19 24 51.06}&\textbf{-29 14 30.12}&\textbf{8}&\textbf{112} \\
\textbf{1925-610}&\textbf{19 30 6.16}&\textbf{-60 56 9.18}&\textbf{29}&\textbf{704} \\
1928-698 & 19 33 31.16 & -69 42 58.91 & 6 & 136 \\
\textbf{1929-457}&\textbf{19 32 44.89}&\textbf{-45 36 37.93}&\textbf{9}&\textbf{67} \\
1933-400 & 19 37 16.22 & -39 58 1.55 & 20 & 541 \\
1934-638 & 19 39 25.03 & -63 42 45.62 & 2 & 6 \\
\textbf{1935-692}&\textbf{19 40 25.53}&\textbf{-69 7 56.97}&\textbf{31}&\textbf{427} \\
\end{tabular}
\end{table}

\begin{table}[hbt!]
\begin{tabular}{|l|l|l|c|c|}

1936-623 & 19 41 21.77 & -62 11 21.06 & 7 & 193 \\
1941-554 & 19 45 24.23 & -55 20 48.84 & 1 & 5 \\
1946-582 & 19 50 37.40 & -58 4 39.75 & 1 & 2 \\
1950-613 & 19 55 10.77 & -61 15 19.14 & 6 & 66 \\
1953-325 & 19 56 59.46 & -32 25 46.01 & 1 & 11 \\
\textbf{1954-388}&\textbf{19 57 59.82}&\textbf{-38 45 6.36}&\textbf{27}&\textbf{578} \\
1959-639 & 20 4 29.48 & -63 47 23.31 & 4 & 95 \\
\textbf{2002-375}&\textbf{20 5 55.07}&\textbf{-37 23 41.48}&\textbf{24}&\textbf{554} \\
2004-447 & 20 7 55.18 & -44 34 44.28 & 1 & 7 \\
\textbf{2008-159}&\textbf{20 11 15.71}&\textbf{-15 46 40.25}&\textbf{2}&\textbf{23} \\
2018-453 & 20 22 26.41 & -45 13 29.55 & 2 & 36 \\
2021-330 & 20 24 35.58 & -32 53 35.91 & 1 & 43 \\
2030-689 & 20 35 48.88 & -68 46 33.84 & 5 & 147 \\
\textbf{2037-253}&\textbf{20 40 8.77}&\textbf{-25 7 46.66}&\textbf{3}&\textbf{17} \\
\textbf{2052-474}&\textbf{20 56 16.36}&\textbf{-47 14 47.63}&\textbf{33}&\textbf{589} \\
2102-659 & 21 6 59.72 & -65 47 43.59 & 6 & 44 \\
2106-413 & 21 9 33.19 & -41 10 20.61 & 18 & 329 \\
2107-105 & 21 10 0.98 & -10 20 57.32 & 1 & 9 \\
\textbf{2109-811}&\textbf{21 16 30.85}&\textbf{-80 53 55.22}&\textbf{5}&\textbf{132} \\
2112-556 & 21 16 29.82 & -55 27 20.44 & 1 & 17 \\
2117-614 & 21 21 4.07 & -61 11 24.62 & 4 & 29 \\
2117-642 & 21 21 55.02 & -64 4 30.04 & 1 & 1 \\
2123-463 & 21 26 30.70 & -46 5 47.89 & 21 & 311 \\
2126-185 & 21 29 21.42 & -18 21 22.79 & 1 & 24 \\
2140-781 & 21 46 30.07 & -77 55 54.73 & 1 & 25 \\
\textbf{2142-758}&\textbf{21 47 12.73}&\textbf{-75 36 13.22}&\textbf{30}&\textbf{818} \\
2142-765 & 21 47 53.15 & -76 21 29.22 & 1 & 9 \\
2146-783 & 21 52 3.15 & -78 7 6.64 & 9 & 139 \\
2152-699 & 21 57 5.98 & -69 41 23.68 & 1 & 22 \\
2200-617 & 22 3 59.64 & -61 30 22.01 & 1 & 33 \\
2204-540 & 22 7 43.73 & -53 46 33.82 & 30 & 557 \\
2205-636 & 22 8 47.24 & -63 25 47.49 & 1 & 1 \\
\textbf{2210-257}&\textbf{22 13 2.50}&\textbf{-25 29 30.08}&\textbf{2}&\textbf{34} \\
2215-508 & 22 18 19.02 & -50 38 41.73 & 1 & 13 \\
\textbf{2220-351}&\textbf{22 23 5.93}&\textbf{-34 55 47.18}&\textbf{23}&\textbf{334} \\
2224-793 & 22 29 18.54 & -79 7 8.52 & 1 & 24 \\
2225-694 & 22 29 0.18 & -69 10 30.27 & 1 & 27 \\
2227-445 & 22 30 56.44 & -44 16 29.89 & 3 & 74 \\
\textbf{2232-488}&\textbf{22 35 13.24}&\textbf{-48 35 58.79}&\textbf{27}&\textbf{505} \\
\textbf{2236-572}&\textbf{22 39 12.08}&\textbf{-57 1 0.84}&\textbf{27}&\textbf{695} \\
2239-631 & 22 43 7.84 & -62 50 57.32 & 1 & 19 \\
2243-563 & 22 46 16.79 & -56 7 46.01 & 3 & 44 \\
\textbf{2244-372}&\textbf{22 47 3.92}&\textbf{-36 57 46.30}&\textbf{24}&\textbf{380} \\
\textbf{2245-328}&\textbf{22 48 38.69}&\textbf{-32 35 52.19}&\textbf{22}&\textbf{492} \\
2255-282 & 22 58 5.96 & -27 58 21.26 & 1 & 14 \\
2300-683 & 23 3 43.56 & -68 7 37.44 & 21 & 316 \\
2311-452 & 23 14 9.38 & -44 55 49.24 & 1 & 12 \\
\end{tabular}
\end{table}

\begin{table}[hbt!]
\begin{tabular}{|l|l|l|c|c|}
2311-477 & 23 13 51.90 & -47 29 11.73 & 1 & 3 \\
2318-087 & 23 21 18.25 & -8 27 21.52 & 1 & 6 \\
2321-065 & 23 23 39.11 & -6 17 59.24 & 1 & 4 \\
2326-477 & 23 29 17.70 & -47 30 19.11 & 25 & 443 \\
\textbf{2331-240}&\textbf{23 33 55.24}&\textbf{-23 43 40.66}&\textbf{2}&\textbf{14} \\
2333-415 & 23 36 33.99 & -41 15 21.98 & 24 & 276 \\
2333-528 & 23 36 12.14 & -52 36 21.95 & 7 & 131 \\
2344-514 & 23 47 19.86 & -51 10 36.07 & 20 & 279 \\
2345-500 & 23 47 43.69 & -49 46 27.88 & 1 & 5 \\
2351-154 & 23 54 30.20 & -15 13 11.21 & 1 & 8 \\
\textbf{2353-686}&\textbf{23 56 0.68}&\textbf{-68 20 3.47}&\textbf{27}&\textbf{648} \\
\textbf{2355-534}&\textbf{23 57 53.27}&\textbf{-53 11 13.69}&\textbf{33}&\textbf{600} \\
2357-318 & 23 59 35.49 & -31 33 43.82 & 20 & 308 \\
\bottomrule
\end{tabular}
\end{table}
%\end{center}

%\section{Appendix B : Session Summary Statistics for CRDS53-116} 
\clearpage
\onecolumn
\section{Session Summary Statistics for CRDS53-116}    
\label{Appendix_B}

Table \ref{tab:session_statistics} summarises the data set used for the overview of the CRDS network performance: the first column is the session code, column 2 is the start date of the session, Column 3 is the number of IVS stations scheduled and in brackets the number that actually observed or provided usable data,  column 4 is the number of scheduled observations in the session, column 5 is the number of observations that were able to be correlated, column 6 are the number of recoverable observations, column 7 are the number of observations actually used for the analysis and column 8 is the post-fit residuals in picoseconds (ps) for the session.

%\begin{center}
\begin{longtable}{|l|l|r|r|r|r|r|r|}
\caption{Session Summary Statistics for CRDS53-116. \label{tab:session_statistics}} \\
\hline
\textbf{Session}&\textbf{Start Date}&\textbf{Stations}&\textbf{Scheduled}&\textbf{Correlated}&\textbf{Recoverable}&\textbf{Obs}&\textbf{Fit}\\
\textbf{}&\textbf{YYYY.DOY}&\textbf{Scheduled}&\textbf{Obs}&\textbf{Obs}&\textbf{Obs}&\textbf{Used}&\textbf{ps}\\ \hline
\endfirsthead

\multicolumn{7}{c}%
{{\bfseries \tablename\ \thetable{} -- continued from previous page}} \\
\hline \textbf{Session}&\textbf{Start Date}&\textbf{Stations}&\textbf{Scheduled}&\textbf{Correlated}&\textbf{Recoverable}&\textbf{Obs}&\textbf{Fit}\\
\textbf{}&\textbf{YYYY.DOY}&\textbf{Scheduled}&\textbf{Obs}&\textbf{Obs}&\textbf{Obs}&\textbf{Used}&\textbf{ps}\\ \hline
\endhead

\hline \multicolumn{8}{|c|}{{Continued on next page}} \\ \hline
\endfoot

\hline \hline
\endlastfoot
CRDS53 & 2011.222 & 5 (5) &      812 &     718  &  462 &      449 & 40.9\\ 
CRDS54 & 2011.278 & 4 (3) &      696 &     217  &  149 &      142 & 45.5\\ 
CRDS55 & 2011.320 & 5 (4) &      956 &     431  &  249 &      231 & 65.3\\ 
CRDS56 & 2012.016 & 3 (3) &      360 &     230  &  148 &      145 & 51.4\\ 
CRDS57 & 2012.032 & 3 (3) &      360 &     146  &   93 &       92 & 25.1\\ 
CRDS58 & 2012.131 & 3 (3) &      861 &     619  &  559 &      557 & 38.0\\ 
CRDS59 & 2012.143 & 3 (3) &      363 &     144  &   98 &       97 & 29.0\\ 
CRDS60 & 2012.207 & 4 (3) &      822 &     665  &  197 &      193 & 42.4\\ 
CRDS61 & 2012.256 & 5 (5) &     1087 &    1007  &  755 &      635 & 42.1\\ 
CRDS62 & 2012.333 & 5 (4) &     1100 &     450  &  282 &      272 & 68.4\\ 
CRDS63 & 2013.014 & 6 (4) &     1616 &     587  &  477 &      470 & 56.9\\ 
CRDS64 & 2013.084 & 4 (4) &      714 &     639  &  490 &      475 & 29.6\\ 
CRDS65 & 2013.128 & 6 (5) &     1616 &     786  &  371 &      353 & 38.3\\ 
CRDS66 & 2013.211 & 5 (5) &     1245 &    1115  &  778 &      771 & 35.6\\ 
CRDS67 & 2013.245 & 6 (6) &     1978 &    1680  & 1402 &     1382 & 38.0\\ 
CRDS68 & 2013.331 & 6 (5) &     1864 &    1116  & 826  &      810 & 36.5\\ 
CRDS69 & 2014.043 & 6 (5) &     1840 &    1444  &  756 &      711 & 27.0\\ 
CRDS70 & 2014.097 & 6 (4) &     1884 &     936  &  523 &      506 & 36.7\\ 
CRDS71 & 2014.177 & 6 (6) &     1915 &    1144  &  849 &      839 & 33.2\\ 
CRDS72 & 2014.233 & 6 (6) &     1911 &    1847  & 1438 &     1393 & 32.9\\ 
CRDS73 & 2014.279 & 6 (6) &     1914 &    1018  &  774 &      762 & 37.5\\ 
CRDS74 & 2014.343 & 6 (6) &     1850 &    1641  & 1243 &     1215 & 40.3\\ 
CRDS75 & 2015.027 & 6 (5) &     1905 &    1245  &  821 &      820 & 40.4\\ 
CRDS76 & 2015.090 & 6 (6) &     1850 &    1728  & 1272 &     1247 & 25.1\\ 
CRDS77 & 2015.139 & 6 (5) &     1917 &    1180  &  704 &      632 & 40.0\\ 
CRDS78 & 2015.224 & 5 (5) &     1244 &    1208  &  922 &      921 & 18.8\\ 
CRDS79 & 2015.265 & 5 (4) &     1220 &     615  &  431 &      431 & 27.7\\ 
CRDS80 & 2015.285 & 5 (4) &     1403 &     610  &  425 &      425 & 13.4\\ 
CRDS81 & 2016.027 & 6 (6) &     2003 &    1740  & 1163 &     1161 & 34.9\\ 
CRDS82 & 2016.097 & 6 (6) &     2109 &    1992  & 1314 &     1312 & 31.1\\ 
CRDS83 & 2016.137 & 4 (4) &      846 &     751  &  442 &      442 & 23.7\\ 
CRDS84 & 2016.202 & 5 (4) &     1383 &     870  &  651 &      651 & 19.4\\ 
CRDS85 & 2016.271 & 6 (6) &     2089 &    1449  &  989 &      988 & 25.0\\ 
CRDS86 & 2016.328 & 6 (5) &     2101 &    1171  &  875 &      872 & 26.4\\ 
CRDS87 & 2017.094 & 6 (6) &     2092 &    1719  & 1214 &     1213 & 25.9\\ 
CRDS88 & 2017.137 & 5 (4) &     1362 &     707  &  531 &      531 & 23.4\\ 
CRDS89 & 2017.207 & 5 (5) &     1360 &    1335  &  984 &      984 & 38.1\\ 
CRDS90 & 2017.227 & 5 (5) &     1317 &    1259  &  872 &      852 & 37.2\\ 
CRDS91 & 2017.270 & 5 (5) &     1341 &    1203  &  897 &      866 & 33.0\\ 
CRDS92 & 2017.327 & 5 (5) &     1384 &    1369  &  937 &      905 & 37.0\\ 
CRDS93 & 2018.024 & 5 (5) &     1716 &    1653  &  696 &      644 & 79.8\\ 
CRDS96 & 2018.171 & 5 (5) &     1756 &    1667  &  995 &      914 & 31.4\\ 
CRDS97 & 2018.226 & 5 (5) &     1779 &    1221  &  767 &      713 & 34.0\\ 
CRDS98 & 2018.269 & 5 (5) &     2140 &    2090  & 1458 &     1299 & 42.4\\ 
CRDS99 & 2019.015 & 5 (4) &     1386 &     620  &  365 &      340 & 47.2\\ 
CRD101 & 2019.086 & 5 (5) &     1711 &    1707  & 1245 &     1186 & 40.3\\ 
CRD102 & 2019.126 & 6 (6) &     2322 &    2264  & 1518 &     1422 & 33.6\\ 
CRD103 & 2019.176 & 6 (6) &     2386 &    1517  &  874 &      844 & 51.4\\ 
CRD104 & 2019.218 & 6 (5) &     2356 &    1482  &  738 &      719 & 34.5\\ 
CRD105 & 2020.036 & 6 (6) &     2480 &    2362  &  500 &      477 & 36.4\\ 
CRD106 & 2020.126 & 6 (6) &     2698 &    1468  &  550 &      527 & 52.2\\ 
CRD107 & 2020.203 & 6 (6) &     2585 &    1063  &  388 &      369 & 39.3\\ 
CRD108 & 2020.217 & 6 (6) &     2798 &    1369  &  333 &      303 & 36.9\\ 
CRD109 & 2020.331 & 6 (6) &     2259 &    1838  &  464 &      381 & 46.6\\ 
CRD110 & 2020.344 & 6 (6) &     2340 &    1592  &  441 &      432 & 46.8\\ 
CRD111 & 2021.033 & 5 (5) &     1611 &    1435  &  860 &      838 & 52.1\\ 
CRD112 & 2021.124 & 6 (4) &     1929 &    1038  &  923 &      910 & 23.4\\ 
CRD113 & 2021.208 & 6 (5) &     1913 &    1038  &  653 &      637 & 44.5\\ 
CRD114 & 2021.230 & 7 (4) &     1665 &     404  &  194 &      186 & 38.2\\ 
CRD115 & 2021.286 & 9 (5) &     1715 &     394  &  232 &      227 & 21.1\\ 
CRD116 & 2021.335 & 8 (4) &     1414 &     564  &  292 &      272 & 27.7\\
\end{longtable}
%\end{center}

\twocolumn

%===================================================================

\end{document}